\documentclass[11pt]{article}
\usepackage{
    fullpage,
    graphicx,
    amsmath,
    amsfonts,
    amssymb,
    amsthm,
    authblk,
    bbold,
    dsfont,
    braket,
    bm
}

\usepackage[colorlinks]{hyperref}
\usepackage{comment}
\usepackage{orcidlink}

\usepackage[T1]{fontenc}

\usepackage{tocloft}
\DeclareMathOperator{\poly}{poly}
\DeclareMathOperator{\polylog}{polylog}

\hypersetup{
    linkcolor=blue,
    citecolor=blue,
    urlcolor=blue
}

\title{The Compendium of Super-Quadratic\\ Computational Quantum Advantages}

\author[1]{Philipp Kammerlander\thanks{\href{mailto:philipp.kammerlander@cern.ch}{\texttt{philipp.kammerlander@cern.ch}}}\,\orcidlink{0000-0002-1379-3092}\,}
\author[2]{Thomas E. O'Brien\thanks{\href{mailto:teobrien@google.com}{\texttt{teobrien@google.com}}}\,\orcidlink{0000-0002-8406-6626}\,}
\author[2,3,4]{Robbie King\,\orcidlink{0000-0002-8152-6340}\,}
\author[2]{Rolando D. Somma\,\orcidlink{0000-0003-4335-2607}\,}
\author[2,5]{M\'aria Kieferov\'a\,\orcidlink{0000-0002-0749-8126}\,}
\author[6]{Brad Lackey\,\orcidlink{0000-0002-3823-8757}\,}
\author[2,7,6]{Alexander Schmidhuber\,\orcidlink{0000-0002-2241-5480}\,}
\author[2,8]{Siddhartha Jain\,\orcidlink{0000-0003-2142-5801}\,}
\author[2,9]{Andi Gu\,\orcidlink{0000-0003-2748-7333}\,}
\author[10]{Yuri Alexeev\,\orcidlink{0000-0001-5066-2254}\,}
\author[11]{Shouvanik Chakrabarti\,\orcidlink{0000-0002-9159-4881}\,}
\author[11]{Ruslan Shaydulin\,\orcidlink{0000-0002-8657-2848}\,}
\author[12]{Pietro M. Posta\,\orcidlink{0009-0004-2997-5538}\,}
\author[2]{Guang Hao Low\,\orcidlink{0000-0002-6934-1052}\,}
\author[2]{Dar Gilboa}
\author[6,13,1]{Matthias Troyer\,\orcidlink{0000-0002-1469-9444}\,}
\author[14,1,15]{Barry C. Sanders\,\orcidlink{0000-0002-8326-8912}\,}
\author[12,6,1]{Matthias Christandl\,\orcidlink{0000-0003-2281-3355}\,}
\author[10]{Elica Kyoseva\,\orcidlink{0000-0002-9154-0293}\,}
\author[16,1]{Catherine Lefebvre\thanks{\href{mailto:catherine.lefebvre@gesda.global}{\texttt{catherine.lefebvre@gesda.global}}}\,\orcidlink{0000-0002-5363-4942}\,}
\author[2,1]{Ryan Babbush\thanks{\href{mailto:babbush@google.com}{\texttt{babbush@google.com}}}\,\orcidlink{0000-0001-6979-9533}\,}

\affil[1]
{\small{\it{Open Quantum Institute (OQI), Geneva, Switzerland}}}
\affil[2]
{\small{\it{Google Quantum AI, Venice, CA 90291, USA}}}
\affil[3]
{\small{\it{Simons Institute, Berkeley, CA 94720, USA}}}
\affil[4]
{\small{\it{Oratomic Inc., Pasadena, CA 91101, USA}}}
\affil[5]
{\small{\it{University of Technology Sydney, NSW 2007, Australia}}}
\affil[6]
{\small{\it{Microsoft, Redmond, WA 98052, USA}}}
\affil[7]
{\small{\it{Massachusetts Institute of Technology, Cambridge, MA 02139, USA}}}
\affil[8]
{\small{\it{The University of Texas at Austin, TX 78712, USA}}}
\affil[9]
{\small{\it{Harvard University, Cambridge, MA 02138, USA}}}
\affil[10]
{\small{\it{NVIDIA, Santa Clara, CA 95051, USA}}}
\affil[11]
{\small{\it{Global Technology Applied Research, JPMorganChase, New York, NY 10017, USA}}} 
\affil[12]
{\small{\it{University of Copenhagen, Copenhagen, Denmark}}}
\affil[13]
{\small{\it{Stellenbosch University, Stellenbosch, South Africa}}}
\affil[14]
{\small{\it{University of Calgary, Calgary, Alberta T2N 1N4, Canada}}}
\affil[15]
{\small{\it{CERN, European Organisation for Nuclear Research, Geneva, Switzerland}}}
\affil[16]
{\small{\it{Geneva Science and Diplomacy Anticipator (GESDA), Geneva, Switzerland}}}

\date{}

\begin{document}

\maketitle

\begin{abstract}
This compendium presents a structured collection of computational problems with credible prospects for realizing meaningful quantum advantage, focused on describing problem characteristics rather than specific algorithms. It is intended for non-quantum researchers and practitioners with strong computational backgrounds, particularly domain experts, seeking to assess the relevance of quantum methods to real-world problems. The compendium focuses on enumerating core problems for which there is good evidence that a quantum computer can realize more than a quadratic speedup in offline non-relativized time-complexity (e.g.,\ as opposed to in oracular settings, streaming models, or communication advantages). Entries characterize the structural constraints that real-world instances must satisfy to realize those advantages in practice. The intention is to provide a reference that can be leveraged to connect quantum algorithms to practical use cases.
\end{abstract}

\tableofcontents

\clearpage
\section{Introduction}
Quantum computing is undergoing steady and tangible progress, evolving from a predominantly theoretical research endeavor into a technology with credible prospects for real-world applications. Multiple hardware platforms have demonstrated increasing numbers of logical qubits, in some cases operating below error-correction thresholds~\cite{Google-Quantum-AI-and-Collaborators2024-dj}. As the field advances toward large-scale, fault-tolerant quantum computers, a parallel and equally important challenge emerges: determining how to most productively harness these emerging computational capabilities.

In parallel with hardware developments, the field of quantum algorithms has seen remarkable growth. Each year, researchers identify more computational tasks for which quantum computers could provide advantages over classical methods. As the pace of discovery accelerates, now is the time to connect these abstract algorithmic advances to practical, real-world problems.

Despite this progress, quantum algorithms remain a highly specialized domain. Developing new algorithms typically requires deep quantum computing expertise and understanding how to deploy existing quantum algorithms in practice demands substantial technical knowledge. At the same time, identifying suitable application domains requires insights that lie beyond the scope of quantum algorithm research. Domain experts often possess a detailed understanding of real-world problems and the limitations of state-of-the-art classical workflows, yet may not necessarily be equipped to assess whether and how quantum computing could offer improvements. The interdisciplinary technical knowledge gap between quantum algorithm experts and domain experts poses a significant barrier to connecting quantum advantages to real-world problems and advancing real-world applications that will benefit society and our planet.


\paragraph{Goal and Motivation}
This compendium aims to help bridge this gap by providing a structured and accessible collection of computational problems for which the scientific community has identified promising prospects for time-complexity quantum advantage. Rather than an exhaustive catalog of all possible quantum-advantaged problems, it is a curated collection highlighting problems where meaningful progress appears plausible.
Our goal is to enable non-specialists in quantum algorithms, particularly researchers and practitioners with strong computational backgrounds in specific domains, to engage with quantum approaches in a rigorous and productive manner. 
To this end, we describe core computational tasks in detail, emphasizing the structural properties that enable advantageous quantum algorithms and, in particular, the constraints that must be satisfied by a problem instance to benefit from this advantage, while avoiding unnecessary focus on the underlying algorithm mechanics. 
In this sense, our work complements existing resources such as the ``Quantum Algorithm Zoo''~\cite{QuantumAlgorithmZoo}, shifting the emphasis from cataloging algorithms to characterizing computational problems and their regimes of advantage. 
While this compendium assumes a solid computational background, it is designed to be accessible to non-quantum experts. 

The motivation for this compendium is to enable the search for quantum applications in computation through an algorithms-first approach~\cite{babbush2026grand}. A common strategy for identifying quantum applications is the problem-first approach, which begins with an important real-world challenge and then attempts to develop a quantum algorithm to solve it. In practice, this strategy is seldom successful because quantum advantages typically rely on highly specific mathematical structure that most real-world problems do not possess. 
Instead, we advocate an algorithms-first perspective: begin with computational tasks for which a quantum advantage is known, and then identify real-world problems that can be mapped onto those tasks. Accordingly, rather than starting from a domain-specific challenge in a particular field or industry, this compendium organizes known problems with known quantum-advantaged algorithmic primitives, enabling domain experts to recognize where these primitives may provide value in practice.

\paragraph{Scope and Focus of the Compendium}
We restrict our attention to quantum advantages in algorithmic time complexity within standard computational models, which we refer to as quantum speedups.
We do not attempt to survey quantum advantages in other computational resources, such as space~\cite{Zhao2026MassiveClassicalData} or communication~\cite{Raz1999-ic}.
Within this scope, a key working assumption is that practical quantum advantage will likely require a super-quadratic speedup~\cite{BMN+21}.
By a super-quadratic speedup, we mean an asymptotic separation for which, as the relevant problem-size parameter grows, the best known classical computational cost grows faster than the square of the corresponding quantum cost, up to qualifications such as logarithmic factors.
Thus, quadratic speedups, such as reducing a cost scaling as $N$ to one scaling as $\sqrt{N}$, are outside our scope, whereas cubic, quartic, super-polynomial, and exponential separations are included. Quadratic speedups can be mathematically significant, but are often unlikely to overcome the substantial overheads of early fault-tolerant quantum computation. Computational tasks admitting stronger speedups are considerably rarer but potentially much more consequential. We therefore focus on super-quadratic speedups relative to the best classical algorithms currently known, while recognizing that these classical baselines are moving targets.

Within this scope, a computational problem should be distinguished from a particular instance of that problem.
A problem specifies a family of instances of increasing size together with the required output; a decision problem is equivalently represented by a language, or, when restrictions are imposed on admissible inputs, by a promise problem. Computational complexity concerns how the resources required to solve such a family scale with instance size or with other explicitly identified parameters, rather than how long one algorithm takes on one particular instance. Unless stated otherwise, complexity results in this compendium refer to worst-case scaling. When an advantage instead holds in the average case, for a restricted or promised family of instances, or under a specified distribution of inputs, we make this qualification explicit. Such distinctions are essential for applications: a strong worst-case separation need not imply an advantage for practically relevant instances, whereas an average-case separation is meaningful only with respect to an input distribution that adequately represents those instances.

We likewise distinguish an asymptotic speedup from a direct finite-size performance advantage~\cite{Sanders2025}. An asymptotic speedup compares how classical and quantum computational costs scale as the relevant size parameters increase; it does not by itself imply that a quantum computer will outperform a classical computer at any particular problem size. Constant factors, input/output costs, state preparation, error correction, compilation, hardware clock rates, parallel classical computation, and other implementation overheads can determine the crossover point or prevent one from being reached at practically relevant sizes. Conversely, demonstrating that a quantum implementation is faster on a particular instance does not establish a super-quadratic speedup. Entries in this compendium therefore concern evidence for asymptotic super-quadratic advantage, while also identifying, where known, the structural and parameter regimes required for that scaling to be relevant in practice.

These distinctions concern computational complexity, but practical advantage additionally depends on end-to-end execution costs. In modern computing systems, the dominant cost is often not the computation itself but the movement of data. Reading data from storage, transferring information across memory hierarchies, and communicating between processors can require more time and resources than the arithmetic operations performed on that data. As a result, a large algorithmic speedup does not necessarily translate into an equivalent reduction in end-to-end execution time when the overall cost is dominated by input/output (I/O) or communication complexity. Similar considerations apply to quantum computing. While complexity analyses often focus on the computational core of a quantum algorithm, practical performance can be limited by the costs of state preparation, oracle construction, data encoding, and measurement~\cite{Hoefler2023PracticalQuantumAdvantage}. 

Throughout this compendium, computational speedups should therefore be understood in the context of the assumptions made about data access and I/O complexity. Quantum advantage ultimately requires that these costs do not overwhelm the underlying algorithmic improvement. Linear algebra provides a useful example. Many quantum algorithms for solving linear systems offer exponential or super-polynomial improvements in computational complexity only when the input data can be loaded in polylogarithmic time through highly structured encodings. For arbitrary data, where loading the input requires time linear in the data size, the asymptotic advantage may disappear. Similar considerations apply to the output. To retain a quantum advantage, the desired result must be extractable efficiently from the quantum computation; if recovering the answer requires expensive procedures such as full quantum state tomography, much of the potential speedup can be lost.

In assessing evidence for such speedups, we distinguish algorithms from heuristics, also known as heuristic algorithms.
An algorithm for a well-posed computational problem comes with a specified procedure and a guarantee,
deterministic or probabilistic,
relating its output, accuracy, success probability, and computational cost. A heuristic is a procedure that may perform well empirically, including for difficult search, optimization, or learning tasks, but lacks a corresponding general performance guarantee. This distinction matters when claiming quantum advantage.
A comparison between quantum and classical algorithms can support a complexity-theoretic speedup when their solution criteria are matched. Comparisons involving heuristics require greater care: empirical superiority on selected instances, or at a particular problem size, does not establish an asymptotic quantum speedup.
We therefore include heuristic methods only when the classical and quantum procedures can be compared under clearly specified and commensurate criteria for solution quality, success probability, instance distribution, and computational cost.

An additional focus of this work is on problems that are efficiently verifiable, including those for which candidate solutions can be checked with high confidence using classical or quantum resources. Verifiability is both practically and methodologically important because it enables benchmarking, validation, and deployment, and we regard some efficient means of assessing the quality of an output as a necessary condition for real-world usefulness. If the quality of a computation's output cannot be efficiently assessed, then establishing that the output produces measurable value in practice is problematic. Conversely, if the output can be replaced by an arbitrary or approximate alternative with no efficiently detectable loss in performance, then the computation itself is unlikely to be essential. Efficient classical verification provides the strongest standard, but, for applications such as quantum simulation, we also include problems whose outputs can be verified using another quantum computation or, ultimately, through comparison with experimental observations of nature.

To further clarify the scope, we exclude oracle-based separations and other quantum-advantage results for which no efficient instantiation of the oracle is known. Such results are invaluable for understanding the fundamental capabilities and limitations of quantum computation, but, without an explicit realization of the oracle in terms of efficiently computable operations or data structures, the theoretical separation cannot readily be translated into a useful computational advantage. Accordingly, this compendium focuses on algorithmic primitives whose assumptions are sufficiently concrete to support the search for practical applications. This criterion excludes several elegant and influential oracle problems and their corresponding quantum algorithms, including the welded-trees problem~\cite{Childs2003}, Forrelation~\cite{aaronson2015forrelation}, Simon's problem~\cite{simon}, and the Bernstein--Vazirani problem~\cite{Bernstein1997a}. These results are of fundamental importance in quantum complexity theory but fall outside the scope of this compendium because no efficient realization of the required oracle relevant to the intended applications is currently known.

\paragraph{The Stages Framework}
In this compendium, we adopt the framework presented by the Google Quantum AI team, which categorized the development of quantum applications into multiple stages~\cite{babbush2026grand}. Our focus is on problems that allow for algorithms known to be in one of the two first stages, namely: 
\begin{itemize}
\item {Stage I}: Primitives with Evidence for Quantum Advantage.
Algorithms in Stage I are quantum computational primitives with evidence for how well they perform. This entails the discovery of new quantum algorithms in an abstract setting, typically with a focus on worst-case asymptotic complexity, concerned with the scaling of the algorithm for the most difficult possible instance.
\item {Stage II}: Concrete Instances Exhibiting Quantum Advantage.
Quantum algorithms advance to Stage II when they allow for efficient generation of specific problem instances where a quantum computer is expected to outperform the best known classical algorithms run on the best classical infrastructure. Typically this entails two parts: on the one hand the classical problem instances need to be shown to be hard; on the other hand the quantum algorithm should give a super-quadratic speedup on average for these instances (or one should at least have a sense of how to generate hard instances with high probability). Importantly, worst-case speedups alone are no longer sufficient for algorithms at Stage II. 
\end{itemize}

In this compendium, we classify entries according to the Stage I or Stage II result discussed here. This does not imply that no corresponding Stage III or more advanced result exists for the same problem family or application area; such developments may lie outside the scope of the particular result summarized in the entry.

Stage III, as discussed in that perspective~\cite{babbush2026grand}, occurs when one matches Stage II instances to real-world problems such that the conditions for quantum advantage are met. For example, the discovery of a quantum algorithm for simulating molecules would be a Stage I result. A Stage II result would involve describing some set of molecules for which the quantum algorithm is efficient and the best known classical algorithms are not. A Stage III result would identify instances of those Stage II molecules for which the simulation actually helps to better understand open chemical problems or develop a new technology.

The purpose of this compendium is thus to support advancing algorithms from Stage I to Stage II, and to assist with Stage III research and development. In particular, the latter requires not only identifying instances on which quantum algorithms outperform classical methods, but also understanding when those instances arise in problems of genuine scientific or technological interest. 
As quantum computing continues to mature, it is increasingly important to focus on the transition from provable or conjectured quantum advantage to applications in which that advantage can have meaningful impact. 

The molecular simulation example above is particularly relevant because quantum simulation is widely expected to provide substantial benefits in chemistry and materials science~\cite{bauer2020quantum}.
More broadly, other areas already illustrate this progression. 
In cryptanalysis, quantum algorithms such as Shor’s algorithm have established clear and transformative advantages, motivating large-scale efforts toward post-quantum security~\cite{NISTPQC}. 
These examples highlight the broader challenge addressed by this compendium: moving from identifying problems for which quantum algorithms offer an advantage to characterizing, finding, and ultimately exploiting the concrete instances in which that advantage matters.

We present Stage II results before Stage I results. This ordering reflects the emphasis of the compendium on concrete regimes in which quantum advantage may be realized: Stage II results are closer to this goal and therefore provide the natural starting point. The subsequent Stage I entries collect promising primitives for which identifying suitable hard instance families remains an important next step.

\paragraph{Related Initiatives}
Finally, a growing ecosystem of initiatives---including large-scale chal\-lenges like XPRIZE~\cite{XPRIZE}, global collaborative platforms like the Open Quantum Institute~\cite{OQI}, and incentive-driven programs such as Q4Bio~\cite{Q4Bio}, as well as other quantum-focused consortia, such as IBM-led Quantum Working Groups~\cite{IBMWG}---reflects a broader recognition of the importance of bridging the gap between theory and application. This compendium is intended to contribute to this col\-lect\-ive effort by providing a common reference point for researchers and practitioners across disciplines.

\paragraph{Structure of the Compendium}
The compendium is divided into two main parts: Part \ref{Part A} starts with the most mature {Stage II} problems that offer concrete instances where quantum advantage could be realized, while Part \ref{Part B} covers select {Stage I} problems that still lie in the discovery, design and analysis stage, with evidence for how the quantum algorithms would perform.

For each selected problem, we provide a succinct \textit{Problem Description} along with \textit{Potential Application Spaces}. We summarize both the \textit{Classical and Quantum Complexities}, and then define the constraints to be satisfied in order to realize a quantum advantage, i.e.,\ the \textit{Regimes of Advantage}. Some entries are concluded with \textit{Notes on Future Directions and Related Advantages}.

Each entry begins with a standardized summary of the \emph{Maximum Quantum Speedup}. This denotes the strongest speedup identified in the current literature against classical baselines for at least one specified regime of the problem. It is not intended to apply uniformly to all instances or parameter choices: depending on the input and access model, required precision and output, problem parameters, and classical comparator, the speedup may be smaller or may disappear entirely. The assumptions and limitations underlying the label are developed in the discussions of \emph{Classical Complexity}, \emph{Quantum Complexity}, and, in particular, \emph{Regimes of Advantage}. The one-word label should therefore be read only as a compact signpost for the more detailed assessment that follows.

Each entry also states the \emph{Evidence Basis} for the claimed speedup. Since a speedup claim depends both on the support for the quantum performance and on the basis of the corresponding classical comparison, this label gives a concise indication of where the main evidence comes from. Its purpose is to help readers distinguish fundamentally different kinds of support at a glance, without reducing them to a common ranking. The label is necessarily a summary and the underlying results, assumptions, and limitations are discussed within the entry.

Neither label is intended to support direct comparisons across entries. 
The relevant quantum complexities, classical baselines, resource measures, and strength of evidence may differ substantially between problems. Depending on the entry, the claimed advantage may rest on rigorous complexity-theoretic results, comparisons with the best-known classical algorithms, average-case or restricted-model evidence, or numerical or heuristic studies. All assessments reflect the literature and the best quantum and classical methods known to us at the time of writing. In particular, classical baselines are moving targets, and comparison with a state-of-the-art classical algorithm should not be interpreted as a classical lower bound unless explicitly stated otherwise.

\paragraph{An Important Note on How to Use the Compendium}
This document is not a regular paper intended to be read with a single thread from beginning to end. Rather, it is a collection of selected problems with concrete instances where quantum computers could outperform state-of-the-art classical computers. Readers interested in a given problem are invited to directly go to the relevant section for concise and structured information on how to tackle it with a specific quantum approach, and consult the references for additional details.

We hope that this work will catalyze further applied research effort within the scientific community, particularly accelerating progress on {Stage II} problems and advancing them toward greater maturity, ultimately enabling impactful quantum applications for society and the planet.

\newpage

\part{Stage II Problems: Concrete Instances Exhibiting Quantum Advantage}\label{Part A}

The problems selected in this Part \ref{Part A} concern concrete instance families for which quantum computers could outperform state-of-the-art classical algorithms. These instances are chosen in part for their connection to structures arising in real-world settings. Such connections motivate potential application spaces, but do not necessarily constitute established end-to-end applications of the corresponding quantum algorithms.
To achieve a quantum advantage in real-world applications, it is essential that the specific constraints outlined in the following sections are strictly met.


\pagebreak
\section{Hidden Subgroup Problem}\label{sec:hidden_subgroup}

\emph{Section contributed by:} Brad Lackey, Robbie King, M\'aria Kieferov\'a, Philipp Kammerlander\\

Let $G$ be a group, and let $H \leq G$ be a ``hidden'' subgroup. The hidden subgroup problem (HSP) asks us to recover a generating set for $H$ given oracle access to a function $f:G \to X$ that is constant on each left coset of $H$ and distinct on different cosets. 

In the standard quantum formulation the oracle is the unitary
\[
U_f:\; \ket{g}\ket{y}\;\mapsto\;\ket{g}\ket{y\oplus f(g)},
\]
where the second register stores an injective encoding of the function value. Starting from the uniform superposition $|G|^{-1/2}\sum_{g\in G}\ket{g}\ket{0}$ and applying $U_f$ produces the entangled state
\[
\frac{1}{\sqrt{|G|}}\sum_{g\in G}\ket{g}\ket{f(g)}.
\]
Measuring the second register, or equivalently tracing it out, leaves the first register in a uniform superposition over a random coset of $H$,
\[
\ket{gH}=\frac{1}{\sqrt{|H|}}\sum_{h\in H}\ket{gh}.
\]
The central algorithmic question is therefore how to extract information about $H$ from one or many such coset states.

In the Abelian setting this extraction step is especially clean. Applying the quantum Fourier transform over $G$ maps a coset state to a superposition over characters of $G$, and only those characters that are trivial on $H$ appear with nonzero amplitude. Thus a measurement after the Fourier transform samples from the annihilator subgroup $H^{\perp}$, and each sample gives a linear constraint on $H$. After sufficiently many samples, a classical linear-algebra calculation reconstructs the hidden subgroup. Standard expositions of Abelian HSP and its Fourier-sampling solution include refs.~\cite{nielsen2010quantum,kitaev2002classical,rieffel2011quantum}.

Period finding serves as the prototypical subroutine in Abelian HSP. By framing the search for a hidden translational symmetry as an algebraic problem, period finding demonstrates how Fourier sampling transforms nonlocal regularity into a directly measurable period lattice. For example, Simon's problem~\cite{simon}---a crucial conceptual precursor to Shor's algorithm---is fundamentally a period-finding problem over $G=\mathbb{Z}_2^n$. Similarly, Shor's algorithm effectively reduces the problem of integer factorization to a period-finding subroutine over the cyclic group $G=\mathbb{Z}_N$ (with discrete logarithms following a similar paradigm over $G = \mathbb{Z}_p^* \times \mathbb{Z}_p^*$). In these instances, the hidden object is often presented directly as a period rather than a formal subgroup generator, but the underlying mechanism is identical: the hidden subgroup captures the periodic structure generated by the oracle~\cite{Shor1994,Shor1997}.

In the non-Abelian setting, the same preparation step still yields coset states, but the information-extraction step becomes far less transparent. The quantum Fourier transform now decomposes the state into higher-dimensional irreducible representations, so the post-transform data are no longer one-dimensional characters but matrix-valued representation labels together with internal indices. Therefore, sampling does not directly return simple linear constraints in $H$, and it becomes much harder to identify which measurements retain enough subgroup information to be useful. This is the basic reason the non-Abelian HSP is substantially harder than its Abelian counterpart.

The HSP over the symmetric group $G=S_n$, for example, is closely related to graph isomorphism, and strong no-go results show that straightforward Fourier-sampling strategies are inadequate in that setting~\cite{MooreRussellSchulman2008,MooreRussellSchulman2005b}. The dihedral case, $G=D_{2n}$, is more forgiving but still delicate: rather than a general polynomial-time algorithm, what exists is a family of subexponential quantum algorithms, most notably Kuperberg's sieve and its refinements~\cite{Kuperberg2005,Kuperberg2011}. From this perspective, non-Abelian HSP marks an important boundary of the Fourier-sampling method rather than a uniformly solved extension of the Abelian theory.

Closely related problems broaden this picture by demanding more from the extraction step. The hidden shift problem acts as a more demanding variation of period finding: rather than extracting the internal periodic symmetry of a single function, one must recover the hidden translation between two distinct functions. Specifically, given oracle access to $f_0,f_1:G\to X$ with the promise that $f_1(x)=f_0(xs)$ (or $f_1(x)=f_0(x+s)$ in additive notation) for some unknown element $s\in G$, the goal is to recover the shift $s$~\cite{vanDamHallgrenIp2006,ChildsVanDam2005}. The quantum strategy again relies on preparing superpositions that encode global correlations between the functions and using Fourier analysis to translate those correlations into readable phase information about the unknown structure. In favorable cases, especially for injective hidden shift instances, the problem can be lifted to a standard HSP over a larger semidirect-product group, such as a generalized dihedral group~\cite{vanDamHallgrenIp2006,ChildsVanDam2005}. When this lifting is less natural, it is cleaner to treat hidden shift as an independent algorithmic primitive. 

Taken together, HSP, period finding, and hidden shift constitute a cohesive family of hidden-structure problems, unified by a common mechanism: each converts a nonlocal algebraic regularity into measurable interference patterns via Fourier sampling. We report here a selection of problem instances, namely factoring, discrete logarithm, Pell's equation, counting points on algebraic curves, exponential congruences, Gauss sums and elliptic curve isogeny.

\subsection{Factoring}\label{sec:factoring}
\emph{Maximum Quantum Speedup: Superpolynomial}\\
\emph{Evidence Basis: Rigorous quantum algorithm against the classical state of the art}

\paragraph{Problem Description}
Multiplying integers is easy, but reversing that process is harder: given an integer $N$, one wants to recover its prime factors. Still, factoring a random integer of a given size is often not difficult, since half of all integers are even and many others have small prime divisors. The hard regime is when the composite has little obvious structure, especially when it is the product of two, or a few, very large primes, for example $N=pq$.

\paragraph{Potential Application Spaces}
This problem is central to public-key cryptography, where the asymmetry between easy forward computation and hard inversion is used as a security resource. Given a composite modulus $N$, anyone can efficiently perform many arithmetic operations modulo $N$, including modular multiplication and exponentiation, but a party that knows the factorization $N=pq$ gains additional power: it can efficiently invert operations that are believed to be hard to reverse from $N$ alone. The standard example is the RSA primitive, where encryption is performed by exponentiation with a public encryption key $e$, while decryption uses a secret exponent $d$ chosen so that $de \equiv 1 \pmod{(p-1)(q-1)}$; knowledge of $p$ and $q$ makes it possible to construct this inverse relation efficiently~\cite{RSA1978}.

\paragraph{Classical Complexity}
Integer factoring is one of the most thoroughly studied problems in computational number theory. For large integers with no small factors, the main classical algorithms are subexponential rather than polynomial. Historically, the quadratic sieve was the fastest practical general-purpose method for many years~\cite{Pomerance1996}, and for sufficiently large inputs it was superseded by the general number field sieve, which remains the best known classical algorithm for factoring general integers~\cite{BuhlerLenstraPomerance1993}. 
The quadratic sieve has heuristic complexity
\begin{align*} 
\exp\!\left((1+o(1))(\log N)^{1/2} (\log\log N)^{1/2}\right),
\end{align*}
while the general number field sieve has heuristic complexity
\begin{align*} 
\exp\!\left(\left((64/9)^{1/3}+o(1)\right)(\log N)^{1/3} (\log\log N)^{2/3}\right).
\end{align*}
Thus the best known classical methods are substantially faster than brute force, but still superpolynomial in $\log N$.

\paragraph{Quantum Complexity}
It has been long known that factoring can be reduced to order finding. 
Shor's algorithm solves the order finding problem in polynomial time, therefore giving an algorithm for factoring with runtime polynomial in $\log N$~\cite{Shor1997}. At a high level, the quantum part prepares a superposition of exponents, computes modular exponentiation in superposition, applies the quantum Fourier transform, and classically post-processes the measured outcome to recover the order. Asymptotically this yields a polynomial-time algorithm, in sharp contrast with the best known classical subexponential algorithms. In practice the dominant cost is the implementation of modular arithmetic, especially modular exponentiation, and there is now a substantial literature on circuit optimization and resource estimation for fault-tolerant realizations of Shor's algorithm. There is a long line of increasingly gate- and qubit-efficient constructions, from Beauregard~\cite{Beauregard2003} and Eker\aa{}--H\aa{}stad~\cite{EkeraHastad2017} to more recent work by Chevignard, Fouque, and Schrottenloher~\cite{ChevignardFouqueSchrottenloher2024}, as well as Gidney~\cite{Gidney2025Million}. 

A distinct recent development is Regev's multidimensional factoring algorithm, which reduces the quantum circuit size relative to standard implementations of Shor's algorithm using lattice-based classical post-processing~\cite{regev2025}.

\paragraph{Regimes of Advantage}
\begin{itemize}
    \item The relevant regime is not small demonstration instances, for which classical methods are overwhelmingly simpler, but cryptanalytic composites that are products of exactly two large primes. The largest public integer factored with the number field sieve to date is RSA-250, a 250-digit (829-bit) RSA challenge number, whose factorization required roughly 2700 core-years of computation~\cite{BoudotEtAl2021RSA250}.
    \item Relative to that benchmark, a sufficiently large fault-tolerant quantum computer running Shor's algorithm would offer a decisive asymptotic advantage for every known cryptanalytic factoring target, because all practically relevant RSA moduli lie in the hard semiprime regime where the best classical attack remains subexponential while Shor stays polynomial in $\log N$. In that sense, once fault-tolerant quantum hardware reaches the scale needed to implement the algorithm, the advantage is not confined to isolated record computations: it applies across essentially the entire range of factoring instances that matter in cryptanalysis.
\end{itemize}

\paragraph{Notes on Future Directions and Related Advantages}
\begin{itemize}
    \item One important current direction is the shift from abstract logical-gate counts to increasingly detailed fault-tolerant resource estimates, since these estimates are now directly feeding back into hardware and architectural design. The 2019 surface-code estimate of Gidney and Eker\aa{} already framed RSA-2048 as a concrete engineering target rather than a purely asymptotic one~\cite{GidneyEkera2019}. 
    \item More recent work ties algorithm cost to specific architectural choices such as approximate arithmetic, qLDPC codes, heterogeneous modules, accelerators, and reconfigurable neutral-atom platforms~\cite{Gidney2025Million,CainEtAl2026Atomic,WebsterEtAl2026Pinnacle,MundadaEtAl2026Heterogeneous}. In other words, for factoring, improved resource estimation is no longer just a bookkeeping exercise: it is increasingly shaping which fault-tolerant architectures look plausible, which components deserve specialized optimization, and how one compares competing hardware roadmaps for utility-scale quantum advantage.
    \item A related direction is Jacobi factoring, which builds on quantum square-free decomposition and the computation of the Jacobi symbol in superposition~\cite{LiPengDuSuter2012,Kahanamoku2025}. The resulting circuits can efficiently factor broad classes of integers with unequal prime multiplicities, notably instances of the form \(N=p^2q\), while achieving near-linear gate complexity and, in relevant parameter regimes, sublinear quantum space and depth.
\end{itemize}

\subsection{Discrete Logarithm}\label{sec:discrete_log}

\emph{Maximum Quantum Speedup: Exponential}\\
\emph{Evidence Basis: Rigorous quantum algorithm against the classical state of the art}

\paragraph{Problem Description}
The discrete logarithm problem is a classic example of an easy forward map with a seemingly much harder inverse. In a finite cyclic group $G=\langle g\rangle$, it is straightforward to compute powers $g^x$, but given a generator $g$ and an element $h \in G$, the task is to recover an integer $x$ such that $g^x=h$, usually modulo the order of $g$. This problem fits naturally into the Abelian hidden subgroup framework. If $h=g^x$ and $r$ is the order of $g$, one may define a function on $\mathbb{Z}_r \times \mathbb{Z}_r$ by
$$f(a,b)=g^a h^{-b} = g^{a-bx}.$$ 
This function is constant exactly on cosets of the subgroup generated by $(x,1)$, so recovering the hidden subgroup reveals the discrete logarithm. In this sense, discrete logarithm is one of the clearest examples of how an algebraic inversion problem can be recast as an Abelian HSP.

\paragraph{Potential Application Spaces}
The hardness of discrete logarithm underlies several major public-key constructions. In multiplicative groups of finite fields, it is the basis for the Diffie--Hellman key exchange and related cryptographic primitives~\cite{DiffieHellman1976}. For example, Alice may publish $g^a$, Bob may publish $g^b$, and both can compute the shared secret $g^{ab}$, while an eavesdropper who sees only $g$, $g^a$ and $g^b$ is believed to face solving a discrete logarithm. 

There is also an elliptic-curve variant: if $P$ is a point of large prime order on an elliptic curve over a finite field and $Q=[x]P$ is obtained by scalar multiplication, one asks to recover $x$ from $P$ and $Q$. This elliptic-curve discrete logarithm problem underlies elliptic-curve Diffie--Hellman and other elliptic-curve cryptosystems, where it plays the same conceptual role that ordinary discrete logarithms play in finite-field cryptography~\cite{Koblitz1987,Miller1986}. 

Beyond cryptography, discrete logarithms also appear naturally in computational algebra and arithmetic over finite groups and fields, for example when one wants explicit coordinates relative to generators, efficient descriptions of multiplicative structure in finite fields, or algorithmic access to cyclic components in related algebraic settings~\cite{LidlNiederreiter1997,Cohen1993}.

\paragraph{Classical Complexity}
The classical picture depends strongly on the group family. For discrete logarithms in finite fields, there is an index-calculus family of algorithms, beginning with Adleman's subexponential method and culminating, for prime fields and many cryptographically relevant cases, in number-field-sieve style algorithms~\cite{Adleman1979,Gordon1993}. 
The best known classical algorithms for discrete logarithms in $\mathbb{F}_q$ are subexponential, with heuristic complexity 
\begin{align*} 
\exp\!\left(\left((64/9)^{1/3}+o(1)\right)(\log q)^{1/3} (\log\log q)^{2/3}\right).
\end{align*}

For elliptic-curve discrete logarithms, however, no analogue of index calculus is known in general. The best general-purpose classical attacks remain Pollard-type methods, such as Pollard's $\rho$ algorithm, which require on the order of the square root of the subgroup size, so for a prime-order subgroup of size $n$ the complexity is roughly $O(\sqrt{n})$ group operations~\cite{Pollard1978,Shoup1997}. Thus the finite-field problem admits substantial structure-exploiting classical speedups, whereas the elliptic-curve variant is still governed by generic-group style algorithms.

\paragraph{Quantum Complexity}
Shor's algorithm solves discrete logarithms in time polynomial in the input size by exploiting the same Abelian hidden-subgroup structure that underlies the factoring algorithm~\cite{Shor1994,Shor1997}. At a high level, one prepares superpositions over exponent pairs, evaluates the group operation coherently, applies the quantum Fourier transform over the relevant Abelian domain, and classically post-processes the measurement outcomes to recover the hidden relation. For finite fields this yields a polynomial-time quantum algorithm in $\log q$, in sharp contrast with the best known classical subexponential index-calculus methods. The elliptic-curve case is also polynomial-time, but the implementation details differ because the dominant cost is reversible elliptic-curve arithmetic, especially controlled point addition and the field operations needed for inversions or projective-coordinate updates. Early detailed circuit analyses for elliptic-curve discrete logarithms were given by Proos and Zalka~\cite{ProosZalka2003}, and later work by Roetteler, Naehrig, Svore, and Lauter developed more explicit resource estimates for prime-field elliptic curves that are directly relevant to cryptographic parameter sizes~\cite{RoettelerNaehrigSvoreLauter2017}. Thus, as in factoring, the asymptotic quantum complexity is polynomial, but the concrete cost depends heavily on how one realizes the underlying arithmetic in a fault-tolerant architecture.

\paragraph{Regimes of Advantage}
The relevant regime is not small demonstration instances, but cryptographically meaningful groups in which discrete logarithms are used to protect key exchange or public-key protocols.
\begin{itemize}
    \item For finite-field discrete logarithms, a sufficiently large fault-tolerant quantum computer running Shor's algorithm would eventually dominate the best known classical index-calculus methods across the full range of cryptanalytic interest, since the classical complexity remains subexponential while the quantum algorithm is polynomial in $\log q$.
    \item For elliptic-curve discrete logarithms, the contrast is even sharper: the best known classical attacks are Pollard-type methods with roughly square-root complexity in the subgroup size, whereas Shor's algorithm remains polynomial-time.
    \item  Once hardware reaches the scale needed for fault-tolerant implementations, the quantum advantage is not confined to specially chosen examples or isolated benchmarks. It applies broadly to the discrete-logarithm instances that matter in practice, including both finite-field Diffie--Hellman systems and elliptic-curve cryptography, with the elliptic-curve setting providing the clearest separation from the known classical state of the art.
\end{itemize}

\paragraph{Notes on Future Directions and Related Advantages}
\begin{itemize}
    \item One important current direction is the same one that now shapes the factoring literature: moving from abstract polynomial-time complexity to detailed, architecture-aware resource estimates for cryptographically relevant instances. This is especially visible for elliptic-curve discrete logarithms, where recent work connects circuit design directly to hardware assumptions such as clock speed, connectivity, qubit budgets, and fault-tolerant compilation strategies~\cite{BabbushEtAl2026Crypto,KimEtAl2026ECDLP,ChevignardFouqueSchrottenloher2026ECDLP}. In particular, recent studies distinguish between fast-clock and slow-clock architectures in assessing real-world vulnerability, develop lower-depth quantum circuits for prime-field elliptic curves, and explore qubit-saving designs that trade space against gate count. As in factoring, the main lesson is that the practical impact of Shor's algorithm is no longer captured by asymptotic statements alone: for discrete logarithms, architecture, arithmetic design, and fault-tolerant resource estimation are becoming inseparable parts of the security discussion.
\end{itemize}

\subsection{Pell's Equation}

\emph{Maximum Quantum Speedup: Superpolynomial}\\
\emph{Evidence Basis: Rigorous quantum algorithm against the classical state of the art}

\paragraph{Problem Description}

Given a positive non-square integer $d$, Pell's equation is the Diophantine equation $x^2-dy^2=1$. The integer solutions are generated by powers of a fundamental unit $\epsilon=x_1+y_1\sqrt d$ in the real quadratic field $\mathbb{Q}(\sqrt d)$. For algorithmic purposes, the key point is that the smallest nontrivial solution can have exponentially many bits in the input length $\log d$, so outputting $(x_1,y_1)$ explicitly is typically not a polynomial-size task. Accordingly, the standard computational formulation is to compute a concise representative of that solution, namely the regulator
\[
R=\log \epsilon=\log(x_1+y_1\sqrt d),
\]
to prescribed precision.

\paragraph{Potential Application Spaces}
Pell-type computations arise naturally in computational number theory and in cryptographic systems built from real quadratic fields. Their primary practical relevance lies in arithmetic over real quadratic number fields, where regulator computations are a fundamental primitive, including in cryptographic proposals such as the Buchmann-Williams key-exchange protocol~\cite{buchmann1989key}. More broadly, Pell's equation serves as the simplest representative of a wider class of computational problems involving regulators, unit groups, and class groups of algebraic number fields.

\paragraph{Classical Complexity}

The best known classical algorithms for this task are subexponential in the discriminant size, and therefore superpolynomial in the input length. In the real quadratic case, index-calculus style methods originating with Buchmann~\cite{buchmann1988subexponential} and refined in subsequent work~\cite{biasse2014subexponential} achieve heuristic or conditional running times of the same $L$-function type familiar from computational number theory,
\[
\exp\!\big(O(\sqrt{\log d\,\log\log d})\big),
\]
under standard mathematical assumptions such as the generalized Riemann hypothesis (GRH)\footnote{See for instance~\cite{bach1990explicit} for a nice explanation of how GRH appears in number-theoretic algorithms.}. Thus, as in factoring and finite-field discrete logarithms, the best known classical approaches are substantially better than brute force but remain asymptotically superpolynomial.

\paragraph{Quantum Complexity}

Hallgren gave a quantum algorithm that computes the regulator (and hence solves Pell's equation in the standard compact-output sense) in time polynomial in $\log d$~\cite{hallgren2007polynomial}. At a high level, the algorithm maps the arithmetic structure of principal ideals and units to a periodicity problem over a lattice-like domain, then uses quantum period-finding techniques to recover that period with polynomial overhead. This yields an asymptotic separation from the best known classical subexponential methods: quantum runtime is polynomial in input size, while classical runtime remains of subexponential $L$-type.

\paragraph{Regimes of Advantage}
\begin{itemize}
\item The relevant regime is not small demonstration values of $d$, but large-discriminant instances at security-parameter scales, including those motivated by cryptographic constructions over real quadratic fields such as the Buchmann--Williams key-exchange~\cite{buchmann1989key}. In that regime, the best known classical attacks rely on subexponential regulator and class-group computations, and hence the regime of advantage mirrors that of Shor's algorithm against factoring or discrete logarithms.
\end{itemize}

\paragraph{Notes on Future Directions and Related Advantages}
\begin{itemize}
    \item An important direction is to view Pell's equation as the degree-two entry point to a broader family of quantum algorithms for algebraic number theory. In particular, Hallgren extended these ideas to polynomial-time quantum algorithms for computing unit groups and class groups of more general number fields~\cite{hallgren2005fast}. This links Pell-type speedups to a larger hidden-structure paradigm in which lattice periodicity and arithmetic in ideals replace the cyclic-group setting of Shor's algorithm. 
    \item On the practical side, the key open questions are analogous to those now central for factoring and discrete logarithms: architecture-aware resource estimation, robust reversible implementations of algebraic number field arithmetic, and identification of application-level workloads where regulator or class-group computations are the genuine computational bottleneck relative to modern classical heuristics.
\end{itemize}

\subsection{Counting Points on Algebraic Curves}
\emph{Maximum Quantum Speedup: Uncertain, potentially exponential}\\
\emph{Evidence Basis: Rigorous quantum algorithm against the classical state of the art}

\paragraph{Problem Description}

Let $C$ be a smooth, projective, absolutely irreducible algebraic curve of genus $g$ defined over a finite field $\mathbb{F}_q$. The \emph{zeta function} of $C$, denoted $Z(C, t)$, is a generating function that encodes the number of points on the curve over the base field and all its finite extensions. It is defined as:
\[ Z(C, t) = \exp \left( \sum_{k=1}^{\infty} \frac{N_k t^k}{k} \right) \]
where $N_k$ is the number of points on $C$ in the extension field $\mathbb{F}_{q^k}$. This zeta function is a rational function of the form $Z(C, t) = \frac{P(t)}{(1-t)(1-qt)}$, where $P(t)$ is a polynomial of degree $2g$ with integer coefficients. The computational problem is to determine the polynomial $P(t)$ given the equation of the curve. Computing the zeta function allows one to determine the number of rational points on the curve (group order).

\paragraph{Potential Application Spaces}

Point counting is most relevant in application areas where one works with structured families of curves rather than isolated low-genus instances. In cryptography, this includes higher-genus Jacobian-based constructions, where security and parameter selection depend on understanding group orders over finite fields. In coding theory, point counting is a core primitive behind algebraic-geometry code constructions based on families such as modular and Shimura curves and towers of function fields, where one tracks rational points and genus together to obtain strong asymptotic parameters~\cite{TsfasmanVladutZink1982,GarciaStichtenoth1995,stichtenoth1993}.

\paragraph{Classical Complexity}

For elliptic curves ($g=1$), Schoof's algorithm~\cite{schoof1985elliptic} computes the number of points in polynomial time $O((\log{q})^{8})$, while its extension, the Schoof–Elkies–Atkin algorithm runs in $\widetilde{O}((\log{q})^{4})$. However, for curves of arbitrary genus $g$, the general problem is much harder. Pila's algorithm runs in time $\poly(\log{q})$ only if the genus $g$ is constant~\cite{pila1990frobenius}. There are $p$-adic algorithms~\cite{kedlaya2001counting} that are polynomial in $g$ but exponential in $\log p$ (the characteristic), making them unsuitable for fields of large characteristic.

\paragraph{Quantum Complexity}

Kedlaya proposed a quantum algorithm that computes the zeta function of a curve of genus $g$ over $\mathbb{F}_q$ in time polynomial in both $g$ and $\log q$~\cite{kedlaya2006quantum}. The algorithm works by representing the Jacobian of the curve over suitable extension fields as a finite Abelian black-box group and using quantum algorithms for Abelian groups to efficiently determine its order. The orders of these Jacobians provide cyclic resultants of the Weil polynomial, from which the zeta function can be reconstructed efficiently. Importantly, the group operations on the Jacobian are explicitly implementable from the input curve.

\paragraph{Regimes of Advantage}
The relevant regime is governed by how genus scales with input size. 
\begin{itemize}
    \item For fixed genus, classical point counting is polynomial in $\log q$~\cite{pila1990frobenius}, so no asymptotic super-quadratic advantage is expected there. 
    \item The intended Stage-II setting consists of efficiently generable families $\{C/\mathbb F_{q}\}$ of smooth projective curves, given by descriptions of size $\poly(n)$, for which both the genus $g$ and $\log q$ grow with $n$~\cite{kedlaya2006quantum}. In this regime, the claimed speedup is relative to the classical state of the art; no average-case classical lower bound is asserted.
    \item In that scaling regime, a quantum algorithm polynomial in both $g$ and $\log q$ can in principle separate from the best known classical methods, whereas small-$g$ demonstrations do not probe the relevant asymptotics.
\end{itemize}

\paragraph{Notes on Future Directions and Related Advantages}
An important next direction is to connect the abstract polynomial-time quantum result more tightly to concrete curve families used in practice. 
\begin{itemize}
    \item On the classical side, fixed-genus and small-characteristic methods remain highly effective in their native regimes~\cite{pila1990frobenius,kedlaya2001counting}, so the central question is to identify realistic workloads where genus growth is intrinsic rather than incidental. The algebraic-geometry coding setting already provides such families through modular and Shimura curves and through function-field towers~\cite{TsfasmanVladutZink1982,GarciaStichtenoth1995,stichtenoth1993}; a natural follow-on is to quantify when point counting is the dominant cost in parameter generation and code construction pipelines.
    \item In parallel, as in factoring and discrete logarithms, a key practical milestone is architecture-aware resource estimation for implementations of Kedlaya-type quantum routines, including reversible finite-field arithmetic, sparse linear-algebra kernels, and precision management in Frobenius-eigenvalue estimation~\cite{kedlaya2006quantum}.
\end{itemize}

\subsection{Exponential Congruences}
\emph{Maximum Quantum Speedup: Polynomial}\\
\emph{Evidence Basis: Rigorous quantum algorithm against the classical state of the art}

\paragraph{Problem Description}

Exponential congruences combine the multiplicative structure generated by exponentiation with additive constraints over a finite field. The basic problem considered here is to find 
nonnegative integers \(x,y\) satisfying
\[
    a f^x + b g^y = c,
\]
where $a,b,c,f,g\in\mathbb{F}_q^*$. Since only the exponents modulo the multiplicative orders of \(f\) and \(g\) matter, writing $s=\operatorname{ord}(f)$ and $t=\operatorname{ord}(g)$, we restrict \(0\leq x<s\) and \(0\leq y<t\).

It is important to distinguish this problem from ordinary discrete logarithm. A one-term exponential equation $g^x=h$ is exactly the discrete-logarithm problem discussed above. More generally, a purely multiplicative relation such as $g_1^{x_1}g_2^{x_2}=h$
can be transformed, after computing suitable discrete logarithms, into linear congruences among the unknown exponents. The new difficulty in an exponential congruence comes from the addition of multiplicative terms: taking logarithms does not linearize $a f^x+b g^y=c$. Equivalently, one is searching for a representation
\[
    c\in a\langle f\rangle+b\langle g\rangle
\]
as a sum of elements from two multiplicative cosets. The orders $s$ and $t$, and not only the field size $q$, are essential parameters governing the complexity of the problem~\cite{VanDamShparlinski2008,Shparlinski2019}.

\paragraph{Potential Application Spaces}

Exponential congruences arise primarily in computational number theory, particularly in problems involving interactions between additive and multiplicative structure in finite fields. They are closely related to cyclotomic classes and exponential and character sums, and have also appeared in cryptographic constructions involving constrained public-key parameters and meaningful hash collisions~\cite{LenstraDeWeger2005,VanDamShparlinski2008}. The $n$-variable generalization
\[
    \sum_{i=1}^n a_i f_i^{x_i}=c
\]
has also been studied algorithmically~\cite{Sasaki2010}.

They are also structurally connected with certain non-Abelian hidden-subgroup problems: efficient solution of special exponential congruences would imply efficient measurements for some semidirect-product hidden subgroup problems, although the known exponential-congruence algorithms remain superpolynomial in $\log q$ and therefore do not themselves yield polynomial-time algorithms for those problems~\cite{VanDamShparlinski2008}.

\paragraph{Classical Complexity}

The classical complexity depends strongly on the multiplicative orders $s$ and $t$. The original analysis of van Dam and Shparlinski gave worst-case deterministic complexity $q^{9/8}(\log q)^{O(1)}$, but this bound has subsequently been substantially improved. Shparlinski showed that one can either find a solution or determine that none exists in time at most $\min\left\{ (st)^{1/2}, q^{3/4} \right\} (\log q)^{O(1)}$~\cite{Shparlinski2019}. Thus the current general worst-case upper bound is $q^{3/4+o(1)}$.

Shparlinski also showed that for fixed $a,b,f,g$ and almost all $c\in\mathbb{F}_q^*$ we can derive a stronger order-sensitive bound $\min\left\{ (st)^{1/2}, q s^{-1/2}, q t^{-1/2} \right\} (\log q)^{O(1)}$. Consequently, we have an upper bound $q^{2/3+o(1)}$ for almost all \(c\)~\cite{Shparlinski2019}.

Special algebraic cases can be substantially easier. In particular, when one base generates the full multiplicative group, the problem essentially reduces to discrete logarithm; when one order is small, enumeration reduces the search space. The interesting regime is therefore not characterized simply by large $q$, but by sufficiently large multiplicative orders together with proper subgroup structure that prevents the equation from collapsing to a single discrete-logarithm computation.
No matching classical lower bounds are known in these regimes. Throughout this entry, the potential speedup therefore refers to a comparison with the best currently known classical algorithms.

\paragraph{Quantum Complexity}

Van Dam and Shparlinski gave quantum algorithms combining two ingredients: Shor's polynomial-time quantum algorithm for discrete logarithms and Grover search over one of the exponent variables. Bounds from character sums imply that only a restricted portion of that variable's range needs to be searched~\cite{VanDamShparlinski2008}. For arbitrary $f$ and $g$, this gives the worst-case bound $q^{3/8}(\log q)^{O(1)}$. For all but $o(q)$ choices of \(c\), this improves to $q^{1/3}(\log q)^{O(1)}$.

More substantial improvements are possible when the multiplicative orders are large. In the worst case, if $st > Cq^{3/2}(\log q)^{1/2}$ there is a quantum algorithm with complexity $q^{1/2}(st)^{-1/4}(\log q)^{O(1)}$. At the threshold $st\approx q^{3/2}$, this is approximately $q^{1/8}$, and the runtime decreases further as the product $st$ grows. For almost all $c$, the same order-dependent quantum complexity applies under the weaker condition $st > q^{4/3}(\log q)^{2/3}$~\cite{VanDamShparlinski2008}.

\paragraph{Regimes of Advantage}

The distinction between generic and large-order instances is essential. For the Stage-II claim, the intended families can be generated efficiently by choosing finite fields with known large proper multiplicative subgroups, constructing $f$ and $g$ with the corresponding orders, and choosing nonzero $a,b$; in the typical-case setting, $c$ is sampled uniformly. The relevant nondegenerate regime has $s$ and $t$ large but proper and excludes small-order cases and parameter choices that collapse to a one-variable or ordinary discrete-logarithm problem. The quantum and classical runtimes quoted below apply to this family.
\begin{itemize}
    \item \emph{Generic instances do not presently give a super-quadratic comparison from the best published bounds.}
    Combining the modern worst-case classical bound $q^{3/4+o(1)}$ with the general quantum bound $q^{3/8+o(1)}$ gives only a quadratic relationship between the two upper bounds. Similarly, for almost all $c$, the general comparison $q^{2/3+o(1)}$ classically versus $q^{1/3+o(1)}$ quantumly is quadratic~\cite{VanDamShparlinski2008,Shparlinski2019}.

    \item \emph{Super-quadratic comparisons occur in large-order regimes.}
    When $st$ is sufficiently large, the order-dependent quantum algorithm becomes faster than $q^{3/8}$, while the best known general classical bounds do not improve comparably. For example, near the worst-case threshold $s,t\approx q^{3/4}$, we have classical complexity $\lesssim q^{3/4}$ versus quantum complexity $\lesssim q^{1/8}$, giving a sixth-power relationship between these particular upper bounds. For typical $c$, balanced orders near $s,t\approx q^{2/3}$ give classical complexity $\lesssim q^{2/3}$ and quantum complexity $\lesssim q^{1/6}$, corresponding to a fourth-power relationship.

    \item \emph{The easy large-order cases must be excluded.}
    Large multiplicative order alone does not imply a new exponential-congruence advantage. If $f$ or $g$ generates all of $\mathbb{F}_q^*$, the problem reduces essentially to discrete logarithm. The distinctive instances are therefore those with large but proper multiplicative subgroups.
\end{itemize}
As emphasized above, no corresponding classical lower bound is known. The claimed advantage should therefore be understood as an order-dependent comparison with the best currently known classical algorithms rather than as a proved classical-quantum separation.

\paragraph{Notes on Future Directions and Related Advantages}
\begin{itemize}
    \item A central open problem is to sharpen the classical complexity for the large-order proper-subgroup regime. Improved classical algorithms could reduce or eliminate the apparent super-quadratic gaps identified above, whereas stronger hardness evidence would make these families substantially more compelling as candidates for quantum advantage.
    \item Exponential congruences with more variables, $\sum_{i=1}^{n}a_i f_i^{x_i}=c$, have also been studied~\cite{Sasaki2010}. Existing analyses indicate that merely introducing additional exponent variables does not automatically strengthen the quantum advantage, so the two-variable problem remains the cleanest setting in which to understand the role of multiplicative orders and additive coupling.
    \item Finally, the connection with semidirect-product hidden-subgroup problems provides a structural motivation for further algorithmic improvement. Known exponential-congruence algorithms remain superpolynomial in $\log q$; finding a polynomial-time algorithm for relevant special cases would have implications beyond the congruence problem itself~\cite{VanDamShparlinski2008}.
\end{itemize}

\subsection{Gauss Sums}

\emph{Maximum Quantum Speedup: Uncertain, potentially superpolynomial}\\
\emph{Evidence Basis: Rigorous quantum algorithm against the classical state of the art}

\paragraph{Problem Description}

Gauss sums are a central number-theoretic primitive defined from additive and multiplicative characters over finite fields and rings. For example, over a finite field they take the form
\[
G(\mathbb{F}_{p^r},\chi,\beta)=\sum_{x\in\mathbb{F}_{p^r}}\chi(x)\,\omega_p^{\mathrm{Tr}(\beta x)},
\]
where \(\chi\) is multiplicative, \(x\mapsto \omega_p^{\mathrm{Tr}(\beta x)}\) is additive, and \(\omega_p=e^{2\pi i/p}\). Van Dam and Seroussi showed that, given an explicit specification of these characters, one can estimate the phase of the corresponding Gauss sum efficiently on a quantum computer~\cite{van2002efficient}. Their construction uses a discrete-logarithm subroutine to prepare character states such as
\[ \ket{\chi} =\frac{1}{\sqrt{p^r-1}}\sum_{x\in\mathbb{F}_{p^r}^*}\chi(x)\,\ket{x},\]
and then maps the Gauss-sum angle to an observable \emph{relative} phase that can be estimated by standard phase-estimation methods.

Given a finite field \(\mathbb{F}_{p^r}\) (or finite ring \(\mathbb{Z}/n\mathbb{Z}\)), a multiplicative character \(\chi\), and an additive-character parameter \(\beta\), the goal is to estimate the phase \(\gamma\) in \(G=|G|e^{i\gamma}\) to additive precision. In this setting, the magnitude \(|G|\) is often comparatively easy to characterize, while the argument is the computationally difficult part.

\paragraph{Potential Application Spaces}

Gauss sums and related Jacobi sums appear broadly in computational number theory, including arithmetic over finite fields and rings and methods connected to primality testing~\cite{Cohen1993,BerndtEvansWilliams1998}. Gauss sums also have applications in coding theory, cryptography, and signal processing~\cite{LidlNiederreiter1997,MacWilliams1977-ci,GolombGong2005}.

\paragraph{Classical Complexity}

There is no unconditional classical hardness theorem for general Gauss-sum phase estimation. However, van Dam and Seroussi give a reduction from discrete logarithm to Gauss-sum phase approximation in finite fields, which provides evidence that efficient classical algorithms for this task are unlikely~\cite{van2002efficient}. The right summary is therefore reduction-based hardness evidence, not a proved separation.

\paragraph{Quantum Complexity}

For finite fields, the algorithm estimates the phase with expected error \(\epsilon\) in time \(O(\epsilon^{-1}\,\polylog(p^r))\). The same framework yields an analogous result over finite rings \(\mathbb{Z}/n\mathbb{Z}\), with complexity \(O(\epsilon^{-1}\,\polylog n)\), and also gives Jacobi-sum estimation~\cite{van2002efficient}. At a high level, the runtime is governed by character-state preparation and repeated phase estimation to reach target precision.

\paragraph{Regimes of Advantage}
\begin{itemize}
    \item The relevant regime consists of large finite fields and nontrivial multiplicative characters outside special low-order cases with explicit classical evaluations~\cite{van2002efficient}. The target phase precision must satisfy $1/\epsilon=\poly(\log q)$ for the quantum runtime to remain polynomial in the input length. If $\epsilon$ is instead polynomially small in $q$, the precision cost alone becomes polynomial in the field size and the advantage must be reassessed.
    \item The classical-hardness evidence comes from the reduction from discrete logarithm to Gauss-sum estimation for the family of characters used in that reduction. It does not establish a general average-case lower bound for arbitrary Gauss-sum instances.
\end{itemize}
As in the other number-theoretic subsections, practical impact is expected only at fault-tolerant scales where asymptotic gains dominate overheads.

\paragraph{Notes on Future Directions and Related Advantages}
\begin{itemize}
    \item An important open direction, already emphasized in the original work, is to sharpen the classical complexity picture: can Gauss-sum estimation be shown hard even when factoring and discrete logarithm are assumed easy, or can stronger reverse reductions be found~\cite{van2002efficient}?
    \item A second direction is to move from asymptotic complexity to architecture-aware cost models, clarifying when Gauss/Jacobi-sum estimation is competitive in practice rather than mainly structurally interesting.
\end{itemize}

\subsection{Elliptic Curve Isogeny}
\emph{Maximum Quantum Speedup: Uncertain, potentially exponential}\\
\emph{Evidence Basis: Rigorous quantum algorithm against the classical state of the art}

\paragraph{Problem Description}

Isogeny-based cryptography uses the arithmetic of elliptic curves over finite fields, where the secret key is encoded into isogeny walks or class-group actions on sets of curves. 
Formally, an isogeny between elliptic curves $E$ and $E'$ over a field $K$ is a nonconstant morphism $\phi:E\to E'$ defined over $K$ that maps the identity point of $E$ to the identity point of $E'$. Equivalently, it is a surjective group homomorphism with finite kernel. The degree of an isogeny measures its algebraic complexity and, for separable isogenies, equals the size of its kernel. Conversely, a finite subgroup $G\subset E$ determines, up to isomorphism, an isogeny $E\to E/G$ with kernel $G$. This gives rise to isogeny graphs whose vertices represent elliptic curves and whose edges represent isogenies of prescribed degree.

A good viewpoint stems from the ``hard homogeneous space'' perspective: one has an efficiently computable group action, but recovering the description of the acting element from the public information is believed hard~\cite{Couveignes2006HHS,RostovtsevStolbunov2006}. In particular, hidden-shift structure exploited by quantum attacks appears from the structure of the protocol (notably class-group-action formulations), rather than being against isogenies in general~\cite{ChildsJaoSoukharev2010}.

\paragraph{Potential Application Spaces}

The main application space is post-quantum public-key cryptography, including key exchange, key encapsulation, and digital signatures. Isogeny-based constructions have been especially attractive where compact public keys are important, in addition to providing mathematical diversity with lattice-based families~\cite{DeFeoJaoPlut2011}. For our purposes, the relevant use cases are cryptographic deployments where security margins can be tied to explicit attack-cost models, including quantum attack models, rather than only to asymptotic hardness labels.

\paragraph{Classical Complexity}

Classically, the core isogeny problems underlying group-action systems are not believed to admit general polynomial-time algorithms, and security analyses are typically parameterized through best known generic and structure-exploiting attacks. By contrast, some supersingular-isogeny protocols were broken by efficient classical cryptanalysis~\cite{CastryckDecru2022,robert2023breaking}. So there is protocol fragility for some isogeny-based schemes, but the broader isogeny landscape cannot be considered ``broken.''

\paragraph{Quantum Complexity}

For cryptosystems that reduce to hidden-shift-like formulations, quantum attacks can leverage dihedral hidden subgroup machinery. This results in a subexponential quantum complexity via the Kuperberg sieve, but not a general polynomial-time algorithm~\cite{ChildsJaoSoukharev2010,Kuperberg2005,Kuperberg2011,Regev2004DHS}.

\paragraph{Regimes of Advantage}
The relevant Stage-II regime consists of families of ordinary elliptic curves over growing finite fields whose shared endomorphism rings have growing discriminants and large ideal-class groups, which determine the size of the relevant isogeny search space. The class groups must be sufficiently large for the classical state of the art to remain exponential in the input length; instances with a known short connecting isogeny or an unusually small class group fall outside this regime. Under the assumptions of the Childs–Jao–Soukharev algorithm, including the Generalized Riemann Hypothesis, the corresponding quantum attack is subexponential~\cite{ChildsJaoSoukharev2010}.

The practically relevant regime is cryptanalytic. One should treat quantum security of such systems similarly to how classical cryptography treats RSA or finite-field Diffie-Hellman: systems are not automatically broken, but key sizes (and related parameters) must be chosen against the best known attack exponents and concrete resource estimates. In other words, advantage appears when quantum subexponential attacks become cheaper than intended security targets at deployed parameter scales.

\paragraph{Notes on Future Directions and Related Advantages}

An important direction is to characterize which isogeny constructions induce hidden-shift structure exploitable by sieve-based quantum algorithms. Another direction is tighter end-to-end cost modeling: mapping asymptotic subexponential attacks to realistic fault-tolerant resources, for the purpose of evaluating concrete parameter sets.

\pagebreak
\section{Optimization}

\emph{Section contributed by:} Andi Gu, Philipp Kammerlander, Ruslan Shaydulin, Shouvanik Chakrabarti

\subsection{Optimal Polynomial Intersection}

\emph{Maximum Quantum Speedup: Exponential}\\
\emph{Evidence Basis: Rigorous quantum algorithm against the classical state of the art}\\

Decoded Quantum Interferometry (DQI) is a quantum algorithm for combinatorial optimization~\cite{Jordan2024-sy}. A problem instance is specified by $m$ constraints on $n$ variables with $m > n$; each constraint is either satisfied or not by a given assignment, and the goal is to find an assignment that satisfies as many as possible. 

DQI reduces optimization to decoding. The $m$ constraints define a linear code $\mathcal{C}$ over $\mathbb{F}_p$ whose codewords are the strings of constraint values produced by the candidate solutions. DQI does not search the solution space directly; it uses the quantum Fourier transform to set up interference that concentrates measurement probability on solutions satisfying many constraints. Arranging this interference reduces to a decoding problem on the dual code $\mathcal{C}^\perp$. This decoding must be done coherently; one way of doing this is to run a classical decoder for $\mathcal{C}^\perp$ reversibly. The number of errors $\ell$ the decoder can correct controls how strongly the output is biased toward good solutions, as quantified by the `semicircle law'. When each constraint admits $r$ of the $p$ field values, the expected fraction of satisfied constraints is
\begin{align*}
\frac{\langle s \rangle}{m} = \left( \sqrt{\frac{\ell}{m}\left(1 - \frac{r}{p}\right)} + \sqrt{\frac{r}{p}\left(1 - \frac{\ell}{m}\right)}\right)^2 .
\end{align*}
At $\ell = 0$ (decoder can correct 0 errors) this equals the random-assignment baseline $r/p$, and it rises as the decoder corrects more errors. When $r/p = 1/2$ the law traces a semicircle, hence its name.

DQI is most effective on optimization problems with algebraic structure, where the dual code is efficiently decodable. The principal example is optimal polynomial intersection (OPI)---polynomial curve fitting over a finite field---which we discuss in more depth below.


\paragraph{Problem Description}

Given a finite field $\mathbb{F}_p$, distinct evaluation points $y_1,\ldots,y_m$, and sets $F_i \subseteq \mathbb{F}_p$ of allowed values, find a polynomial $P$ of degree less than $n$ that maximizes $|\{i : P(y_i) \in F_i\}|$. Here the $m$ evaluation points play the role of the $m$ constraints, and the $n$ coefficients of $P$ are the $n$ variables of the general formulation above.

\paragraph{Potential Application Spaces}

OPI is polynomial regression over a finite field: one seeks the low-degree polynomial that best fits a collection of observations, where at each evaluation point only a restricted set of field values counts as a match. Problems of this form arise across several areas. In coding theory, OPI is a special case of the list-recovery problem for Reed-Solomon codes, so it is closely related to the design and analysis of error-correcting codes~\cite{GuruswamiSudan1998}. In cryptography, the difficulty of reconstructing a polynomial from noisy or ambiguous evaluations underlies primitives such as oblivious polynomial evaluation~\cite{naor1999} and noisy polynomial interpolation~\cite{bleichenbacher2000}, where fitting a polynomial through adversarially chosen value sets serves as a hardness assumption.

\paragraph{Classical Complexity}

The best known classical polynomial-time algorithm is Prange's infor\-mation-set decoding~\cite{Prange1962}, which fixes a subset of constraints to satisfy exactly and then completes the assignment, satisfying a fraction of the remainder at the random baseline. Beyond this, one must resort to exponential-time search methods. General-purpose local-search heuristics such as simulated annealing do not help on OPI: because the problem is dense (every variable appears in every constraint), the objective landscape is nearly flat around the random baseline, and on the balanced instances of interest simulated annealing performs no better than Prange's algorithm~\cite{Jordan2024-sy,GuJordan2025AGDQI}. No polynomial-time classical method is known to exceed Prange's satisfaction fraction, and the complexity of doing so is open.

\paragraph{Quantum Complexity}

The quantum circuit complexity is $O(T + m^2)$ gates, where $T$ is the cost of implementing the classical decoder reversibly; for polynomial curve fitting with Reed-Solomon codes this is polynomial in the problem size. The $m^2$ term, which arises from preparing the weighted Dicke state, can be reduced to nearly linear given quantum random access to the input~\cite{Rosmanis2026linearDQI}. The space complexity is $O(m\log p)$ qubits (the error and syndrome registers), polynomial in the input size and exponentially smaller than the $p^{n}$ candidate solutions.

\paragraph{Regimes of Advantage}

For suitably chosen instances, DQI satisfies a strictly larger fraction of OPI constraints than any of the classical algorithms discussed above. As with the speedups for factoring and discrete logarithm, this separation is established relative to the best known classical algorithms rather than by a proof of classical hardness. Realizing the advantage requires the following conditions to hold simultaneously:
\begin{itemize}
    \item The constraint matrix must have algebraic structure that makes the dual code decodable in polynomial time; in polynomial curve fitting the Vandermonde structure yields Reed-Solomon codes~\cite{ReedSolomon1960} that admit such a decoder, whereas generic unstructured problems lead to NP-hard syndrome decoding~\cite{berlekamp1978} and hence no advantage.
    \item The solution space must be exponentially large even though the field itself stays polynomial in size; for polynomial curve fitting, field sizes of at least several hundred already make the space of candidate polynomials far too large for any classical exhaustive or algebraic enumeration.
    \item Balanced constraint sets $r/p \approx 1/2$ make the gap between the semicircle law and Prange's algorithm the widest.
    \item The system must be overconstrained ($m/n\to1$ becomes classically easy), with the margin over the best classical algorithm largest at moderate ratios, $m/n$ between roughly 2 and 10~\cite{Jordan2024-sy}; more overconstrained, lower-rate instances give a smaller margin but can remain classically intractable at lower quantum cost, so resource-optimized proposals for verifiable advantage run at substantially larger ratios (e.g.,\ $m/n\approx58$)~\cite{khattar2025verifiablequantumadvantageoptimized}.
\end{itemize}

This advantage is subject to known limits. For unstructured max-LINSAT, surpassing the random-assignment fraction $r/p$ by any constant is NP-hard, so no efficient algorithm, classical or quantum, can do better in the worst case~\cite{Kramer2026LINSAT}. It is conjectured in~\cite{Anschuetz2025DQIstructure} that, on generic instances and when restricted to classical decoding algorithms, DQI cannot surpass the overlap-gap property and can be matched or outperformed by classical message-passing algorithms. OPI avoids these obstructions because it is not a worst-case instance: its algebraic structure separates it from the instances arising in NP-hardness reductions, and no classical algorithm is known that exploits this structure as effectively as DQI~\cite{Kramer2026LINSAT}.

DQI for OPI can be generalized to algebraic geometry codes. While Reed-Solomon codes are restricted to block length $p$ over $\mathbb{F}_p$, replacing the affine line by the Hermitian curve $y^q + y = x^{q+1}$ over $\mathbb{F}_{q^2}$, which carries $m = q^3$ rational points $P_1,\ldots,P_m$, gives the Hermitian Optimal Polynomial Intersection (HOPI) problem~\cite{GuJordan2025AGDQI}: given constraint sets $F_i \subset \mathbb{F}_{q^2}$, find a low-pole-order rational function $f \in \mathcal{L}_t$ maximizing $|\{i : f(P_i) \in F_i\}|$. This is polynomial regression on an algebraic curve rather than a finite field and, because the dual of a Hermitian code is again Hermitian, is equivalent to approximate list recovery for Hermitian codes.

\paragraph{Notes on Future Directions and Related Advantages}

Since DQI reduces optimization to decoding, its performance is governed entirely by the decoder: the semicircle law fixes the satisfaction fraction in terms of the number of errors $\ell$ that the decoder can correct. Improving DQI is therefore equivalent to improving its decoder, which can be approached in two complementary ways. The first is to employ a stronger \emph{classical} decoder, implemented reversibly within the algorithm. Here DQI can draw on the existing coding-theory literature; soft-decision decoders, for example, use reliability information to correct additional errors and thereby raise the satisfaction fraction DQI achieves on OPI~\cite{chailloux2026quantumadvantagesoftdecoders}. Conversely, the instances on which DQI performs poorly are those whose dual codes are difficult to decode classically.

In the second approach the decoder need not be classical. DQI supplies its decoder with a coherent superposition of error patterns, and a reversible classical decoder discards this coherence, treating each syndrome as an independent classical inference problem and thus remaining bounded by the Shannon capacity of the corresponding channel. An intrinsically \emph{quantum} decoder can instead act on the superposition directly~\cite{ChenLiuZhandry2022,ChaillouxTillich2024}, exploiting information that measurement would destroy and so bounded only by the (generally larger) Holevo quantity. Quantum message-passing decoders such as belief propagation with quantum messages (BPQM) realize this in practice, decoding beyond the Shannon limit~\cite{PiveteauRenes2022}. This has the potential to extend DQI's advantage to regimes inaccessible to any classical decoder, in particular the denser, non-algebraic codes (e.g.,\ high-degree max-$k$-XORSAT) for which classical decoders fall well short of the information-theoretic limit~\cite{Jordan2024-sy}.

\ \\
\emph{Generalization to other codes.}
DQI for OPI can be generalized to algebraic geometry codes. While Reed-Solomon codes are restricted to block length $p$ over $\mathbb{F}_p$, replacing the affine line by the Hermitian curve $y^q + y = x^{q+1}$ over $\mathbb{F}_{q^2}$, which carries $m = q^3$ rational points $P_1,\ldots,P_m$, gives the Hermitian Optimal Polynomial Intersection (HOPI) problem~\cite{GuJordan2025AGDQI}: given constraint sets $F_i \subset \mathbb{F}_{q^2}$, find a low-pole-order rational function $f \in \mathcal{L}_t$ maximizing $|\{i : f(P_i) \in F_i\}|$. This is polynomial regression on an algebraic curve rather than a finite field and, because the dual of a Hermitian code is again Hermitian, is equivalent to approximate list recovery for Hermitian codes.

As with Reed-Solomon codes for OPI, Hermitian codes have efficient decoders~\cite{SAKATA1988321}, so the semicircle law applies to HOPI as well. Substituting the Hermitian parameters into the semicircle law (whose per-constraint alphabet size $p$ is here $q^2$, so the ratio $r/p$ becomes $r/q^2$), DQI dominates Prange's algorithm at every rate, with the margin growing with field size and peaking near $r/q^2 \approx 0.28$, where DQI satisfies roughly $40\%$ more constraints in expectation~\cite{GuJordan2025AGDQI}. Moreover, since Hermitian codes reach block length $q^3$ over $\mathbb{F}_{q^2}$, they cost only $\lceil (2/3)\log_2 m \rceil$ qubits per field element rather than the $\lceil \log_2 m \rceil$ required by a Reed-Solomon code of the same length. This saving makes HOPI a candidate for verifiable quantum advantage on moderate-scale fault-tolerant hardware. Whether the framework extends usefully to other algebraic geometry code families, and how far the field-size reduction can be pushed while preserving efficient dual decoding, remain open.

The Hermitian extension illustrates a broader strategy suggested by the duality between optimization and decoding.
Rather than starting from an optimization problem, one can begin from a code family equipped with a particularly strong decoder and a well-understood dual, and then read off the optimization problem to which DQI applies. Each such code family yields a distinct family of structured optimization problems with an associated quantum speedup: Reed-Solomon codes give OPI, their multivariate Reed-Muller generalization gives multivariate OPI~\cite{Jordan2024-sy}, and Hermitian and other algebraic geometry codes give HOPI~\cite{GuJordan2025AGDQI}. The essential requirement is that the dual code remain efficiently decodable in the large-error, list-recovery regime. Surveying the coding-theory literature for code families that meet this criterion, and designing new ones for it, is therefore a route to identifying further optimization problems with quantum advantage~\cite{Jordan2024-sy}.

\ \\
Finally, as with factoring, attention is shifting from asymptotic complexity to concrete fault-tolerant resource estimates. A recent optimized compilation of DQI for OPI estimates that a classically intractable instance can be solved with roughly $5.72$ million Toffoli gates~\cite{khattar2025verifiablequantumadvantageoptimized}, making OPI a candidate for verifiable quantum advantage on early fault-tolerant hardware.

\subsection{$k$-SAT}
\label{sec:ksat}

\emph{Maximum Quantum Speedup: Improved exponential scaling constant}\\
\emph{Evidence Basis: Numerical and analytical scaling comparison against the classical state of the art}\\

Discrete optimization problems have long been regarded as promising candidates for super-quadratic quantum advantages. However, only a small number of concrete problems and instance regimes are currently believed to provide evidence strong enough to qualify as Stage II quantum advantage candidates.
An important example is the combinatorial optimization problem $k$-SAT, an NP-complete satisfiability problem for $k\geq3$.

\paragraph{Problem Description}
$k$-SAT is a satisfiability problem over $n$ boolean variables $\{x_i\}_{i=1}^n$ with $m$ logical clauses, each clause containing exactly $k$ literals. For instance, a clause in $3$-SAT could be $x_2\lor x_4\lor \lnot x_5$. In random $k$-SAT, the clauses are chosen at random. 
The computational task is to decide whether a given instance of $k$-SAT is satisfiable and, if so, to find an assignment that satisfies all clauses.
The focus here is on the exact satisfiability problem, i.e.,\ no partial credit is given for assignments satisfying most clauses. 
Instances are studied in the random ensemble near the satisfiability threshold, where the clause-to-variable ratio $\alpha=m/n$ is tuned close to the phase transition between almost certainly satisfiable and almost certainly unsatisfiable instances~\cite{Friedgut1999, Achlioptas2004}.

\paragraph{Potential Application Spaces}

Generic SAT-based workflows in the exact satisfiability regime are used where a satisfying assignment is only useful if all constraints are satisfied simultaneously. Heuristic near-optimal assignments are insufficient because even a single violated constraint may invalidate the entire output.
As such, highly structured industrial SAT instances, which often benefit strongly from specialized classical heuristics, are less likely to be among the direct applications~\cite{MarquesSilva2021CDCL}. 
Instead, model checking, hardware and software verification, cryptanalysis and key recovery, and automated theorem proving are tailored to this regime~\cite{Biere2009Handbook}. Another possible application space is constraint-based configuration problems in safety-critical settings, e.g.,\ in aerospace or medical control.

\paragraph{Classical Complexity}
Worst-case $k$-SAT is NP-complete for $k\geq 3$ and is widely believed to require exponential time in the number of variables. For random $k$-SAT, instances near the satisfiability threshold are empirically among the hardest, with state-of-the-art classical solvers exhibiting exponential scaling. However, exponential average-case hardness in this regime is not known in general and remains conjectural.
Far below the satisfiability threshold, where many satisfying assignments exist, exact classical solvers based on local search and message-passing methods work well with a near-linear practical scaling. Heuristic solvers are highly effective in this regime, too, solving industrial instances easily.
Far above the satisfiability threshold, heuristic approaches based on local search rapidly detect frustration or inconsistency, and unsatisfiability certificates are relatively easy to obtain.
The interesting regime is near-threshold, where the runtime is believed to scale exponentially in $n$ on average (not only worst-case). This regime is believed to contain some of the hardest instances for both classical and quantum algorithms~\cite{kirkpatrick1994}, 
as the solution space structure becomes highly nontrivial: assignments satisfying all constraints become exponentially rare and are organized into a clustered structure~\cite{Mezard2002}.
Among the best classical heuristic solvers are stochastic local search SAT solvers, e.g.,\ WalkSATlm with a time-to-solution that grows as $1.25^n$~\cite{Selman1994,Boulebnane2022-sj}. Their performance is empirical rather than worst-case guaranteed.

\paragraph{Quantum Complexity}
Focusing on the near-threshold regime, the Quantum Approximate Optimization Algorithm (QAOA) with sufficiently large depth $p$ is suggested to achieve a better exponential scaling constant than strong heuristic baselines on random $k$-SAT for large $n$ and high $k$~\cite{Boulebnane2022-sj}. That is, in the asymptotic scaling $O(\exp(\tau n))$, 
the factor $\tau$ is suggested to be smaller for large-depth QAOA than for the best classical heuristics.
Importantly, the comparison is heuristic vs.\ heuristic, i.e.,\ not against provably optimal classical algorithms or worst-case complexity separations, but against a strong, non-exhaustive classical baseline.
The analysis is based on an analytical argument for large $n$,
a derived scaling of the success probability with depth $p$, and is further supported by numerical experiments for small and medium $n$, where limiting theoretical bounds are matched closely.
For a specific example, in random $8$-SAT QAOA with depth $p=14$ outperforms the classical baseline WalkSATlm with a time-to-solution that grows as $1.23^n$, which can be improved to $1.11^n$ by combining it with amplitude amplification~\cite{Selman1994,Boulebnane2022-sj}.

\paragraph{Regimes of Advantage}
\begin{itemize}
    \item The strongest evidence for quantum advantage is concentrated where the clause-to-variable ratio $\alpha=m/n$ is near the satisfiability threshold, where instances are believed to be classically hard on average, with scaling exponent/hardness increasing with $k$.
    In this regime the observed advantage is a reduction in the effective exponential runtime base compared to classical heuristics, i.e.,\ a smaller scaling constant. 
    \item As mentioned above, the scaling advantage achieved by QAOA can be improved by combining it with amplitude amplification. If QAOA prepares a state with overlap $p_{\mathrm{opt}}$ on the satisfying configuration, amplitude amplification can be used to amplify that overlap to a constant with cost proportional to $1/\sqrt{p_{\mathrm{opt}}}$.
    The practicality of the resulting combined scaling advantage has been analyzed by compiling the algorithm to a planar surface code, yielding a crossover time of approximately $15$ hours for $8$-SAT at $n=179$ with amplitude amplification on roughly $74$ million physical qubits and a physical error rate of $10^{-3}$~\cite{2504.01897}.
\end{itemize}

\paragraph{Notes on Future Directions and Related Advantages}
A next step is to determine whether similar improvements in exponential scaling constants can be obtained for broader classes of constraint satisfaction and optimization problems, including Max-SAT and graph coloring, for instance. A related example is the low autocorrelation binary sequences (LABS) problem, an exact optimization problem in which one seeks a binary sequence minimizing its sidelobe energy. Numerical results indicate improved exponential scaling of QAOA combined with quantum minimum finding relative to the best known general-purpose classical heuristic for LABS~\cite{ShaydulinEtAl23LABS}. These results provide further evidence that QAOA-based approaches may yield nontrivial scaling improvements across structurally different hard optimization problems.

It will be important to understand how such advantages depend on problem structure and whether they extend beyond specific or random ensembles to practically relevant instance families. For LABS in particular, the numerical results of ref.~\cite{ShaydulinEtAl23LABS} exhibit a plateau with increasing QAOA depth, which may reflect limitations of the parameter-setting procedure and motivates further investigation of improved parameter optimization and larger depths. More generally, these results contribute to a growing body of evidence that quantum algorithms may achieve meaningful improvements for hard optimization problems without requiring superpolynomial complexity separations, potentially opening a wider landscape of practically relevant quantum advantages in combinatorial optimization.

\pagebreak
\section{Planted Inference}\label{sec:planted-inference}

\emph{Section contributed by:} Alexander Schmidhuber \\

Planted inference problems are average-case statistical tasks in which an instance is drawn either from a null distribution with no hidden structure, or from a planted distribution containing a latent signal, assignment, cluster structure, or low-rank component. The algorithmic goal may be \emph{detection}, i.e.,\ distinguishing the planted and null distributions; \emph{weak recovery}, i.e.,\ producing an estimator nontrivially correlated with the hidden signal; or \emph{strong/exact recovery}, i.e.,\ reconstructing the signal up to its intrinsic symmetries.

The value of planted inference models is that they provide a convenient abstraction of recurring
mathematical motifs in noisy data analysis: a weak global signal is encoded in many individually ambiguous local measurements, and the algorithmic question is whether their correlations can
be aggregated efficiently. Planted inference problems generically exhibit so-called statistical-computational gaps: regimes in which detection and recovery are possible information-theoretically, while broad families of polynomial-time classical algorithms---spectral methods, approximate message passing, low-degree polynomials, and the
sum-of-squares hierarchy---are known or conjectured to fail~\cite{HSS2015,HKPPRS2017,WeinElAlaouiMoore2019,Kunisky2025LCD}.

The three planted-inference quantum speedups discussed here---Tensor PCA, Planted Noisy \(k\)-XOR, and hypergraph community detection---all arise from the same high-level template. The detection (recovery) problem is encoded in the leading eigenvalue (eigenvector) of a so-called Kikuchi matrix of dimension $O(n^\ell).$ On a quantum computer this matrix can be represented using only $O(\ell \log(n))$ qubits. The eigenvalue or eigenvector is computed classically in time $O(n^\ell)$ via the power method or related algorithms. The core algorithmic ingredient behind the quartic
speedups is the efficient preparation of a \emph{guiding state}: a simple state assembled directly from the instance data whose overlap with the planted eigenspace is quadratically enhanced, by a factor of roughly $O(n^{\ell/2})$, precisely
because the planted eigenvector is itself correlated with the data. This allows quantum phase estimation combined with amplitude amplification to estimate the relevant eigenvalue using roughly the fourth root of the classical cost, while using super-polynomially less space:
\[
    T_{\rm classical} \approx O( n^\ell),
    \qquad
    T_{\rm quantum} \approx O(n^{\ell/4}\operatorname{poly}(n)),
\]
up to problem-dependent polynomial and polylogarithmic factors~\cite{Hastings2020,SchmidhuberODonnellKothariBabbush2025,SchmidhuberZlokapa2025}. Notably, the same guiding state is efficiently accessible classically but does not yield a similar improvement there. 

The significance of a quartic speedup is best assessed against the overheads
of fault-tolerant quantum computation. Resource analyses indicate that
quadratic speedups are unlikely to yield a practical advantage once the
overheads of quantum error correction are accounted for, and that
substantially larger polynomial speedups are required for a realistic
crossover~\cite{BMN+21}; quartic speedups are among the first
candidates to clear this bar
\cite{BMN+21,SchmidhuberODonnellKothariBabbush2025,2510.07273}. The algorithms
discussed here additionally have unusually small space requirements of order
$\widetilde{\mathcal O}(\ell \log n)$ qubits, making them natural candidates
for early fault-tolerant hardware.

These quartic speedups exhibit three properties---average-case hardness, efficiently generatable hard instances, and classically verifiable outputs---that distinguish planted inference from many quantum machine-learning proposals whose input model or output verification is more fragile~\cite{babbush2026grand}.

\subsection{Tensor PCA}

\emph{Maximum Quantum Speedup: Quartic}\\
\emph{Evidence Basis: Rigorous quantum algorithm with average-case lower bounds for restricted classical methods}

\paragraph{Problem Description}

A canonical example is Tensor Principal Component Analysis (Tensor PCA). Also called the spiked tensor model, it asks one to detect or recover a low-rank signal hidden in a random order-\(p\) tensor. The standard rank-one-plus-noise model is
\[
    Y = \beta x^{\otimes p} + G,
    \qquad
    \|x\| = \sqrt{n}
    \quad\text{or}\quad
    x\in\{\pm 1\}^n,
\] where $G$ is a random symmetrized Gaussian noise tensor and $\beta$ is the Signal-to-Noise ratio (SNR).
The detection task is to distinguish the spiked distribution from pure Gaussian noise, while recovery asks for an estimator correlated with \(x\), up to the sign symmetry when \(p\) is even~\cite{MontanariRichard2014,HSS2015,WeinElAlaouiMoore2019}.
Finally, the tensor completion task takes as input a small number of observations of the tensor's entries and asks to complete the missing entries~\cite{2510.07273}.
The associated maximum-likelihood problem is the optimization of a random degree-\(p\) polynomial,
\[
    \max_{u\in \{\pm 1\}^n}
    \sum_{S\in \binom{[n]}{p}} Y_S \prod_{i\in S} u_i,
\]
or its spherical analogue. 

\paragraph{Potential Application Spaces}

Tensor PCA is best viewed as a controlled abstraction of high-order signal extraction. It captures the setting in which the informative statistics are not pairwise covariances but \(p\)-way correlations. This can be relevant to tensor decomposition methods in psychometrics, chemometrics, signal processing, numerical linear algebra, computer vision, neuroscience, data mining, and graph analysis~\cite{KoldaBader2009}. It is also closely related to method-of-moments algorithms for latent-variable models, where moments of data from Gaussian mixtures, hidden Markov models, and topic models have structured tensor decompositions~\cite{AnandkumarGeHsuKakadeTelgarsky2014}.

\paragraph{Classical Complexity}

For \(p=2\), the maximum-likelihood problem associated with Tensor PCA reduces to the spiked Wigner/matrix PCA problem, where spectral methods are essentially optimal. The qualitatively new regime begins at \(p\ge 3\), where 
Tensor PCA exhibits a large gap between information-theoretic and known efficient thresholds. Detection and recovery are information-theoretically possible down to roughly
\[
    \beta \gg n^{(1-p)/2},
\]
while the best polynomial-time classical algorithms are known to succeed only if
\[
    \beta \gtrsim n^{-p/4},
\]
up to constants and logarithmic factors, with matching evidence from low-degree and sum-of-squares lower bounds~\cite{MontanariRichard2014,HSS2015,HKPPRS2017,WeinElAlaouiMoore2019}. The $\ell$-th order Kikuchi method has classical time complexity $\widetilde{\mathcal O}(n^\ell)$
and succeeds at signal strengths of order
\[
    \beta \gtrsim n^{-p/4}\ell^{1/2-p/4},
\]
up to constants and, in earlier analyses, logarithmic factors~\cite{WeinElAlaouiMoore2019,KothariXu2026}. Thus constant \(\ell\) gives polynomial-time algorithms at the usual computational threshold, while \(\ell\) growing with \(n\) gives increasingly expensive algorithms that move toward the information-theoretic regime.

\paragraph{Quantum Complexity}

Hastings showed that the same high-dimensional spectral method for Tensor PCA can be quantized, yielding a quartic speedup and a superpolynomial space saving over the fastest classical spectral algorithm~\cite{Hastings2020}. When the level-\(\ell\) Kikuchi method has classical cost \(\widetilde{\mathcal O}(n^\ell)\), the corresponding quantum cost is
\[
    \widetilde{\mathcal O}\!\left(n^{\ell/4}\operatorname{poly}(n)\right),
\]
while using only \(\widetilde{\mathcal O}(\ell\log n)\) qubits~\cite{Hastings2020,SchmidhuberODonnellKothariBabbush2025}. 

\paragraph{Regimes of Advantage}

\begin{itemize}
\item For small constant $\ell$, the $\poly(n)$ overhead in the quantum runtime reduces the speedup below quartic. The choice of $\ell$ for which the speedup in time is largest is therefore a large constant or slowly growing (say, $\ell \sim \log(n)$). 
    \item The speedup applies only in the hard-but-detectable regime where \(\beta\) is below the best poly\-nomial-time threshold but above the information-theoretic threshold. Due to the dependency on the guiding state overlap, the speedup is largest (quartic) near the computational threshold $\beta \lesssim n^{-p/4}$; as one moves closer to the information-theoretic threshold the speedup decays smoothly, eventually reaching a minimal value that is still super-quadratic. 
    \item The case \(p=2\) is not in scope for this speedup. Ordinary matrix PCA is already efficiently solved by classical spectral methods at the statistical threshold; the quartic speedup is a higher-order phenomenon.
\end{itemize}

\paragraph{Notes on Future Directions and Related Advantages}

\begin{itemize}
\item A major open direction is whether the quantum speedups for Tensor PCA can be improved beyond quartic. Recent results suggest promising approaches towards degree-6 speedups~\cite{hastings2026accelerating}, but currently require conditions on the spectral density of the Kikuchi Hamiltonian that do not hold generically.

    \item A second major open direction is robustness: real tensors may be sparse, non-Gaussian, structured, multi-spiked, or contaminated. Establishing analogous guided-state constructions for such models would make the result more relevant to real-world data analysis.

\end{itemize}

\subsection{Planted Noisy \(k\)-XOR / Sparse LPN}

\emph{Maximum Quantum Speedup: Quartic}\\
\emph{Evidence Basis: Rigorous quantum algorithm with average-case lower bounds for restricted classical methods}

\paragraph{Problem Description}

In planted \(k\)-XOR, one observes noisy parity measurements of an unknown assignment $z \in \{\pm 1\}^n$.
For a set \(A\subseteq[n]\), define the set exponent
$
    z^A := \prod_{j\in A} z_j .
$
The instance consists of \(m\) independently sampled \(k\)-sets
\[
    A_1,\ldots,A_m \in \binom{[n]}{k}
\]
and signs
\[
    b_i = z^{A_i}\eta_i,
    \qquad i=1,\ldots,m,
\]
where the noise signs \(\eta_i\in\{\pm1\}\) are independent with
$
    \mathbb E[\eta_i]=\rho .
$
Thus each constraint is correct with probability \((1+\rho)/2\). The null model keeps the same random \(k\)-sets \(A_i\) but replaces the labels \(b_i\) by independent unbiased random signs. The detection task is to distinguish the planted and null distributions; the recovery task is to output an assignment correlated with \(z\). This problem is also called Sparse Learning Parity with Noise in the cryptography community. The associated (average-case and distributional) optimization problem is the planted MAX \(k\)-XOR objective
\[
    \max_{x\in\{\pm1\}^n}
    \sum_{i=1}^m b_i x^{A_i}.
\]

\paragraph{Potential Application Spaces}

The main application space is average-case cryptography and random-CSP complexity. Sparse LPN and planted noisy \(k\)-XOR appear in constructions and assumptions related to XOR codes, pseudorandomness, lossy and code-based cryptography, indistinguishability obfuscation, and secure computation primitives such as multiparty homomorphic secret sharing and sublinear MPC~\cite{BSV2019,JLS22,DIJL23,DJ24}. More broadly, the model is a clean abstraction of inference on sparse hypergraphs: a hidden global assignment is weakly encoded in many noisy local parity checks. The same mathematical structure also appears in coding theory, graphical models, statistical physics, and the study of semidefinite and sum-of-squares hierarchies.

\paragraph{Classical Complexity}

Let
$
    \Delta := \frac{m}{n}
$
be the constraint density. For constant \(k\) and constant noise bias \(\rho\), detection is information-theoretically possible only if the constraint density exceeds 1, $\Delta > 1$. Efficient classical algorithms require substantially larger density \[
    \Delta \gtrsim n^{(k-2)/2}.
\] The standard spectral approach is the Kikuchi method: at level \(\ell\), it solves Planted Noisy $k$-XOR using time and space $ \widetilde{\mathcal O}(n^\ell)$
provided 
\begin{align*}
\Delta \gtrsim \left(\frac{n}{\ell}\right)^{(k-2) / 2}\ . 
\end{align*}
 This hard intermediate regime is supported by a substantial body of classical evidence from random-CSP refutation, Lasserre/sum-of-squares lower bounds, and related average-case hardness results \cite{AR01,Sch08,OW14,MW16,KMOW17,WeinElAlaouiMoore2019}. As for Tensor PCA, the above SNR-runtime trade-off is tight up to constants; previous loose logarithms were recently removed~\cite{SchmidhuberHastings2026kXOR,SchmidhuberHastings2026Moore}.

\paragraph{Quantum Complexity}
The quantum algorithm~\cite{SchmidhuberODonnellKothariBabbush2025} achieves a quartic quantum speedup. Informally, whenever the level-\(\ell\) Kikuchi method solves the problem at a given density using time
\[
    \widetilde{\mathcal O}(n^\ell),
\]
the quantum algorithm solves the same planted detection problem with runtime
\[
    n^{\ell/4}\operatorname{poly}(n),
\]
and qubit count 
$
    \widetilde{\mathcal O}(\ell\log n).
$ This quantum algorithm was recently sharpened and extended to recovery and odd values of $k$ in ref.~\cite{SchmidhuberHastings2026kXOR}.

\paragraph{Regimes of Advantage}

\begin{itemize}
    \item In analogy to Tensor PCA, the natural advantage regime is 
    \[
        1 \ll \Delta \lesssim n^{(k-2)/2},
    \]
    where the signal is statistically present but the classical Kikuchi level \(\ell\) is a large constant or grows slowly with \(n\). In this regime, the speedup is near-quartic. At the information-theoretic threshold, the speedup is sub-quartic but super-quadratic.

    \item A recent classical algorithm gives a quadratic improvement over the previous best classical planted \(k\)-XOR algorithm for sufficiently large constant \(k\), reducing the time advantage in that large-\(k\) regime from quartic to quadratic, while preserving a space advantage for the quantum algorithm~\cite{GuptaHeODonnellSinger2025}. For small fixed \(k\), especially the low-arity regimes most often used as clean benchmarks, the Kikuchi-based quartic comparison remains the central one (see also the discussion on future directions).
\end{itemize}

\paragraph{Notes on Future Directions and Related Advantages}

\begin{itemize}
\item An important open direction is to verify whether the classical algorithm in ref.~\cite{GuptaHeODonnellSinger2025} extends to the practically relevant regime of small $k$ or whether it can be quantized by also incorporating a guiding state, hence re-establishing the quartic speedup for all $k$. 

\item Most standard post-quantum cryptosystems are based on dense LPN or lattice assumptions such as LWE. Sparse LPN is a nearby sparse planted primitive that has also been used in cryptographic constructions. It remains an interesting direction for further research to try to generalize these speedups for sparse problems to dense cryptographic primitives.

\end{itemize}
\subsection{Community Detection in Higher-Order Hypergraph SBMs}

\emph{Maximum Quantum Speedup: Quartic}\\
\emph{Evidence Basis: Rigorous quantum algorithm with average-case lower bounds for restricted classical methods}

\paragraph{Problem Description}

Community detection asks one to infer latent vertex labels from relational data. The central task is to infer communities---roughly speaking, subsets of vertices that contain unusually many hyperedges---given the adjacency tensor of a hypergraph. The quartic speedup applies specifically to higher-order hypergraph community detection, where the informative correlations live in \(p\)-body interactions of vertices~\cite{SchmidhuberZlokapa2025}.

A representative \(p\)-uniform hypergraph stochastic block model (SBM) has labels
\[
    x_i \in [q],
    \qquad
    i=1,\ldots,n,
\]
and for every \(p\)-subset \(S=\{s_1,\ldots,s_p\}\) one observes a hyperedge indicator
\[
    Y_S \sim \operatorname{Bernoulli}\!\left(\theta_0+\epsilon f(x_{s_1},\ldots,x_{s_p})\right)
\]
under the planted model, while under the null model
\[
    Y_S \sim \operatorname{Bernoulli}(\theta_0)
\]
independently. The function \(f\) is chosen so that lower-order marginals vanish: no statistic of fewer than \(p\) labels reveals the communities. The detection task is to distinguish the planted hypergraph from the null hypergraph or, more practically, recover the labels~\cite{SchmidhuberZlokapa2025}.

\paragraph{Potential Application Spaces}

Higher-order community detection abstracts clustering problems in which pairwise relationships do not capture all relevant structure. Examples include hypergraph clustering, simplicial-complex data analysis, group interactions in social and collaboration networks, co-purchase and recommendation data, multiway biological interactions, gene regulation, protein-complex data, neural co-activation, and higher-order relational learning~\cite{SchmidhuberZlokapa2025}. In these settings, the basic signal may be that certain groups of \(p\) entities co-occur or interact in a way that cannot be decomposed into pairwise affinities.

\paragraph{Classical Complexity}

For the \(p\)-marginal hypergraph SBM, define the normalized signal-to-noise ratio
\[
    \operatorname{SNR}
    =
    \frac{\epsilon}{\sqrt{\theta_0(1-\theta_0)}}.
\]
Ref.~\cite{SchmidhuberZlokapa2025} proves low-coordinate-degree lower bounds, which include the usual low-degree-polynomial evidence for average-case hardness. Informally, degree-\(\ell\) methods fail below
\[
    \operatorname{SNR}
    \lesssim
    \ell^{1/2-p/4} n^{-p/4}.
\]
They also give a matching Kikuchi upper bound up to a logarithmic factor: for even \(p>2\), the \(\ell\)-th order Kikuchi method solves the strong-detection problem in
\[
    \mathcal O(n^\ell)
    \quad\text{time and space}
\]
provided
\[
    \operatorname{SNR}
    \gtrsim
    \ell^{1/2-p/4}n^{-p/4}\sqrt{\log n}.
\]
Thus, as in Tensor PCA, there is a smooth SNR-computation tradeoff: larger lifted dimension \(n^\ell\) allows detection at lower signal strength~\cite{SchmidhuberZlokapa2025}.

\paragraph{Quantum Complexity}

There is an explicit quantum algorithm solving the same detection problem in time
\[
    n^{\ell/4}\,\widetilde{\mathcal O}(n^p),
\]
using
\[
    \widetilde{\mathcal O}(\ell\log n)
\]
qubits and
\[
    \widetilde{\mathcal O}(\ell n^p)
\]
classical space~\cite{SchmidhuberZlokapa2025}. Compared with the classical \(\mathcal O(n^\ell)\) time and space cost, this gives a quartic time speedup in the dominant lifted-dimension factor and a superpolynomial space advantage.
Here, community detection refers specifically to the strong detection problem: distinguishing with vanishing error probability whether an observed hypergraph was drawn from the planted Hypergraph SBM or from the corresponding null distribution. The algorithm does not currently establish recovery of the underlying community labels. While the relevant eigenstate is conjectured to contain sufficient information for label recovery via tomography and rounding, establishing such a recovery algorithm remains an open problem.

\paragraph{Regimes of Advantage}

\begin{itemize}
 \item As before, the speedup is largest if \(\ell\) is a large constant or grows slowly with \(n\). 
 
    \item The result applies to higher-order models with marginal order greater than two. Ordinary pairwise SBMs are not the target: they have efficient classical algorithms in many regimes and do not exhibit this Kikuchi-based quartic speedup.

\end{itemize}

\paragraph{Notes on Future Directions and Related Advantages}

\begin{itemize}
\item The rigorous result is for detection, not for recovery. By analogy with Tensor PCA, the prepared eigenstate is expected to encode information about the planted community labels, and a tomography-and-rounding procedure may yield recovery. Establishing this recovery guarantee remains an open direction~\cite{SchmidhuberZlokapa2025}.

    \item The paper suggests that a quantity called marginal order may provide a general framework for characterizing when Kikuchi-type quantum speedups~\cite{SchmidhuberZlokapa2025} exist, unifying the three examples in this section. Expanding on this remains an important direction for further research.

\end{itemize}
\pagebreak


\section{Quantum Simulation}

\emph{Section contributed by:} Thomas E. O'Brien\\

Quantum simulation is the sub-area of quantum applications research concerned with calculating properties of physical systems in quantum mechanical regimes.
(This excludes, e.g.,\ work on classical oscillators, and methods to solve the Poisson or wave equations, which calculate properties of physical systems in classical regimes.)
Quantum simulation stands out from all of the other sections in this compendium by simultaneously being one of the largest fields of quantum applications research~\cite{mcardle2020quantum,bauer2020quantum,Cao2019QuantumComputing}, but one of the fields with the fewest rigorous results for Stage II-type quantum advantage. 
While it is known that many Hamiltonians have BQP-Complete dynamics~\cite{cubitt2018universal}, verifiable quantities such as local observables often concentrate to predictable values after durations of time-evolution that would be asymptotically difficult to simulate~\cite{Wild2023}.
And, while significant and seminal work has gone into analyzing specific problems in chemistry~\cite{Reiher2017,Li2019,vonBurg2020,Goings2022-oj}, condensed matter~\cite{Wecker2015b,Kivlichan2020ImprovedTrotterization, yoshioka2024hunting,wu2024variational} and materials science~\cite{Bauer2015,Rubin2023-rm,clinton2024towards,rubin2024quantum}, these works have overwhelmingly considered specific instances of problems rather than ensembles.
This makes robust competition with classical methods (and thus a clear claim of quantum advantage) challenging, as one cannot simply `draw a new problem from the ensemble' to determine whether targeted heuristic classical methods continue to perform well.

Given recent questions about the claim of an exponential quantum advantage in chemistry~\cite{lee2023evaluating} and continuing classical advances in solving strongly-correlated quantum systems~\cite{SimonsHubbard,zhai2025classical}, we believe that more Stage II research is highly necessary to better understand the prospects for quantum computing in one of its most high profile application spaces, and we hope the distinction drawn in this review helps to motivate that work.

In this section we describe the (currently) singular example of a topic in quantum simulation where specific Stage II ensembles of problem instances with evidence for classical hardness have been determined.
We refer to Sec.~\ref{sec:stageIqsim} for a longer list of problems in quantum simulation that lie within Stage I, because no specific ensembles with typical quantum advantage have been formally defined as of this time.

\subsection{Out-of-Time-Order Correlators}
\label{sec:OTOC}

\emph{Maximum Quantum Speedup: Exponential}\\
\emph{Evidence Basis: Rigorous quantum algorithm with an experimental benchmark against the classical state of the art}\\

This problem class captures the set of simulation experiments requiring both forward and backward time evolution.
Most notably, recent experiments measuring out-of-time-ordered correlators (OTOCs) have shown strong evidence for being typically classically hard in the correct parameter regimes~\cite{google2025observation,king2025simplified}.
However, this includes also proposals for quantum gravity experiments involving many-body teleportation~\cite{brown2023quantum,schuster2022many}, as well as experiments in nuclear magnetic resonance such as Loschmidt echoes~\cite{Levstein1998} and multiple-quantum coherence spectroscopy~\cite{baum_1985}.

\paragraph{Problem Description}

Let $H$ and $\rho$ be a Hamiltonian and state on an $N$-dimensional quantum system with $\|H\|\leq 1$. Fix $O_1, O_2\in\mathbb{C}^{N\times N}$ with $\|O_1\|\leq 1$, and let $O_2(t)=e^{iHt}O_2e^{-iHt}$ for some time $t\in\mathbb{R}$. Estimate $\mathrm{Trace}[\rho (O_1 O_2(t))^{2k}]$ for some integer $k\geq 1$ to additive precision $\epsilon$.

Alternatively, let $U$ be a quantum circuit on an $N$-dimensional quantum system, and estimate $\mathrm{Trace}[\rho (O_1 U^{\dagger}O_2U)^{2k}]$ to additive precision $\epsilon$. 

When $k>1$, we say that the above formulas give $k$th order OTOCs, or OTOC$^{k}$.

\paragraph{Potential Application Spaces}

OTOCs were first explored in the context of scrambling in gravitational systems~\cite{maldacena2016bound}; they allow one to estimate the rate of information propagation in a system (as they track whether the operator $O_2$ has reached the sites where $O_1$ has non-trivial support).
This can be used to distinguish different universality classes~\cite{yao2016interferometric}; e.g.,\ between fast scrambling models and those with Lieb-Robinson bounds~\cite{zhou2020operator}, or between Anderson and many-body localized phases~\cite{alet2018many}, or between random circuits with~\cite{rakovszky2018diffusive} and without~\cite{nahum2018operator,vonkeyserlingk2018operator} conservation laws.

More recently, some interest has been garnered in measuring the distribution of $O_2(t)$ over weight-$k$ Pauli operators (i.e.,\ as a function of $k$)~\cite{qi2019measuring}.
This concept can be further used in many-body teleportation protocols which are relevant to the quantum gravity community for the purposes of better understanding holographic models~\cite{brown2023quantum,schuster2022many}.

Time reversal can be implemented in certain naturally-found spin systems, most notably for the dipolar coupling found in nuclear magnetic resonance.
This allows experiments such as multiple-quantum coherence spectroscopy~\cite{baum_1985}, Loschmidt echoes~\cite{Levstein1998}, and out-of-time-ordered correlators to be measured in natural systems, where they can contain information about molecular structure and many-body spin decoherence.
Recent works~\cite{schuster2023learning,zhang2025quantum} have suggested these may allow measurement of longer distances than conventional NMR techniques.

\paragraph{Classical Complexity}

One can encode an arbitrary target circuit in the output of an OTOC circuit, by making $O_2$ a CNOT gate between the circuit's output bit and an ancilla qubit, and $O_1$ the $Z$ operator on the ancilla.
Thus, OTOC circuits are BQP-complete.
Numerical evidence strongly suggests OTOCs of random circuits are typically challenging to classically simulate using standard techniques such as matrix product state, neural quantum state, mean field, or Pauli pathing techniques~\cite{google2025observation}, and similar results exist for Hamiltonian systems~\cite{zhang2025quantum}.

For many systems, the challenge in classically simulating OTOCs occurs only for intermediate times $t$.
Typically, a local operator $O_2(t)$ propagates out with a diffusive front; when $O_1$ lies outside this front, $\mathrm{Trace}[\rho (O_1 O_2(t))^{2k}]=1$, but once the front has completely passed $O_1$, $\mathrm{Trace}[\rho (O_1 O_2(t))^{2k}]=0$ (up to $\mathcal{O}(1/N)$ corrections); in both cases classical solutions are trivial.

A class of methods that have had some success in simulating OTOCs are Monte Carlo Pauli pathing algorithms in the doubled operator space~\cite{google2025observation}.
These can simulate OTOCs in random circuits with sufficiently low error to prevent claims of quantum advantage.
The effectiveness of this approximation can be tracked by the sensitivity of the circuit to correlated Pauli noise (i.e.,\ where Pauli operators are inserted randomly into the forward evolution step $U$ and the backward evolution step $U^{\dagger}$ as correlated pairs).
Higher-order OTOCs are significantly more sensitive to the insertion of correlated Pauli noise than $k=1$ OTOCs, making this classical method infeasible~\cite{google2025observation}.
This sensitivity is also significantly larger for Hamiltonian OTOCs than in random circuits~\cite{zhang2025quantum}, suggesting that classical algorithms for these problems may also suffer from exponentially scaling time complexity.

\paragraph{Quantum Complexity}

Given access to circuit encodings of $U$ (or $e^{iHt}$), $O_1$ and $O_2$, one can compute an OTOC on a quantum computer to precision $\epsilon$ with $\epsilon^{-2}$ circuit repetitions, or with a single circuit of length $\epsilon^{-1}$ using amplitude estimation.
One can measure exponentially many OTOCs simultaneously for states $\rho$ near the identity, by working with a shadow Hamiltonian simulation~\cite{somma2025shadow,chiew2026quantum} (see Sec.~\ref{sec:shadowH}).

The circuit depth requirements for an OTOC are typically dominated by the cost of simulating $U$ or implementing $e^{iHt}$.
As we are often interested in intermediate times $t\sim \log(N)^{1/d}$ for a $d$-dimensional system, this can be quite low for, e.g.,\ local spin systems or random circuits.
Measuring out-of-time-ordered correlators requires either access to $U^{\dagger}$ ($e^{-iHt}$) or the ability to interfere multiple copies of the state~\cite{cotler2022information,schuster2023learning}; this is typically not a challenge for digital quantum computers, but it can make analog OTOC experiments challenging.

\paragraph{Regimes of Advantage}

As soon as exact classical methods break down (and for systems where Monte Carlo methods fail), OTOCs can achieve quantum advantage.
This has already been achieved for a $65$ qubit $\mathrm{OTOC}^{2}$ experiment using random iSWAP circuits on a 2D square lattice~\cite{google2025observation}.
The leading classical method for simulating these circuits is direct tensor network circuit contraction, which implies that deeper circuits may yet require fewer qubits to achieve advantages.
Work is still required in this space to determine specific instances of problems of external relevance (e.g., in nuclear magnetic resonance), and elevate this problem to Stage III research.


\clearpage
\part{Stage I Problems: Primitives with Evidence for Quantum Advantage}\label{Part B}

The problems selected in Part~\ref{Part B} are subject to quantum algorithms in a less advanced stage than those discussed in the previous Part. The analysis provides here evidence of how these quantum algorithms may operate in abstract settings. With continued development by the scientific community, these {Stage I} results could advance to {Stage II}, helping to identify more concrete problem instances. This is a critical step before their practical relevance can be meaningfully assessed, or their applicability to real-world settings considered.


\pagebreak

\section{Quantum Simulation}\label{sec:stageIqsim}

\emph{Section contributed by:} Thomas E. O'Brien, Rolando D. Somma, Siddhartha Jain, Philipp Kammerlander, Yuri Alexeev, Elica Kyoseva\\

Quantum simulation stands out from all of the other examples in this compendium by being one of the most extensive areas of study within quantum-algorithm applications research~\cite{mcardle2020quantum,bauer2020quantum,Cao2019QuantumComputing}, but one that often lacks substantial rigorous evidence of a practical quantum advantage.
Consequently, aside from the special case of out-of-time-ordered correlators (Sec.~\ref{sec:OTOC}) being in Stage II~\cite{google2025observation,bermejo2026tensor}, most other quantum simulation problems remain limited to Stage I.
This limitation is partly due to the difficulty of specifying natural ensembles of quantum systems that can provably result in classically-intractable dynamics (cf.~\cite{quek2025quantum}).
It does not lessen the substantial progress made in developing quantum simulation algorithms and applications in recent years, and quantum simulation continues to be a---if not the---leading application for quantum computing. Rather, it highlights the critical need for further interdisciplinary collaboration bridging theoretical physics and computer science.

Research in quantum simulation is often focused on quantum algorithms for state-preparation problems.
A main example is Hamiltonian simulation~\cite{lloyd1996universal} to study dynamics, where the time-evolved state is prepared by implementing a quantum circuit that ``mimics'' the evolution operator. Another example 
is the preparation of eigenstates or thermal Gibbs states
to study systems in equilibrium~\cite{AspuruGuzik2005,McClean2015,holmes2022quantum,Berry2024}. 
Because these problems are fundamentally quantum in nature, exploring quantum advantage requires adapting them to produce classical outputs (or sampling distributions), such as the expected value of a physical property at a specific time. 
This classical data can be efficiently computed from the quantum state using methods like quantum state tomography~\cite{cramer2010efficient}, expectation value estimation~\cite{knill2007expectation}, amplitude estimation~\cite{Brassard2002}, or phase estimation~\cite{kitaev1995quantum}. 
Our list of problems below is concerned with such 
computations.

We note that recent work, often referred to as ``Shadow Hamiltonian Simulation'',
has explored quantum simulation problems without requiring access to the full Hilbert space~\cite{somma2025shadow,stroeks2024solving,Chen2025}.
As 
we discuss explicitly in Sec.~\ref{sec:shadowH},
 such approaches offer an alternative and compact way of representing quantum states, with the potential to efficiently model systems with significantly more degrees of freedom than traditional quantum simulation.
 Nevertheless, since such approaches track only specific information about the system, they naturally impose constraints on the types of properties that can be computed.


\subsection{Ground-State Energy}\label{sec:GSE}

\emph{Maximum Quantum Speedup: Exponential}\\
\emph{Evidence Basis: Rigorous quantum algorithm against the classical state of the art}

\paragraph{Problem Description}

Understanding the ground-state configuration is a foremost problem in quantum mechanics, as it allows 
for studies of the low-energy physics of molecules, solid state systems, and beyond. Consequently,
estimating the ground state energy of a quantum system is a well studied computational problem in quantum physics, both via classical and quantum algorithms.

\emph{Formal:} Consider a quantum system of exponential dimension $N=2^n$. Given a succinct representation of a Hamiltonian 
$H\in\mathbb{C}^{N\times N}$ 
with spectral norm $\|H\| \in \poly(n)$, the goal is to estimate the lowest eigenvalue $E_0$ of $H$ to additive precision $\epsilon \ge 0$. 

\paragraph{Potential Application Spaces}

Ground-state energy estimation of the electronic structure problem has a key role within computational chemistry and materials science, as this quantity dictates molecular structure, reaction rates, and reaction enthalpies~\cite{mcardle2020quantum, bauer2020quantum}.
A well-studied problem is that of the FeMoco cluster within nitrogenase~\cite{Reiher2017}, which has been a standard benchmark for much of quantum simulation algorithm development~\cite{su2021faulttolerant,Lee2020}, despite recent classical progress in solving this problem~\cite{zhai2025classical}.
The target here is to map out the potential energy surface of the molecule in order to understand molecular geometries and reaction mechanisms.
This is a common workflow across transition metal chemistry~\cite{bauer2020quantum}, catalytic chemistry~\cite{mcardle2020quantum}, pharmaceutical chemistry~\cite{Goings2022-oj}, and solid state materials chemistry~\cite{clinton2024towards, Rubin2023-rm, berry2024quantum}.
Ground-state energy calculations are also relevant for determining properties like adsorption energies for carbon capture~\cite{montgomery2025multiscale, rocca2025quantum} and determining energy densities for battery materials~\cite{Rubin2023-rm}. 
In physics, ground-state energies and configurations are crucial for understanding high-temperature superconductivity~\cite{Lee2006, Fradkin2015, dallaire2016method} and strange metal / non-Fermi liquid phases of matter~\cite{Phillips2022,Lee2018}, spin liquids~\cite{balents2010spin,Kitaev2006, Kasahara2018}, topological states of matter~\cite{google2023nonabelian, will2025probing}, and determining vacuum energies in lattice gauge theories~\cite{banuls2020simulating, ciavarella2023quantum}.
Ground-state energies also allow for the computation of order parameters (quantum phases) and other expectations in the ground state via the Hellmann-Feynman theorem~\cite{Feynman1939}.
Additionally, since some quantum algorithms for ground-state energies can be directly extended to estimating the energy of any other eigenstate, they unlock further applications in the excited space (see Sec.~\ref{sec:DOS}).

\paragraph{Classical Complexity}\label{subsec:gse_classical}

The worst-case complexity of estimating the lowest eigenvalue of a $2^n \times 2^n$ Hamiltonian is exponential in $n$. 
This holds even when the matrix is presented in a succinct form, such as a Hamiltonian acting on $n$ qubits with nearest-neighbor interactions, and when the error $\epsilon \in 1/\poly(n)$ is inverse polynomial in $n$.\footnote{Estimating the energy to a constant error per site (i.e.,\ with an error growing linearly in $n$) is known to be classically easy for a range of models, including all local spin models~\cite{bansal2009classical,brandao2013product}.}
Classical exact diagonalization methods, including Lanczos, Davidson, and Krylov subspace methods~\cite{lanczos1950, davidson1975, sleijpen1996}, are rendered inefficient in this context. While each sparse matrix-vector multiplication scales as $\mathcal{O}(\text{poly}(n) 2^n)$, the vector dimension $N = 2^n$ rapidly exceeds memory bandwidth, and the number of Krylov iterations required to resolve small spectral gaps compounds this exponential scaling. By comparison, dense eigensolvers require $\mathcal{O}(2^{3n})$ operations and $\mathcal{O}(2^{2n})$ memory.

Indeed, ground-state energy estimation with inverse polynomial error is a quintessential QMA-Complete problem in quantum complexity~\cite{kempe2006complexity}, and no generic efficient classical or quantum algorithm for this problem is believed to exist.
Moreover, systems are known for which no low-energy states can be prepared with short-depth quantum circuits, which suggests that even approximating energies for these systems classically should incur exponential cost~\cite{anshu2023nlts}. 

While the complexity can be exponential in the worst case, in practice classical methods perform better for ground state energies than other simulation tasks for two reasons: 1) ground state wavefunctions of physical systems can have low entanglement, and 2) ground state energies can be variationally bound using many methods.
Indeed, problem instances exist for which classical methods have proven polynomial-time complexity in the system size $n$ (or inverse error $1/\epsilon$), including non-interacting fermions~\cite{terhal2002classical}, certain stoquastic Hamiltonians that avoid the sign problem~\cite{bravyi2008complexity}, and gapped one-dimensional systems~\cite{landau2015polynomial}.
Heuristics like density matrix renormalization group (DMRG)~\cite{white1992density}, tensor network methods~\cite{verstraete2008matrix}, coupled-cluster methods~\cite{cizek1966correlation}, quantum Monte Carlo methods~\cite{ceperley1980ground}, reduced density matrix methods~\cite{nakata2001variational}, Selected-CI methods~\cite{huron1973iterative}, and machine learning methods~\cite{carleo2017solving} have achieved impressive accuracy in problems in chemistry~\cite{chan2011density} and materials science~\cite{kotliar2006electronic}.
Ground state energy calculations also often require only qualitative accuracies, which can often be achieved by classical methods such as mean field theory~\cite{roothaan1951new}, dynamical mean field theory~\cite{georges1996dynamical}, density functional theory~\cite{hohenberg1964inhomogeneous}, and perturbation theory~\cite{moller1934note}.
These thus remain popular for low-energy calculations in quantum chemistry~\cite{szabo2012modern} and materials science, being typically easy to implement and interpret.

\paragraph{Quantum Complexity}
Quantum algorithms can provide the ground-state energy efficiently provided access to a ``good'' state that has large overlap with the ground state~\cite{gharibian2023dequantizing}.
However, since the problem is QMA-hard in general~\cite{kempe2006complexity}, ground-state preparation is inefficient in the worst-case. 

In practice, strategies for preparation of good initial states include quantum adiabatic approaches~\cite{farhi2000,AspuruGuzik2005,Reiher2017}, variational quantum eigensolvers~\cite{McClean2015}, DMRG-based~\cite{Berry2024}, and dissipation-based~\cite{zhan2026rapid} algorithms.
Whether these strategies can prepare good initial states efficiently for physical or chemistry systems of practical interest remains an open question~\cite{lee2023evaluating}.
Given access to a good initial state, the standard quantum algorithm for estimating energies is quantum phase estimation (QPE)~\cite{kitaev1995quantum,Abrams1999,Higgins2007}.
For precision $\epsilon$, QPE either requires implementing the evolution operator $e^{-{\rm i} t H}$ for times $t =\mathcal{O}(1/\epsilon)$ or requires $\mathcal{O}(\lambda/\epsilon)$ uses of a unitary that block-encodes $H/\lambda$, for $\lambda \ge \|H\|$.
Both options yield an overall circuit cost scaling as $O(1/\epsilon)$ in the worst case.
Some classes of Hamiltonians can be fast-forwarded or partially fast-forwarded, which yields an exponential reduction in the cost of phase estimation~\cite{AtiaAharonov2017,GuSommaSahinoglu2021,Somma2016OneDimensional}.
However, these classes often admit efficient classical methods for simulation as well, and it is known that generic Hamiltonians cannot be fast-forwarded~\cite{berry2007efficient}.

Ground-state energies can also be obtained via expectation estimation provided access to the exact ground state (i.e.,\ as the expectation of $H$), potentially obtained using quantum circuits of lower depth compared to QPE.
However, this strategy might require many repetitions to overcome statistical noise, which can render it impractical~\cite{mcardle2020quantum,Wecker2014}.
\footnote{Recent works~\cite{kanno2023quantum,robledo2025chemistry} have attempted to circumvent this issue by sampling from approximate ground states to obtain a reduced Hilbert space in which to diagonalize the Hamiltonian, which imitates classical selected-CI methods.
However, the bottleneck of selected-CI methods is diagonalization and not the identification of a down-sampled Hilbert space~\cite{yuan2013truncated}, and so these methods do not appear to yield a path towards an asymptotic quantum advantage~\cite{reinholdt2025critical}.}

The prior approaches might not be suitable for simulating large systems with, say, tens of thousands of particles, as their complexity depends on the system size. 
For these cases, embedding~\cite{rubin2016hybrid,Knizia2012} or active space techniques~\cite{Angeli2001,Takeshita2019} can be useful.
These techniques can require the calculation of ground state expectation values beyond energies (see Sec.~\ref{sec:static_obs}).

\paragraph{Regimes of Advantage}

In practice, quantum computers are expected to achieve advantage in estimating ground-state energies for systems that satisfy the following conditions:

\begin{itemize}
\item A representation of the system on a sufficiently small Hilbert space ($100$s - $1000$s of qubits for the foreseeable future).

\item Access to an efficient quantum circuit for preparing good initial states (i.e.,\ that have large overlap with the ground state).

\item Access to an efficient quantum circuit for accurately simulating the evolution operator $e^{-{\rm i} tH}$ or a block encoding of $H$.

\item Problem instances that lack efficient classical methods.
 
\end{itemize}

\paragraph{Notes on Future Directions and Related Advantages}

Eigenenergy estimation is fundamentally linked to quantum phase estimation---the problem of estimating eigenphases $e^{i\phi}$ of a unitary $U$---and within the quantum computing literature phase estimation is considered the more fundamental problem.
However, unitaries are less-well known outside of the quantum computing field (and eigenphases are less of a focus), so we focused on the eigenenergy problem here.

Estimating ground-state energies is widely assumed to be typically classically intractable (i.e.,\ in Stage II) for certain problem classes, including gapless systems, spin liquids, systems near quantum phase transitions, and other cases featuring volume law entanglement or strong multi-reference character.
Indeed, estimating ground state energies for the electronic structure problem is commonly cited as one of the few examples of a problem in Stage III or Stage IV~\cite{babbush2026grand}.
However, rigorous proofs for these claims are currently lacking, and the specific classes of Hamiltonians that could exhibit typical classical hardness have not been formally defined.
Establishing these formal results is a crucial direction for future research in the field.
This leaves an open debate within the community regarding the precise boundaries of quantum advantage in ground-state estimation (cf.~\cite{lee2023evaluating}).
Similarly, devising efficient quantum algorithms that can prepare good initial states for a large class of Hamiltonians will be vital to understand the regimes of quantum advantage in practice.

Work has been undertaken in recent years to reduce the ancilla overhead~\cite{Svore2014,kimmel2015robust,obrien2019quantum,lin2022heisenberg}, improve initial state preparation~\cite{Berry2024}, improve Hamiltonian simulation methods~\cite{vonBurg2020,Lee2020,low2025fast}, and make phase estimation more robust for early fault tolerance~\cite{kshirsagar2024proving,zhang2022computing,dutkiewicz2025error}.
Due to the importance of this problem to the field of quantum simulation, we expect all of these lines of optimization to continue for the foreseeable future.

A related quantity to the ground state energy is the free energy, which includes the entropy term from molecular motion.
This is relevant for many of the above applications where this entropy is large (e.g.,\ for alchemical free energy calculations in computational drug design~\cite{huang2025fullqubit}).
Free energies can be calculated on a quantum device and used within classical computational workflows~\cite{OBrien2022}, but this comes with relatively high costs. Improving quantum algorithms for free energies is also an interesting direction for future research.


\subsection{Density of States and Spectral Density}

\label{sec:DOS}

\emph{Maximum Quantum Speedup: Exponential}\\
\emph{Evidence Basis: Rigorous quantum algorithm, complexity-theoretic evidence for classical hardness}

\paragraph{Problem Description}
Beyond ground states, the density of states (DOS) describes the distribution of energy levels across the spectrum of the Hamiltonian, whereas the spectral density of a specific state determines the probabilities or support of finding an eigenstate with a particular energy and quantum number (e.g.,\ momentum). 
Both quantities are key to studies of quantum many body physics, and the DOS can be thought of as a special case of a spectral density for the maximally mixed state $\rho=I/N$.

Since the spectrum of a Hamiltonian can be exponentially dense, a quantum algorithm for DOS or spectral density does not provide a list of eigenvalues, but rather estimates the support within certain energy regions.
To this end, we follow the \emph{Quantum Eigenvalue Estimation Problem} (QEEP) of ref.~\cite{somma2019quantum}, which can be reduced to the problem of estimating single phases, (see e.g.,\ refs.~\cite{lin2022heisenberg, dutkiewicz2022heisenberg}). 
The QEEP can be solved using a time-series analysis as in ref.~\cite{somma2019quantum} or via quantum phase estimation. 

\emph{Formal:} 
Let $H$ and $\rho$ represent the Hamiltonian and state of an $N$-dimensional quantum system, $N=2^n$. Let $\delta > 0$ be a spectral width and $\epsilon \ge 0$ an estimation error. 
The goal of the QEEP is to output a probability vector 
$\boldsymbol{p}=(p_0,p_1,\ldots)$ of dimension $M \sim 2 \|H\|/\delta$ and to additive error $\epsilon$, where each entry $p_j$ can be roughly interpreted
as the support of $\rho$ in a bin of energies of size $\delta$.

\paragraph{Potential Application Spaces}
DOS and spectral densities can be probed by spectroscopy, so one of the most prominent applications is in determining electronic~\cite{ollitrault2020quantum}, IR/Raman~\cite{mcardle2019digital}, vibronic~\cite{Huh2015}, or nuclear~\cite{ciavarella2023quantum} spectra.
For vibronic Hamiltonians, the vector $\boldsymbol{p}$ can be used to approximate the Franck-Condon profile~\cite{sawaya2019quantum} (using the vibrational vacuum as initial state $\rho$), which contains information about vibrational energy transfer probabilities.
Spectral densities also play a significant role in photochemistry, as under the Born-Oppenheimer approximation one can treat molecular dynamics classically on top of these surfaces.
(This, however, breaks down near conical intersections or other crossing points, requiring more detailed treatment~\cite{Veis2016}.)
DOS and spectral densities corresponding to different symmetry sectors than the ground state are relevant across various areas of physics.
Crossing points determine (first-order) phase transitions~\cite{Sachdev2011}, while superconducting energy gaps~\cite{dallaire2016method}, spin gaps (energy gaps between spin sectors) in high-$T_c$ materials~\cite{Fradkin2015}, and mass gaps (energy gaps between mass sectors) in lattice gauge theories~\cite{banuls2020simulating} all have significant experimental consequences.

When considering the density of states, the output of the QEEP---the vector $\boldsymbol{p}$---can be used to estimate quantities such as the partition function ${\rm tr}(e^{-\beta H}) \approx \sum_{j=-M/2}^{M/2} p_j e^{-\beta E_j}$ (where $E_j=j \|H\|/M$ is the position of the energy bin), a key tool of study in physics~\cite{brandao2008entanglement,harrow2020classical,rouze2026optimal}.
It can also be used for other computations, such as estimating entropies via Boltzmann's law.
The functional form of the full distribution can further indicate whether a system has a spectral density following a random matrix ensemble~\cite{wigner1993characteristic}, or the SYK model~\cite{Sachdev1993} (and to calculate corrections thereto), which is relevant to fundamental questions regarding the thermalization of quantum systems~\cite{srednicki1994chaos,DAlessio2016}.

\paragraph{Classical Complexity}
Highly accurate estimations of the DOS of a quantum system of dimension $N=2^n$ is also exponential in $n$ in the worst case.
Such calculations could provide, for example, an estimate of the ground state energy, which is QMA-hard.
Moreover, when the precision requirements are relaxed so that $\epsilon \in 1/\poly(n)$ and $\delta \in 1/\poly(n)$, a version of the QEEP where $\rho$ can be prepared efficiently can be shown to be BQP-Complete, which implies efficient classical algorithms are not expected.

In many cases, the same classical methods described in Sec.~\ref{sec:GSE} might be used for DOS or spectral density calculations.
However, excited states generally lack variational bounds.
This complicates the use of methods reliant on fixed points or variational principles, although workarounds (with additional overhead) exist in cases such as quantum Monte Carlo~\cite{Ceperley1988}, coupled cluster~\cite{Stanton1993}, and DMRG~\cite{Dorando2007}.

When $\rho$ describes a high-temperature state, where the focus on individual energy levels is no longer relevant, cluster methods~\cite{Rigol2006, Oitmaa2006} and Metropolis methods~\cite{Metropolis1953, Sandvik1999} are the most common tools for these computations, including partition function and energy estimation.
These often have provably polynomial-time costs (in the system size) above a certain energy density or temperature~\cite{harrow2020classical,Helmuth2021}.
Moreover, many quantum systems converge to the predictions of random matrix theory in the large $n$ limit, with small (inverse polynomial or inverse exponential) corrections~\cite{DAlessio2016}.
Many statistical properties of such random matrices can be calculated analytically, negating the need for large-scale simulations whenever this precision is sufficient.

\paragraph{Quantum Complexity}

Given a succinct representation of $H$ (e.g.,\ as a sparse matrix), and assuming $\|H\|\in \poly(n)$, an efficient quantum algorithm for solving the QEEP exists when $\delta \in 1/\poly(n)$, $\epsilon \in 1/\poly(n)$ and the state $\rho$ can be prepared efficiently.
Ref.~\cite{somma2019quantum} achieved a query complexity of $1/(\epsilon^{2}\delta^{3})$\footnote{The original result of ref.~\cite{somma2019quantum} does not separate $\epsilon$ and $\delta$ as done here, however this scaling can easily be derived from the work.}, and more recent work provides some improvements~\cite{lin2022heisenberg}.

When $\rho=I/N$ is the maximally mixed state, the QEEP can be solved within the one-clean-qubit model~\cite{knill1998power}, as it only involves estimating the traces of unitary operators. 
This provides an efficient method for the above precision requirements.
However, if more accurate estimates are needed, the QEEP can be hard even for quantum computers. 
This follows from hardness results including the estimation of the partition function.
For example, estimating the partition function is QMA-hard to multiplicative precision if the temperature is below some critical value~\cite{kempe2006complexity}.

\paragraph{Regimes of Advantage}

The requirements to achieve advantage in computing the density of states, and state-dependent spectral densities, are similar to those of Sec.~\ref{sec:GSE}:

\begin{itemize}

\item A representation of the system on a sufficiently small Hilbert space (100s-1000s of qubits).

\item Access to an efficient quantum process to prepare the initial state $\rho$ (for the density of states, $\rho$ should be the identity or an approximate projector onto an energy window).

\item Access to an efficient quantum circuit for accurately simulating the evolution operator $e^{-{\rm i}tH}$ for time $t\sim 1/\epsilon$.

\item The probability vector $\boldsymbol{p}$ must not concentrate to classically calculable predictions of a random ensemble as $n$ increases. (This feature appears often in strongly correlated systems.)

\item A lack of efficient classical methods for these problems.
    
\end{itemize}

Meeting these requirements can be challenging in practice.
For example, simulating IR/Raman and vibronic spectra requires accuracies of $1\,\mathrm{cm}^{-1}\approx 0.004\,\mathrm{mHa}$, which is a few hundred times smaller than chemical accuracy and may be prohibitive for quantum computers for the foreseeable future.
Moreover, while classical methods are expected to provide poor convergence when the state of the system corresponds to a thermal state of sufficiently low temperature, this can also preclude an efficient quantum algorithm for preparing the state in general, since such problems can be computationally hard.

\paragraph{Notes on Future Directions and Related Advantages}

It is generally believed that estimating densities of states and excited states is as hard as estimating ground state energies, suggesting similar classes of Hamiltonians for a potential Stage II advantage as described in Sec.~\ref{sec:GSE}.
However, defining these classes and giving strong evidence for a typical quantum advantage therein remains an open question.
Additionally, more work is needed to optimize algorithms for the QEEP, and to determine lower bounds on estimating densities of states.

Most representations of physical systems on quantum computers encode the entire Hilbert space on the device, so the spectra that we refer to in this section are many-body spectra and not the single-particle spectral function often probed by, e.g.,\ ARPES or other methods.
We discuss extracting spectral functions in the dynamic observables and response properties section (Sec.~\ref{sec:dynamic_observables}), as they can be derived through the Green's function.

The problem of partition function estimation has some overlap with the problem of thermal state expectation value estimation that we consider in Sec.~\ref{sec:static_obs}.
On a quantum device, it is non-trivial (but possible~\cite{rouze2026optimal}) to go from a preparation of the state $e^{-\beta H}/\mathrm{Trace}[e^{-\beta H}]$ to the partition function, which is a notable difference compared to classical methods.


\subsection{Expectation Values in Equilibrium and Order Parameters}

\label{sec:static_obs}

\emph{Maximum Quantum Speedup: Exponential}\\
\emph{Evidence Basis: Rigorous quantum algorithm, complexity-theoretic evidence for classical hardness}

\paragraph{Problem Description}

One of the most important tasks in quantum physics and quantum computing is that of quantum metrology, where the goal is to measure properties of quantum states as precisely as possible.
In quantum simulation, this problem considers the estimation of expectation values of simple, local operators $O$ (e.g.,\ particle or spin density).
Here, we consider this problem for arbitrary states, but emphasize stationary states including thermal states $\propto e^{-\beta H}$ and eigenstates $\ket{E_j}$ of a Hamiltonian, as they are most important in physics simulation.

\emph{Formal:}
Let $\rho=\rho(H) \in\mathbb{C}^{N\times N}$ be a stationary state of a Hamiltonian $H\in\mathbb{C}^{N\times N}$ of dimension $N=2^n$, and let $O\in\mathbb{C}^{N\times N}$ be a local observable.
The goal is to estimate the expectation $\mathrm{tr}[O\rho]$ to additive precision $\epsilon \ge 0$.

\paragraph{Potential Application Spaces}

Example problems are the measurement of local densities of states and local energies, and other quantum numbers such as occupation number, charge, current, and spin.
These quantities are relevant across all of quantum chemistry and materials science problems~\cite{mcardle2020quantum,bauer2020quantum}.
Particle densities in molecules and materials allow for the classification of chemical bonds, the identification of lone pairs and radical electrons~\cite{bader1990atoms}, the calculation of dipole (and higher-order) moments and the estimation of electrostatic potentials~\cite{atkins2011molecular}.
Charge densities can also be measured in the solid state via methods such as scanning tunneling microscopy~\cite{Tersoff1984}, scanning transmission electron microscopy~\cite{Pennycook2011}, or x-ray diffraction~\cite{Coppens1997}, allowing comparison between simulation and experiment.
In condensed matter physics, these observables detect order, allowing for the identification of phases of matter, the detection of phase transitions, and the measurement of critical exponents.
This in turn has applications across all areas of condensed matter physics~\cite{ashcroft1976solid,Fradkin2015,Lee2006,Si2010,balents2010spin,alet2018many,abanin2019many,Phillips2022}, placing such quantities on the same level of importance in this field as ground state energies themselves.
Susceptibilities of expectation values, including derivatives and variances, play a significant role in determining response properties of a quantum system; due to the Hellmann-Feynman theorem~\cite{Hellmann1937,Feynman1939} and fluctuation-dissipation theorem~\cite{Callen1951,Kubo1966} respectively, both of these can be measured as expectation values themselves.

Expectation values of local quantities play a further significant role as input into classical routines to either improve approximate methods or add corrections to ground state energies.
These are often summarized by the $k$-reduced density matrix (RDM), which contains expectation values of all $k$-body fermionic operators (in the ground state of the system).
Some density functional approximations require as input accurate energy densities for the uniform electron gas (jellium)~\cite{Babbush2017LowStructure}, density matrix embedding theory requires the full $1$-RDM~\cite{rubin2016hybrid}, and methods to capture correlations outside a small active space require either $2$-RDMs (MCSCF) or even $3$- and $4$-RDMs (CASPT2, NEVPT2)~\cite{Takeshita2019}.

\paragraph{Classical Complexity}
Like many other problems in quantum simulation, given a succinct representation of the Hamiltonian $H$ and a succinct specification of a stationary state $\rho(H)$ (e.g.,\ an eigenstate), computing the expectation value of local observables within sufficiently good accuracy is a QMA-hard or QMA-Complete problem. 
For example, local expectations of the ground state can be used to measure the ground-state energy (see Sec.~\ref{sec:GSE}), since $H$ is a sum of local operators.
 
For many classical methods, the challenge in estimating an expectation value is dominated by the challenge in preparing a representation of the target state $\rho(H)$.
If Gaussian states, free-fermion states, cluster states, perturbative approximations, or states with low entanglement can represent the target state well, then the problem can likely be solved in polynomial time on a classical computer.
Such states can be classically stored in a compact form, which allows for efficient calculations of local observables.
This is almost always true for diagonal quantities such as particle density, which is often obtained ``for free''.
\footnote{Obtaining static properties from coupled cluster methods requires solving the $\Lambda$-equations~\cite{handy1984evaluation} to normalize the state, which has a similar cost to energy estimation itself but needs to be completed only once.}
However, truncated methods such as DFT, DMFT, coupled cluster, RDM, and perturbation theory methods can only approximate expectation values that lie outside the truncated operator space (e.g.,\ densities for DFT, 4-RDM elements for CCSD, v2RDM, or MP2, and local correlators for DMFT).
For all such methods, non-local order parameters such as Wilson loops are not practical to estimate.
Monte Carlo methods suffer a blow-up in variance when targeting operators that significantly perturb the state of a walker (such as long-range two-body operators)~\cite{foulkes2001quantum}; however in some cases this can be reduced by the use of replica tricks~\cite{overy2014unbiased}.

One challenge in estimating higher-order operators (such as $3$- or $4$-body correlations) comes from their sheer number: one often wants to estimate a significant fraction, which brings overhead in both calculation and storage (unless one is satisfied with calculations on-the-fly).
In practice, this overhead makes 3- and 4-RDM calculations impractically expensive beyond small-scale systems that can be treated by exact diagonalization / full configuration interaction methods~\cite{knowles1984new}.
In chemistry, tensor network methods calculate the full 2-RDM explicitly during their energy estimation, effectively obtaining it for free.
However, this is not the case for tensor network approaches to local lattice models, where long-range correlators can have significant additional overhead to obtain compared to the energy~\cite{schollwock2011density}.
Monte Carlo methods also incur parallelization costs whenever the number of variables to be estimated becomes larger than the additive complexity of the Monte Carlo step (which costs around $\mathcal{O}(N^3)$~\cite{foulkes2001quantum}).

\paragraph{Quantum Complexity}

Estimating expectations of stationary states on a quantum computer requires first preparing $\rho(H)$ and then performing expectation estimation (cf.~\cite{knill2007expectation}).
In contrast with quantum phase estimation where the initial state is only required to have high overlap with the eigenstate, expectation estimation might require preparing $\rho(H)$ to higher fidelity than the final error. 
For eigenstates, this can be achieved using quantum phase estimation or eigenstate filtering~\cite{ge2019faster,lin2020near} at a cost that is inversely proportional to the overlap, the energy gap to nearby eigenstates, and polylogarithmic in inverse precision.
Thermal states may be prepared by simulating a Lindbladian with the state as a fixed point~\cite{Chen2023,Ding2025};
this can be achieved with a cost polylogarithmic in the inverse precision, polynomial in the inverse temperature, and linear in the mixing time.
A potential exponential overhead in the system size is hidden in both cases, in the overlap and the mixing time respectively, which is to be expected as both problems are worst-case QMA-hard.

In terms of $\epsilon$, estimating a single expectation in a pure state can be performed at a cost of $\mathcal{O}(1/\epsilon)$ calls to a state preparation unitary and its inverse (the Heisenberg limit), following the method of ref.~\cite{knill2007expectation}.
This has a similar cost to quantum phase estimation, but it requires independent execution, and cannot be obtained ``for free'' for particle density or any other operator.
However, quantum computers incur a surprisingly large overhead in the parallel estimation of multiple observables at the Heisenberg limit: $M$ separate observables require at least $\Omega(M^{1/2})$ copies of a single state to be estimated at the Heisenberg limit~\cite{apeldoorn2021quantum,Huggins2021}, which was achieved in ref.~\cite{Huggins2021} using the gradient estimation algorithm of ref.~\cite{gilyen2019optimizing}.
Moreover, expectation value estimation requires reflecting around the target state, which is not possible for mixed states.
The cost of state preparation can be circumvented for local observables using catalytic tomography~\cite{chen2025catalytic}, but with a large constant factor overhead.

A different approach uses shadow state tomography~\cite{Huang2020}, requiring $\mathcal{O}(1/\epsilon^2)$ copies of $\rho(H)$ (the sampling noise limit).
In this case, exponentially many commuting operators can be measured simultaneously, allowing parallel measurement of local spin operators with logarithmic overhead~\cite{cotler2020, bonetmonroig2020}.
The situation is worse for fermionic systems and operators, since these anticommute rather than commute~\cite{Zhao2021}.
Lower bounds in this case imply $\Omega(n^k/\epsilon^{2})$ copies of the state are required to measure the full $k$-RDM on $n$ fermionic modes to $\epsilon$ error per term using single-copy measurements~\cite{bonetmonroig2020,King2024-lj}; this can be avoided with $2$-copy measurements, but current methods have additional polynomial overheads in $\epsilon$~\cite{King2024-lj}.

\paragraph{Regimes of Advantage}

The quantum advantage in this section comes from the preparation of $\rho$. If we allow $\rho$ to be an arbitrary state prepared with a quantum circuit of $\poly(n)$ size, expectation value estimation is a BQP-Complete problem.
To achieve quantum advantage, the following is required:
\begin{itemize}
\item The preparation of $\rho$ must be efficient, e.g.,\ via a quantum circuit of polynomial size.

\item The state $\rho$ must correspond to a quantum state that classical methods cannot efficiently represent. An example is the low-temperature state of a strongly correlated system in $D>1$ dimensions.

\item We need to consider families of Hamiltonians and corresponding states $\rho(H)$ that avoid ``concentration''.
In particular, we seek families for which the expectations vary significantly across instances, or for which the expectation of a large-$n$ state cannot be extrapolated from systems with small $n$.
\end{itemize}

\paragraph{Notes on Future Directions and Related Advantages}

As with other quantum simulation problems, specifying instances and distributions over instances that can be proven to be classically hard to simulate is crucial to understand quantum advantage in practice and shift this problem into Stage II.
This effort requires the collaboration between the computer science, physics, and chemistry communities, which can provide insight about the hardness of classical simulations gained from experience.

The fundamental bounds on parallel measurement of multiple observables have led to proposals that calculations requiring many such observables can be executed on the quantum device itself (cf.~\cite{huang2025fullqubit, gunther2024more}), which may provide a source of new quantum simulation problems.

The term ``energy density'' as used in, for example, battery chemistry~\cite{Urban2016}, materials science~\cite{bauer2020quantum}, lattice gauge theories~\cite{banuls2020simulating}, or warm dense matter~\cite{Bonitz2019}, typically refers to the ground state energy divided by the system size, rather than a local energy (the expectation value of the Hamiltonian restricted to one or a few sites), and so should be considered a part of Sec.~\ref{sec:GSE}.

\subsection{Dynamical Observables and Response Functions}\label{sec:dynamic_observables}

\emph{Maximum Quantum Speedup: Exponential}\\
\emph{Evidence Basis: Rigorous quantum algorithm, complexity-theoretic evidence for classical hardness}

\paragraph{Problem Description}

This section summarizes the broad category of time-dependent observables and their dynamics. It includes both quenches or scattering processes where a system is far from equilibrium, 
and response properties such as the response of a system to an external field, where the initial state $\rho$ might be near thermal equilibrium.
The difference between these two problems is only in the initial state $\rho$, and the cost of its preparation.
Dynamical observables encompass:
\begin{enumerate}
    \item The Green's function or propagator (and derivative quantities such as self-energies) when $\rho \propto e^{-\beta H}$ is the Gibbs state and the observable involves creation and annihilation operators.
    Though this is often considered in the frequency domain, many proposals for obtaining this on a quantum computer~\cite{somma2002simulating,Wecker2015b} are formulated by processing the time domain Green's function using a classical Fourier transform.
    \item The quenched dynamics of a local or global observable $O(t)$ evolving under some Hamiltonian, and its expectation with respect to $\rho$.
    \item Scattering problems or wavefunction overlaps by setting $O(t)=\ket{\psi(t)}\!\bra{\psi(t)}$ to be the projector onto some time-evolved state.
\end{enumerate}

\emph{Formal:} Let $H$ and $\rho$ be a Hamiltonian and state of a quantum system of dimension $N=2^n$.
Let $O_1,O_2\in\mathbb{C}^{N\times N}$ be two operators, and define $O_2(t):=e^{iHt}O_2e^{-iHt}$ for some time $t\in\mathbb{R}$. The goal is to estimate the dynamical expectation $\mathrm{tr}(\rho O_1 O_2(t))$ to additive precision $\epsilon \ge 0$.

More generally, if $U$ is some unitary quantum circuit acting on the system, the goal is to estimate $\mathrm{tr}(\rho O_1 U^{\dagger}O_2U)$ to additive precision $\epsilon \ge 0$.

\paragraph{Potential Application Spaces}

Simulating far-from-equilibrium dynamics is a task of significant importance in condensed matter and high-energy physics.
Scattering processes are ubiquitous in high-energy physics to simulate collider physics~\cite{bauer2021simulating}, study confinement and quasiparticle formation~\cite{banuls2020simulating}, and otherwise probe the Standard Model~\cite{preskill2018simulating}.
The potential for quantum computers to calculate stopping power in inertial confinement fusion has been particularly highlighted in recent work~\cite{rubin2024quantum}.
In condensed matter physics, the dynamics of a system following a quench probes prethermalization~\cite{neyenhuis2017observation} and the eigenstate thermalization hypothesis~\cite{andersen2025thermalization}, and can detect many-body localization~\cite{alet2018many,abanin2019many}. 
In chemistry, excitation transport plays a significant role in understanding biological energy harvesting~\cite{mohseni2008environment}, as well as developing improved solar cells~\cite{green2014emergence} and batteries~\cite{Urban2016}.

A significant body of work studying quench dynamics in local spin and fermionic systems has been achieved in recent years on NISQ hardware, including work studying bound state formation in photons~\cite{morvan2022formation}, thermalization and the Kibble-Zurek mechanism~\cite{andersen2025thermalization}, localization~\cite{smith2015manybody,roushan2017spectroscopic}, charge-spin separation in Fermi-Hubbard systems~\cite{arute2020observation}, topological order~\cite{will2025probing}, and spin glass formation~\cite{lunkin2026hilbert}.
This is largely due to the ease of simulating local spin Hamiltonians on physical architectures. 
Although most of these works were not performed at sizes beyond the capabilities of classical computers, it is expected that this will be one of the earliest areas for quantum computers to achieve practical advantage.

Green's functions and their derivatives play a central role in physics in determining response and susceptibilities of systems.
The imaginary part of a Green's function (the spectral function) determines the presence, group velocities, and lifetimes of quasiparticles that drive spin and particle transport~\cite{mahan2000many}, yields measurements of pseudo- and regular energy gaps, and is directly measured via ARPES in electronic systems~\cite{damascelli2003angle}.
The current-current correlation function is directly related to the optical conductivity of a material in linear response theory via the Kubo formula, allowing for the estimation of resistance and optical absorbance of a material~\cite{kubo1957statistical}.
Similarly, the magnetic susceptibility can be obtained from the spin-spin correlation function~\cite{white2007quantum} (and directly compared to neutron scattering experiments), and superconducting gaps and order parameters can be detected from the anomalous Green's function (in the Nambu-Gor'kov formalism)~\cite{nambu1960quasi}.
Green's functions of impurity models are also required as input to dynamical mean field theory, which has driven proposals for quantum-computing-assisted DMFT calculations~\cite{Bauer2015,rungger2019dynamical}.

\paragraph{Classical Complexity}

This problem class contains simulating both a quantum circuit, as well as simulating any physical implementation of a quantum computer, making it the quintessential BQP-complete problem (with an expected worst-case exponential classical cost).
Moreover, time-evolved states are typically more challenging to represent classically than ground states: time evolution grows entanglement and multi-reference character, and maintains weight on eigenstates in the bulk of a system that follow volume-law entanglement~\cite{schuch2008entropy}.
However, time-evolution also tends to concentrate expectation values to thermal averages~\cite{srednicki1994chaos,rigol2008thermalization}, and late-time dynamics tends to relax to hydrodynamic behaviors~\cite{leviatan2017quantum, white2018quantum}, which can reduce the cost of classical estimation.
In the extreme example of a random quantum circuit (and more generally typical ensembles of circuits under local $SU(2)$ symmetry), it has even been proven that no quantum advantage exists in estimating expectation values, due to this concentration of measure~\cite{angrisani2024classically}.

Many methods for time evolution exist, some of which are variations of methods built for ground state estimation, and some of which solely target dynamics.
For small systems, many exact codes for open- and closed-system dynamics exist~\cite{johansson2012qutip}, including many optimized for simulating quantum circuits~\cite{isakov2021simulations,suzuki2021qulacs}.
Time-dependent versions of coupled cluster~\cite{dalgaard1983some}, self-consistent field~\cite{mclachlan1964time}, density functional theory~\cite{runge1984density}, and variational Monte Carlo have been constructed, as well as various methods for evolving matrix product states~\cite{vidal2004efficient} and other tensor network state representations~\cite{haegeman2011time}, while Krylov subspace / truncated Lanczos methods are as applicable to dynamics~\cite{hochbruck1997krylov} as to ground state properties.
Approximate dynamics-focused methods on the other hand include mean-field approaches~\cite{auerbach2012interacting}, truncated Wigner expansions~\cite{polkovnikov2010phase}, hierarchical equations of motion~\cite{tanimura1989time}, truncated or stochastically unraveled Lindblad evolution~\cite{dalibard1992wave}, or truncated or stochastically unraveled Pauli pathing techniques~\cite{bravyi2018simulation,google2025observation}.
A significant body of work separately calculates Green's functions and derivatives using, e.g.,\ the GW approximation~\cite{hedin1965new}, diagrammatic Monte Carlo~\cite{prokofev1998exact}, numerical renormalization group~\cite{wilson1975renormalization}, or perturbative diagrammatic expansions in loop order~\cite{stefanucci2013nonequilibrium}.

\paragraph{Quantum Complexity}
The problem of computing a time-dependent expectation can be shown to be BQP-Complete even in the case where the initial state is, for example, the trivial all-zero state, the Hamiltonian is sparse, satisfies $\|H\|\le 1$, and can be accessed via an efficient quantum circuit, the evolution time is $t={\rm poly}(n)$, the error $\epsilon$ is a small constant, and the operators are simple local operators~\cite{Aharonov2007}. 
Efficient quantum algorithms for these problems were formerly given in refs.~\cite{lloyd1996universal,Ortiz2001,somma2002simulating} and extended to achieve the Heisenberg limit in ref.~\cite{knill2007expectation}.
Since then, significant work has gone into Hamiltonian simulation---developing circuit implementations for $e^{iHt}$---across a wide range of systems.
These methods fall into three main categories: Trotter-Suzuki (product) formulas~\cite{suzuki1990fractal} including randomized versions~\cite{campbell2019random, childs2019faster}, linear combination of unitaries~\cite{Childs2012,Berry2015}, and quantum signal processing using block encodings~\cite{low2017optimal,Gilyen2019}.
(See ref.~\cite{childs2017lecture}, chapter V or ref.~\cite{lin2022lecture}, chapter 5--7 for detailed expositions on these topics.)
The optimal complexity of simulation depends on the Hamiltonian. 
The relevant parameter is the effective action $\tau=\|H\||t|$, where $\|H\|$ is some norm of the Hamiltonian. 
This normalization prevents an apparent speedup obtained by merely rescaling the Hamiltonian or $t$.
A circuit complexity of $\Omega(\tau)$ is typically required to simulate a Hamiltonian for time $t$, due to the impossibility of generic fast-forwarding~\cite{berry2007efficient,AtiaAharonov2017,Haah2018}.

However, for certain ``fast-forwardable'' Hamiltonians, this complexity may be significantly improved, unlocking other quantum speedups.
Examples are near-diagonal and one-sparse Hamiltonians~\cite{QiangEtAl2016,Ahokas2004}, 
sums of mutually commuting terms~\cite{AtiaAharonov2017,GuSommaSahinoglu2021}, permutation-invariant local spin Hamiltonians~\cite{AtiaAharonov2017,GuSommaSahinoglu2021}, free-fermion and free-boson Hamiltonians~\cite{AtiaAharonov2017,GuSommaSahinoglu2021,JainEtAl2026QHT,IyerEtAl2026SUN}, and more generally Hamiltonians associated with semi-simple Lie algebras of polynomial dimension. In these cases the circuit's complexity can be polynomial in the number of qubits $n$, but the dependence on $\tau$ can be logarithmic, and hence an exponential improvement. In contrast, frustration-free Hamiltonians for initial states in low-energy subspaces are an example of {\em polynomial} fast-forwarding~\cite{Zlokapa2024hamiltonian}.

A key limitation in using quantum computers to estimate expectation values---such as time-dependent observables---arises from the estimation error $\epsilon$. Typically, quantum computers are constrained by the Heisenberg limit $T \sim 1/\epsilon$. This presents a significant challenge when resolving values that decay exponentially with system size. While exponentially fast-forwarding a simulation can exponentially reduce error dependence in quantum phase estimation~\cite{AtiaAharonov2017,JainEtAl2026QHT}, this fast-forwarding is rarely applicable and does not improve the $1/\epsilon$ scaling in general quantum metrology. Consequently, even with fault-tolerant quantum hardware, determining rare scattering processes in nuclear physics, decay rates of the Loschmidt echo in dynamic quantum phase transitions, or exponential tails of local integrals of motion remains a major hurdle.

\paragraph{Regimes of Advantage}

Quantum algorithms perform efficiently when the initial state can be prepared efficiently, the Hamiltonian can be accessed efficiently, and $t={\rm poly}(n)$, $\epsilon=1/{\rm poly}(n)$.
As discussed, this is a BQP-Complete problem and classical computers would take exponential time to solve the problem in the worst case.
The quantum advantage question falls to whether this exponential advantage is manifest in practice, where the Hamiltonians and states are structured. 

As described above, classical algorithms such as Monte Carlo or Tensor Networks can perform efficiently under certain conditions, such as when Hamiltonians are integrable, evolved states exhibit low entanglement entropy, or dynamics are localized within a small spatial region (i.e.,\ a small lightcone).
Thus, candidates for quantum advantage include systems in higher spatial dimensions, those with long-range interactions that drive the spread of entanglement during evolution, or the study of prethermal plateaus~\cite{abanin2019many} and glassy dynamics~\cite{carleo2011localization}, which avoid these conditions.
However, quantum systems that are sufficiently chaotic will concentrate to simple, classically predictable values, preventing quantum advantage (except potentially within a small time window).

In addition, the no-fast-forward theorem~\cite{berry2007efficient} establishes that the quantum complexity of time evolution scales at least linearly with $\tau=\|H\||t|$, suggesting that a quantum advantage may not be achievable for instances with very large evolution times for arbitrary Hamiltonians. Furthermore, quantum simulations require identifying a basis capable of representing the dynamics on a quantum device, which poses significant challenges for high-accuracy simulations of models such as the complete standard model~\cite{jordan2012quantum}.

\paragraph{Notes on Future Directions and Related Advantages}

Several avenues remain open for optimizing the quantum complexity of quantum simulations. Specifically, examples of key areas requiring further investigation include:

\begin{itemize}
\item Determining specific problem classes for which typical quantum advantage can be shown analytically or numerically, and shifting this problem to Stage II.

\item Developing efficient quantum singular value transformation (QSVT) methodologies that exploit the inherent structure of Hamiltonians.

\item Determining optimal structure and term ordering for product formula methods.

\item
Refining error bounds of Hamiltonian simulation techniques, especially constant factor analysis for specific states, operators, and Hamiltonians, or for average case instead of worst-case analysis.

\item
Extending quantum evolution fast-forwarding methods to encompass more generalized cases.

\item Developing means to~\emph{a priori} determine which simulation technique is optimal for a given problem.
\end{itemize}

Recent studies have made strides in many of the above directions (see e.g.,\ refs.~\cite{Low2018,Childs2019,csahinouglu2021hamiltonian,holmes2022quantum,clinton2024towards,king2026quantum}).


\subsection{Shadow Hamiltonian Simulation} \label{sec:shadowH}
\emph{Maximum Quantum Speedup: Exponential}\\
\emph{Evidence Basis: Rigorous quantum algorithm, complexity-theoretic evidence for classical hardness}\\

Shadow Hamiltonian simulation is a framework that considers quantum simulation problems by using ``shadow states'', which encode only partial information about the quantum system. Because this approach is distinct from traditional quantum simulation, its details are elaborated in a separate subsection.
However, the shadow Hamiltonian formalism is extensible to any of the quantum simulation problems described above (Secs.~\ref{sec:OTOC},~\ref{sec:GSE},~\ref{sec:DOS},~\ref{sec:static_obs},~\ref{sec:dynamic_observables}).

\paragraph{Problem Description}
Consider a quantum system of $N=2^n$ sites or modes
modeled by a Hamiltonian $H$ and initialized in a state $\rho$. Let $S=\{O_1,\ldots,O_M\}$ be a set of relevant operators 
and consider the time-dependent expectations $\langle O_m(t)\rangle = {\rm tr}(\rho(t) O_m)$, where $\rho(t)$ is the evolved state. The goal of shadow Hamiltonian simulation is to provide a property of the quantum system that depends on the relevant expectations $\{\langle O_m(t)\rangle\}$ only.

\paragraph{Potential Application Spaces}
Shadow Hamiltonian simulation
can simulate non-interacting fermionic and bosonic systems with a large number of sites $N=2^n$ efficiently. 
Such systems arise
as the mean-field approximations of more generally interacting systems, and are thus of paramount importance in solid state physics and beyond. 
For example, they can model physical features like 
localization and disorder.
Extensions of shadow Hamiltonian simulation can also be used to simulate classical systems of coupled oscillators or the classical wave equation via
 a quantum-classical correspondence (Sec.~\ref{sec:oscillators}).

\paragraph{Classical Complexity}
For system size $N$ and evolution time $t$, 
classical methods such as the Kernel Polynomial Method
(KPM) have complexity $\Omega(Nt)$ and space complexity $\Omega(N)$~\cite{weisse2006kernel}. Essentially, such methods
simulate the time evolution via matrix-vector multiplication. The complexity is worse when the matrix is not sparse.

\paragraph{Quantum Complexity}

Ref.~\cite{somma2025shadow} presents an efficient
quantum algorithm for shadow Hamiltonian simulation when certain assumptions are met.
To this end, let $\ket{\rho(t) ; S} \propto \sum_{m=1}^M \langle O_m(t)\rangle \ket m$ be a (pure) shadow state, where $\ket m$ are basis states. Preparation of this state 
solves the shadow Hamiltonian simulation problem using
standard quantum metrology techniques, such as amplitude or overlap estimation. These techniques can provide time-dependent correlation functions or time-dependent energies of the system, which are BQP-Complete problems~\cite{somma2025shadow}.

Assume that: 
 i) the
 Hamiltonian $H$ leaves the space spanned by $S$ invariant (i.e.,\ they satisfy the invariance property),
 ii) the initial shadow state can be efficiently prepared, and iii) the graph underlying $H$, which is of exponential size, is sparse or represented succinctly. Then, under these assumptions,
 ref.~\cite{somma2025shadow} shows how to prepare
$\ket{\rho(t) ; S}$ from $\ket{\rho(0) ; S}$
with query complexity ${\cal O}(t J + \log(1/\epsilon))$,
where $J$ is the largest interaction strength in $H$ and $\epsilon$ is the error. The gate complexity
has an additional factor ${\cal C}$ that depends logarithmically on $N$, responsible for the block-encoding of the shadow Hamiltonian. The overall complexity has an additive term ${\cal C}_0$
to prepare the initial shadow state, which is often
polylogarithmic in $N$ and $1/\epsilon$ for relevant applications (e.g.,\ thermal states).

A related strategy 
concerns encoding the expectation values $\{\langle O_m(t) \rangle \}_m$ in the entries of a correlation matrix rather than a shadow state. This can be useful when the operators $O_m$ correspond to two-point correlators~\cite{stroeks2024solving}, where the two indices specify the matrix entry. The matrix can be block-encoded, it can be evolved similarly to shadow states, and its entries can also be efficiently computed using standard quantum metrology. 
Also, expectation values of superoperators on shadow states $|\rho(t);S\rangle$ correspond to out-of-time-ordered correlators of the original problem~\cite{somma2025shadow}, which allows exponentially many OTOCs to be measured in parallel using shadow simulation~\cite{chiew2026quantum}.

\paragraph{Regimes of Advantage}
\begin{itemize}
    \item To preserve the mild scaling in system size, ${\cal C}$
    cannot be polynomial in $N$. As a consequence, 
    there must be a succinct presentation of the Hamiltonian $H$ of $N$ sites or an oracle that specifies the interactions efficiently. This occurs in sparse lattices where some uniformity of interactions is readily available. Similarly, shadow Hamiltonian simulation can address cases where ${\cal C}_0$ has a polylogarithmic scaling in $N$, constraining the initial states to be states that can be succinctly represented, such as the vacuum state or the thermal Gibbs state. 

\item The problem of preparing the shadow state
can be shown to be BQP-Complete, even for sparse Hamiltonians. Nevertheless, physical systems 
often involve geometrically-local interactions that imply a light-cone feature, in which the relevant system size $N$ and evolution time are polynomially related as $N = {\cal O}(t^D)$, where $D$ is the spatial dimension. For these cases the complexities of both classical KPM and shadow Hamiltonian simulation are polynomial in $N$ (or $t$)~\cite{somma2025shadow,stroeks2024solving}. 
Considering the quantum complexity $\tilde {\cal O}(t)$ and the classical complexity $ {\cal O}(t^{1+D})$, the asymptotic speedups are polynomial,
being more pronounced for larger $D$.

\item Shadow states can be used to measure a number of physical quantities, such as energies in a semiclassical approximation, energies of subsets of modes, and related global quantities. These guarantee
that the expectation to be measured is not vanishingly small. A related encoding based on matrices allows for the efficient computation of more local properties as long as the precision requirements are not too stringent. That is, since extracting an expectation with precision $\epsilon$
requires ${\cal O}(1/\epsilon)$ state preparations, 
efficiency is preserved if $\epsilon$
is not inverse polynomial in $N$~\cite{stroeks2024solving}.

\item The advantage of shadow Hamiltonian simulation appears when one is dealing with very large system sizes. In physics, this occurs in systems in $D=3$ spatial dimensions, where correlation lengths involve a large number of sites like $10^5$ or more, for which $N=10^{15}$, and classical methods fail due to memory bandwidth problems. 

\end{itemize}

\paragraph{Notes on Future Directions and Related Advantages}

\begin{itemize}
\item The current formalism for shadow Hamiltonian simulation focuses on systems such as free fermions or bosons. The next immediate generalization regards
interacting systems under the mean-field approximation, on which initial work has recently appeared~\cite{chakraborty2026efficient}.

\item While in $D$ spatial dimensions the asymptotic speedup is polynomial, it can become more pronounced for long-range power-law interactions $1/r^\alpha$, particularly when $\alpha$ is small enough that the usual linear-light-cone picture no longer applies.
\end{itemize}

\pagebreak

\section{Linear Algebra}
\emph{Section contributed by:} Rolando D. Somma, Yuri Alexeev, Elica Kyoseva, Guang Hao Low, Ryan Babbush, M\'aria Kieferov\'a

\subsection{Linear Systems}\label{sec:lin_equations}

\emph{Maximum Quantum Speedup: Exponential}\\
\emph{Evidence Basis: Rigorous quantum algorithm, complexity-theoretic evidence for classical hardness}

\paragraph{Problem Description}\label{sec:5.1}

A linear system solver takes matrix $A \in \mathbb{C}^{N \times N}$ and a vector $b \in \mathbb{C}^{N}$ and computes a vector $x \in \mathbb{C}^{N}$ such that $x = A^{-1} b$. However, we do not require the full vector $x$, only its selected properties that can be expressed as expectation values of operators.

The matrix $A$ is assumed to be invertible, with sparsity $s$, and well‑conditioned, with condition number defined as $\kappa = \|A\| \cdot \|A^{-1}\|$. 

In the quantum version, the vector $b$ is encoded as a quantum state $\ket{b}$, and the objective is to prepare a quantum state $\ket{x}$ proportional to the solution: 
$\ket{x} \;\propto\; A^{-1} \ket{b}$. Unlike a classical linear system solver, a quantum algorithm does not provide direct access to the solution vector and only outputs it as a quantum state $\ket{x}$. The quantum state $\ket{x}$ can be useful for computing expectation values of the solution of the system but not for outputting the full solution vector, as that would require $\Omega(N)$ measurements and thus at least linear time in the system dimension $N$.

There are three known quantum methods to solve this problem: direct inversion methods, inversion by adiabatic evolution, and trial state preparation and filtering. However, only direct inversion methods provide exponential speedup over the best classical methods. The first 
direct inversion quantum algorithm was proposed by Harrow, Hassidim, and Lloyd (HHL)~\cite{HHL2009} with subsequent improvements based on 
a linear combination of unitaries~\cite{Childs2017} and then quantum singular value transformation~\cite{Gilyen2019}. 

\paragraph{Potential Application Spaces}\label{sec:5.2}

Direct inversion quantum algorithms are particularly relevant for solving problems where linear sets of equations yield matrices that are very large, sparse (or otherwise highly structured), invertible, and well-conditioned. These algorithms can be used in several areas, such as solving differential equations (see Sec.~\ref{sec:dif_equations}), quantum machine learning~\cite{rebentrost2014quantum}, and the computation of Green’s functions in quantum many-body systems~\cite{Morales2024QuantumLinear}. Since equation solvers are widespread, it is likely that there are other applications where they might be applicable; however, the restrictions on a potential advantage are very limiting.

\paragraph{Classical Complexity}\label{sec:5.3}

For sparse, well-conditioned, and invertible matrices, the best classical algorithms for solving linear systems, such as the conjugate gradient method, are based on the Krylov subspace method~\cite{Krylov}. Their asymptotic runtime scales as $\mathcal{O}(N \kappa \log(1/\varepsilon))$ in the worst case, where $N$ is the system size, $\kappa$ is the matrix condition number, and $\varepsilon$ is the precision of the solution. The space complexity is $\mathcal{O}(N)$, which is required for the storage of the matrix and vectors. Unlike quantum solvers, these algorithms output the full solution vector and have less restrictive requirements on the output model. When $A$ is a low-rank matrix and one only requires samples from the outcome, the runtime is $\text{polylog}(N)$~\cite{chia2022sampling}. 

\paragraph{Quantum Complexity}\label{sec:5.4}

Direct inversion quantum algorithms offer asymptotic improvements over classical Krylov subspace methods. The HHL algorithm achieves complexity of $\mathcal{O}(\log (N) \kappa^2 s^2 / \varepsilon)$ using sparse matrix oracles, where $s$ is the matrix sparsity~\cite{HHL2009}. Linear combination of unitaries (LCU)~\cite{Childs2017} and finally quantum singular value transformation (QSVT) improved the scaling by circumventing phase estimation and leading to an optimally efficient polynomial transformation framework for block-encoded matrices. 
Given a block-encoding of the input matrix, QSVT achieves query complexity $\mathcal{O}(\kappa \log(\kappa/\varepsilon))$, matching the best known theoretical bounds for inverse approximation over the well-conditioned spectral range~\cite{Gilyen2019, Morales2024QuantumLinear}.
More recently, ref.~\cite{Costa2022} used the discrete adiabatic theorem and gave a simplified algorithm with asymptotically optimal query complexity in both $\kappa$ and $\varepsilon$.

A QSVT-based solver with optimal $\mathcal{O}(\kappa\log\frac{1}{\epsilon})$ complexity was also developed using a block-preconditioning technique~\cite{Low2024QLSP}. In the regime where preparing the input state is significantly more costly than the block-encoding, it is also possible to use variable-time-amplification~\cite{Low2024QLSP} to perform only $\mathcal{O}(\frac{1}{\sqrt{p_\text{succ}}})$ queries to input state preparation (where $p^{-1/2}_\text{succ}\le\kappa$ is the solution norm), together with $\mathcal{O}(\kappa\log\frac{1}{p_\text{succ}}\log\frac{\log{1}/{p_\text{succ}}}{\epsilon})$ queries to the block-encoding.

More recently, Dalzell, Li, and Su introduced quantum linear-system solvers whose complexity can go beyond the worst-case condition-number dependence, replacing $\kappa$ by an instance-dependent effective condition number that can be substantially smaller~\cite{Dalzell2026}.

\paragraph{Regimes of Advantage}\label{sec:5.5}
Linear system solvers come with strict requirements on the input model, properties of the linear system and the form of the output. The QSVT-based linear solver offers the best performance out of quantum solvers listed here but comes with the following requirements:

\begin{itemize}
\item For standard QSVT-based solvers, the complexity is polynomial in the condition number $\kappa$, and $\kappa\gtrsim N^\alpha$ eliminates the exponential speedup. Recent beyond-$\kappa$ solvers show, however, that a large worst-case condition number need not preclude quantum advantage when the relevant instance-dependent effective condition number remains small~\cite{Dalzell2026}.

\item QSVT requires that $A$ can be block-encoded in time $\polylog N$. This requires $A$ to be $s$-sparse with $s = \polylog N$, or to have other special structure that can be leveraged for block encodings.

\item The quantum state $\ket{b}$ must be efficiently encoded in time $\polylog N$. This requires $b$ to be very sparse, highly structured or otherwise an output of an efficient quantum algorithm.

\item The solution is prepared only as a quantum state $\ket{x}$. To use it to obtain expectation values with respect to an observable $M$, one must repeatedly measure $\braket{x|M|x}$.
The number of measurements required to achieve precision $\varepsilon$ scales as $\mathcal{O}(1/\varepsilon^2)$. Thus, high precision requires a large number of measurements, leading to diminishing returns for quantum advantage. 
\item Finally, for QSVT to retain a runtime polynomial in $\log N$ the precision must satisfy $\log\frac{1}{\varepsilon} = \poly(\log N)$. In contrast, for $\varepsilon \leq 2^{-\Omega(N)}$, the $\log(1/\varepsilon)$ overhead becomes $\Omega(N)$, eliminating exponential speedup.
\end{itemize}

\paragraph{Notes on Future Directions and Related Advantages}\label{sec:5.6}

The main future directions are focused on trying to fix the deficiencies of direct inversion quantum algorithms.

\begin{itemize}
  \item An important direction for improving QSVT-based solvers is the development of quantum-compatible preconditioners. They could reduce the condition number $\kappa$ while preserving an efficient block-encoding structure, enabling quantum advantage for larger classes of linear systems.

  \item Practical use of these algorithms will also require addressing state preparation bottlenecks. Approaches based on quantum memory may help solve data loading issues once it becomes available in the future.

  \item New techniques are required to reduce the number of measurements, such as classical shadow tomography or low-rank observable estimation.
\end{itemize}

\subsection{Linear Differential Equations}\label{sec:dif_equations}

\emph{Maximum Quantum Speedup: Exponential}\\
\emph{Evidence Basis: Rigorous quantum algorithm, complexity-theoretic evidence for classical hardness}

\paragraph{Problem Description}

Consider an inhomogeneous, linear differential equation of the form
$\frac {{\rm d} \vec x(t)}{{\rm d} t}= A(t) \vec x(t) + \vec b(t)$, where $A(t) \in \mathbb C^{N \times N}$ is an arbitrary time-dependent matrix, the dimension $N=2^n$ is exponential,
$ \vec b(t) \in \mathbb C^N$ is the inhomogeneity, and
 the solution $ \vec x(t) \in \mathbb C^N$ satisfies $\vec x(0):=\vec x_0$. Given $T>0$ and succinct expressions for $A(t)$ and $\vec b(t)$,
 the goal is to compute a property of $\vec x(T)$, such as an entry, expectation value, or overlap.
 A quantum approach reduces this to the problem of preparing a quantum state $\ket{x(T)}$ proportional to $\vec x(T)$.

\paragraph{Potential Application Spaces}
Linear differential equations are ubiquitous in science and industry; particularly, they play a key role in physics. 
Examples of equations that fit with this framework are: the classical wave equation, the heat equation, dissipation in quantum systems, and partial differential equations like the Poisson equation. Related equations appear in quantum machine learning.

\paragraph{Classical Complexity}
The worst-case complexity of exact classical algorithms
is linear in space-time, or $\Omega(T N)$
for the sparse case. Some methods output the full final vector $\vec x(T)$ from which the relevant outputs are obtained. For the time-independent case,
Krylov subspace methods~\cite{saad2003iterative} are efficient with complexity ${\cal O}(kN)$, where $k$ is the dimension of the subspace. The Kernel polynomial method (KPM)~\cite{weisse2006kernel} also address this case using Chebyshev expansions and with similar complexity. These complexities 
get significantly worse in the time-dependent, inhomogeneous case, where the propagator has to be computed. Runge-Kutta~\cite{hairer1993solving}
can be used in this case,
with complexity ${\cal O}(p s N)$ for the sparse case, where $p$ is the number of steps (usually $p=T/h \gg T$) and $s$ denotes the per step order of accuracy. 
Another approach is based on Magnus expansions
of the exponential~\cite{blanes2009magnus}, where 
the propagator is approximated by a truncated Magnus series. While this approach requires computing commutators and such computations are expensive (${\cal O}(N^2)$), it is often combined with others to improve the overall complexity per step to ${\cal O}(N)$.

\paragraph{Quantum Complexity}

Under common assumptions on the matrices ($A$ is sparse and $\|A\|\le 1$), 
and if the norm of the solution $\vec x(t)$ does not vary widely, the quantum complexity for preparing the quantum state $\ket{x(T)}$ is only polylogarithmic in the dimension, which is the key improvement over classical methods. If the goal is to output a property of $\vec x(T)$, additional overheads are incurred due to, e.g.,\ amplitude amplification. To quantify quantum complexities for preparing the normalized state, we define $u:=\max_t \|\vec x(t)\|/\|\vec x(T)\|$. 
The algorithm of ref.~\cite{berry2017quantum} runs in time almost linear in $T$, $u$, the condition number $\kappa \ge 1$ of the matrix that diagonalizes $A$, and is
polylogarithmic in the precision and other parameters; the overall quantum query complexity is $\tilde {\cal O}(T u \kappa \polylog(1/\epsilon))$. 
Refs.~\cite{an2023linear,jin2024quantum,low2025optimal} provide improved quantum complexity $\tilde {\cal O}(T u \log(1/\epsilon))$. Gate complexities can bring an additional $\polylog(N/\epsilon)$ factor.

\paragraph{Regimes of Advantage}

\begin{itemize}
    \item The preparation of a quantum state $\ket{x(T)}$ proportional to $\vec x(T)$ assumes efficient coherent access to the coefficients of the differential equation. In particular, $A(t)$ must admit an efficient sparse-access oracle or block-encoding, while $\vec b(t)$ and the initial condition $\vec x_0$ must be efficiently preparable as quantum states. 
    
    \item Ref.~\cite{berry2014high} presents an efficient quantum state preparation algorithm, with improvements in refs.~\cite{berry2017quantum,krovi2023improved}. Preparation of $\ket{x(T)}$ involves the preparation of a history state $\sum_t \ket{x(t)}\ket t$, producing the correct state after postselection. The history state is obtained via a reduction to the quantum linear systems problem (Sec.~\ref{sec:lin_equations}). If the matrix $A$ is diagonalizable, sparse and bounded, with eigenvalues having non-positive real part (i.e.,\ a stability condition), history-state preparation is efficient.
    
    \item More recently, methods based on linear combination of Hamiltonian simulation~\cite{an2023linear,low2025optimal} and the related ``Schr\"odingerization'' method~\cite{jin2024quantum} achieve optimal query cost. In contrast, these methods aim at preparing $\ket{x(T)}$ directly by implementing the exponential operator that solves the differential equation. They require a slightly stronger assumption $A(t)+A(t)^\dagger \preceq 0$.
    
    \item For instances where $u$ is polynomial in $N$, the exponential quantum speedup disappears. This can occur in dissipative models, where $\|\vec x(T)\|$
    can be exponentially small.

    \item Preparation of $\ket{x(T)}$ allows for the efficient computation of expectation values in the state, with overhead ${\cal O}(1/\epsilon)$ for precision $\epsilon$. This implies that $\epsilon$ cannot be inverse polynomial in $N$ to preserve efficiency, ruling out computations involving exponentially small quantities.

\end{itemize}

\paragraph{Notes on Future Directions and Related Advantages}

\begin{itemize}

    \item There is still room for improvement for 
    the fundamental quantum algorithms, from unifying assumptions across different approaches to leveraging other structures of the matrices, vectors, and solutions.

    \item In the time-dependent case, the quantum algorithms have a very mild overhead due to the rate of change of $A(t)$, often polylogarithmic in $\|\frac{\rm d}{\rm dt}A(t)\|$, which can yield significant improvements over classical algorithms. Use cases that leverage this property, beyond unitary dynamics and where $A(t)$ can be rapidly changing, are unexplored. 

 \item Many problems in quantum simulation can be reduced to simulating time dynamics. Examples are the preparation of thermal Gibbs states or performing quantum phase estimation, which can be useful to compute properties in equilibrium. For these cases, claims on quantum advantage and the comparison to classical techniques are less clear.

    \item The quantum algorithm can be generalized to address weak non-linearities with methods such as Carleman linearization~\cite{liu2021efficient,krovi2023improved, jennings2025quantum}. However, their assumptions are often
    not met in practice. Open areas of research are then finding improved quantum algorithms, relaxing some assumptions while preserving efficiency, and finding practical applications where these algorithms are suitable. Note that for generic non-linear differential equations, efficient quantum algorithms cannot be expected without additional structural assumptions~\cite{liu2021efficient}.
    
\end{itemize}

\subsection{Classical Oscillators and Wave Equation}\label{sec:oscillators}

\emph{Maximum Quantum Speedup: Exponential}\\
\emph{Evidence Basis: Rigorous quantum algorithm, complexity-theoretic evidence for classical hardness}

\paragraph{Problem Description}

Imagine a network (graph) of $N$ coupled classical oscillators where the edges of the graph between nodes $i$ and $j$ represent a spring with spring constant $k_{ij}$ and the $i^{\textrm{th}}$ node of the graph represents a mass $m_i$. The dynamics of the system is described by the equations of motion governed by Hooke's Law: $m_i \frac{d^2 x_i(t)}{d t^2} = - \sum_{j=1}^{N} k_{ij} (x_i(t) - x_j(t))$. 
The quantum algorithm described in ref.~\cite{Oscillators23} propagates these equations forwards in time to produce a quantum state with amplitudes proportional to the weighted displacements $\sqrt{k_{ij}} (x_i(t) - x_j(t))$ and the weighted velocity of those oscillators $\sqrt{m_i} \dot{x}_i(t)$. One can sample these weighted displacements and velocities with a probability proportional to the potential or kinetic energies. 

\paragraph{Potential Application Spaces}

The problem of simulating coupled classical oscillators manifests in many areas of classical physics and engineering. Harmonic systems arise wherever a system in a conservative potential is perturbed only slightly. For example, the vibrations of atoms in a solid, circuits with inductors and capacitors, many first order mechanical models of structures and spans, discretizations of the classical wave equation, certain models of neuronal activity, acoustics, seismology, and more are in scope.

\paragraph{Classical Complexity}

The classical gate complexity of the best known methods for sampling these oscillators' momenta and displacements appears to scale as $\Omega(N \, t)$. The space complexity of those methods is $\Omega(N)$.
The specific example of determining whether a subset of the oscillators has total kinetic energy larger than some threshold value is known to be BQP-complete~\cite{Oscillators23}.

\paragraph{Quantum Complexity}

Suppose that the values $m_i$ are output by a quantum circuit applied to the input $i$ requiring ${\cal C}_m$ gates and the values $k_{ij}$ are output by a quantum circuit applied to the input $i,j$ requiring ${\cal C}_k$ gates. Then, up to polylog factors, the quantum algorithm of~\cite{Oscillators23} prepares the target state with amplitudes having error at most $\epsilon$ with space complexity $\widetilde{\cal O}(\log N)$ and gate complexity $\widetilde{\cal O}({\cal C}_m {\cal C}_k t \, \textrm{polylog}(N, 1/\epsilon))$.

\paragraph{Regimes of Advantage}

\begin{itemize}
\item The quantum complexity scales with ${\cal C}_k$ and ${\cal C}_m$, the size of quantum circuits required to output the value of the masses and spring constants. So if either of those circuits requires a number of gates that scales polynomially in $N$ it would kill the prospect for an exponential quantum speedup. This eliminates any application where a database of size $N$ is required to specify the masses and spring constants (for example, the vibrations of a protein having a random sequence of amino acids). But fortunately, the problem remains BQP-Complete for just two values of $k_{ij}$ and just a single value of the mass. One can easily imagine applications where there is spatial structure in the problem that gives rise to values of the spring constants and masses (e.g.,\ modeling a bridge where suspension beams have one value of $k_{ij}$ and the road has another value, etc.). Or one can imagine problems where it does not matter what the particular spring constants are, but what the distribution is (e.g., in neuronal models). An interesting special case is where the graph is coupled locally in a finite number of spatial dimensions and there is a low dispersion relation in the spring constants, in which case the simulation is related to simulating a discretized classical wave equation.
\item The cost of initial state preparation on the quantum computer is additive to the cost of time evolution. Thus, the initial state must have some structure. If one wishes to retain exponential speedup, the initial state must be preparable in time polylogarithmic in $N$.
\item If the initial state is local then one must choose $t$ sufficiently long that entanglement has propagated across the system. If the graph of the oscillator network is non-local (e.g.,\ fully connected) this is not much of a restriction. However, if the oscillator network is a local graph in $D$ dimensions then one must choose $t=\Omega(N^{1/D})$ to ensure this. Choosing $t = \Theta(N^{1/D})$ and assuming ${\cal C}_k$, ${\cal C}_m$, and initial state preparation all scaling polylogarithmically in $N$, one ends up with classical complexity $\widetilde{\cal O}(N^{(D+1)/D})$ and quantum complexity $\widetilde{\cal O}(N^{1/D})$, resulting in a speedup of only order $D+1$. Thus, in three dimensions $D=3$ we would have a quartic speedup but for two dimensions $D=2$ it would be only a cubic speedup.
\item A key challenge will be finding quantities that can be efficiently extracted from the quantum simulation. Because the quantum algorithm enables sampling from the output dynamics with probabilities proportional to the kinetic and potential energies of the oscillators, one must focus on extracting ``global'' quantities. For example, one can ask a question like ``is a large fraction of the energy in a particular subset of the modes?'' at the end of the simulation but one cannot ask a question like ``what is the displacement of a particular oscillator?'', unless one expects that oscillator to contain an inverse-polynomial fraction of the total system energy.
\item It is worth remarking that this algorithm seems most useful for simulating a \emph{very large} number of oscillator modes. For example, electrical grids are LC circuits, modeled by these equations. However, for the quantum computation to be much faster than a simulation of the quantum computation, one will likely need $N > 2^{50}$ and it seems unlikely that one would be interested in an electrical grid with that many elements. On the other hand, it is not hard to imagine a simulation of the classical wave equation discretized over that many modes.
\end{itemize}

\paragraph{Notes on Future Directions and Related Advantages}

\begin{itemize}
\item An important generalization of the coupled oscillators problem would include effects of dissipation or driving. One could imagine doing this in part by adding ancilla modes that either absorb or feed energy into the system. A partial solution to this is discussed in ref.~\cite{Krovi24}.
\item An important special case of the oscillators simulation is the simulation of the classical wave equation. This corresponds to simulating a network of oscillators that is coupled locally in $D$ dimensions. Other quantum algorithms for simulating the wave equation have been proposed that recover the same scaling as the oscillator-based approach~\cite{Costa2019}. Note that the approaches of~\cite{Costa2019} and~\cite{Oscillators23} are not fully equivalent in that they use different encodings which can have substantial implications (i.e.,\ the scaling can differ by exponential factors) for the preparation of certain initial states, or the extraction of certain quantities.
\end{itemize}

\pagebreak
\section{Optimization}

\emph{Section contributed by:} Ruslan Shaydulin, Shouvanik Chakrabarti

\subsection{Structured Exact Optimization}\label{sec:shortpath}

\emph{Maximum Quantum Speedup: Polynomial}\\
\emph{Evidence Basis: Rigorous quantum algorithm with analytical and numerical comparisons against restricted classical methods}

\paragraph{Problem Description}

This entry is concerned with structured exact optimization with short-path guiding states. 
Despite the name, short-path algorithms are not algorithms for the graph shortest path problem but quantum algorithms for exact discrete optimization. The problem addressed is that of \emph{combinatorial optimization}: minimize a cost function $H(x)$ over a finite but large set of feasible configurations $\mathcal X$. The goal is to find an exact global minimizer $x^\star$. In the original setting of Hastings and later Dalzell, Pancotti, Campbell, and Brand\~ao, $\mathcal X=\{\pm 1\}^n$ and the natural local geometry is the hypercube bit-flip walk~\cite{Hastings2018,DalzellPCB23}. In the generalized setting, $\mathcal X$ is equipped with a reversible Markov chain $P$ with stationary distribution $\pi$, and the mixer is the discriminant matrix $D(P)$ of that chain~\cite{ChakrabartiEtAl24}.

A specific setting corresponds to \emph{unconstrained} problems such as max-CSP, spin-glasses, and max-SAT. The generalized setting accommodates problems where $\mathcal{X}$ is a non-trivial combinatorial set specified by constraints. These include the Maximum Independent Set problem, as well as the computation of an algebraic relaxation of Integer Linear Programs known as the Gomory Group Relaxation~\cite{AugustinoEtAl26GroupRelaxation}.

\paragraph{Potential Application Spaces}

Combinatorial optimization algorithms have broad real-world applications across logistics, scheduling, network design, statistical inference, and operations research. MAX-$k$-CSP and MAX-E$k$-LIN abstract the task of finding solutions that satisfy as many combinatorial constraints as possible, making them useful abstractions for problems in resource allocation, circuit design, scheduling, and consistency checking in databases or distributed systems. Maximum Independent Set models the problem of selecting a largest collection of mutually non-conflicting items, with applications in wireless network scheduling, portfolio or resource selection under incompatibility constraints, register allocation in compilers, identifying non-overlapping tasks or entities in conflict graphs, and structural analysis of graphs that model social media networks and financial transaction data. 
Ground-state problems for antiferromagnetic Ising models arise naturally in statistical physics, but also connect to probabilistic graphical models, Bayesian inference, and image segmentation. 
Integer linear programs (ILPs) are even more broadly used to formulate practical decision problems in supply-chain optimization, transportation, production planning, finance, energy systems, and workforce scheduling. It has been found that for several ILPs of practical importance the Gomory group relaxation provides a better relaxation than continuous LP relaxations, for methods such as branch-and-bound that use a tree of incrementally restricted relaxations to solve ILPs exactly.

\paragraph{Classical Complexity}

The natural baseline for the generalized framework is Markov chain search: sample from $\pi$ until an optimum is observed. Up to polynomial factors this costs $1/\pi(E^\star)$, where $\pi(E^\star)$ is the stationary probability mass on optimal configurations. For the uniform hypercube this reduces to exhaustive search over $2^n$ assignments, or $2^n/|\mathcal X^\star|$ if there are multiple optima. For group relaxations of integer programs, the analogous local-search cost is $|K|/|K^\star|$, where $K$ is the kernel subgroup and $K^\star$ is the set of optimal kernel elements.

This baseline is not always the best classical algorithm. For standard problems such as SAT, maximum independent set, and spin-glass optimization, specialized exponential-time algorithms may beat naive search by a large constant in the exponent.

In many applications of combinatorial optimization, it is desired to solve the problems exactly using exponential time algorithms such as branch-and-cut and lazy constraint generation, or heuristics such as simulated annealing. These algorithms may be favored over polynomial time approximation algorithms. This arises due to the moderate size of problems encountered in practice (often in the range of 100-1000 variables), and the often mild exponential dependence of algorithms for practical problems. For example, it is believed that there is no polynomial time algorithm for Maximum Independent Set that offers a better than 1/2 approximation for Erd\H{o}s-R\'{e}nyi random graphs, while numerical investigations of branch-and-bound solvers indicate a ``mild'' $2^{0.03 n}$ exponential dependence. The short-path algorithms operate in this exact solving regime.

\paragraph{Quantum Complexity}

The quantum algorithm then has three conceptual steps: First, prepare the easy state $|\sqrt\pi\rangle$, the ground state of the Hamiltonian $-D(P)$. Second, use phase estimation together with amplitude amplification to jump to the ground state $|\psi_b\rangle$ of $H_b=-D(P)+b\,g_\eta(H/|E^\star|)$, where $E^\star$ is the optimal value and $g_\eta$ is a truncated cost function that keeps only a well near the optimum. The parameter $b$ is chosen so that the ground state $|\psi_b\rangle$ is still close enough to the initial state $|\sqrt\pi\rangle$ to be prepared efficiently, but already has larger overlap with the optimal configurations than $|\sqrt\pi\rangle$ itself. 
Third, project from $|\psi_b\rangle$ onto the optimal subspace of the classical Hamiltonian $H$. If the first jump has only polynomial cost and the optimum overlap obeys
\begin{align*}
\|\Pi^\star |\psi_b\rangle\|^2 \gtrsim \pi(E^\star)^{1-c},
\end{align*}
then the overall runtime scales as $\widetilde O(\pi(E^\star)^{-1/2+c/2})$ up to polynomial factors. In the uniform hypercube case this is written as $O^*(2^{(1/2-c')n})$ for a related constant $c'>0$.

The rigorous constants currently proved for the original CSP settings are small, but numerical studies suggest that the useful values of $b$ can be much larger than the proof allows. For example, in random 3-spin instances the inverse overlap was fit to roughly $2^{0.43n}$ over the numerically accessible range, instead of the Grover-like $2^{0.5n}$ scaling~\cite{DalzellPCB23}. In the generalized framework, constrained MaxCut-type numerics similarly find overlap exponents below the square-root-search exponent for some choices of $b$~\cite{ChakrabartiEtAl24}, achieving a near-cubic speedup over brute-force-search over the in-constraint subspace. 

Quantum short-path algorithms are likely to provide speedups over the state of the art in settings where \emph{local search} algorithms are competitive with the best classical algorithms. The local search algorithms can be unstructured search, as in the case of~\cite{DalzellPCB23}, or more generally, Markov Chain Search~\cite{ChakrabartiEtAl24}, where the search is guided by the stationary state of an efficiently samplable Markov Chain. Examples of these regimes include MAX-$k$-CSP and MAX-E$k$-LIN, where under the Strong Exponential Time Hypothesis, no classical algorithm can improve over brute-force search in the limit of large $k$. 
The super-quadratic speedup provided by Dalzell for these problems was therefore over the best classical algorithms known at the time. Le Gall and Tamaki have since developed a classical search-and-condition procedure that obtains similar improvements, leaving the status of super-quadratic speedups for these problems in question~\cite{LeGallTamaki26}. In the generalized setting, super-quadratic speedups have been shown to exist over Markov Chain Search algorithms for finding the ground states of the hardcore model (or Maximum Independent Set), and the antiferromagnetic Ising model on graphs with bounded degree. Particularly, for Maximum Independent Set on regular graphs, a setting where classical algorithms have been studied extensively, the short-path algorithm provides a super-quadratic speedup over all known algorithms when the graph degree is sufficiently high. Finally, we note that the restriction to regular graphs here is due to the need to analytically determine the scaling of Markov Chain search and the speedup offered by the short-path algorithm does not require regularity.

The short-path algorithm also provides a speedup for computing the \emph{group relaxation} of integer linear programs (ILPs). The group relaxation is generally exponential time to compute but is known to provide significant downstream reductions in the complexity of ILP solvers for some classes of problems. The short-path algorithm provides a super-quadratic speedup over a classical algorithm that searches (potentially guided by Glauber dynamics at positive temperature) over all feasible solutions of the relaxation. In the most general case, this classical algorithm is the best possible.

\paragraph{Regimes of Advantage}

\begin{itemize}
\item The local search geometry must be efficiently usable. The chain $P$ should have efficiently computable transitions, an efficiently preparable coherent stationary state $|\sqrt\pi\rangle$, and an efficient block-encoding of $D(P)$. If specifying or implementing the chain costs exponential time, the advantage is lost.
\item The short-path Hamiltonian must remain in a pre-localized regime for a useful value of $b$. Equivalently, $H_b$ should have an inverse-polynomial spectral gap and the state $|\psi_b\rangle$ should retain non-negligible overlap with $|\sqrt\pi\rangle$. If the cost well creates a competing localized state too early, the short jump becomes hard.
\item The guiding state must increase the overlap with optima by an exponential factor. In the hypercube analyses this is controlled by depolarizing or subdepolarizing properties of the cost function. In the Markov-chain formulation it is controlled by $\Delta_P$-stability or by a small $P$-pseudo-Lipschitz norm.
\item The low-energy threshold set should not by itself give an equally strong classical search method. If near-optimal points contain a large explicit Hamming shell, restricted ball, or other easily searchable region around an optimum, then the same structure may yield a classical conditioning-and-search algorithm. This is the main lesson of the dequantization result for the original CSP proof regimes~\cite{LeGallTamaki26}.
\item For constrained problems, the mixer should preserve the feasible set whenever possible. Penalty encodings often enlarge the search space and make the relevant stability parameter scale with the penalty, which can remove the super-quadratic regime. This is why transposition, Glauber, and Cayley walks are more natural than unconstrained bit-flip walks for the constrained ex\-amples.
\end{itemize}

Existing results establish advantages over specified classical baselines and identify promising concrete families, but do not yet establish a super-quadratic advantage against the classical state of the art on an efficiently generable average-case family. Establishing such a family is the principal remaining step toward Stage II.

\paragraph{Notes on Future Directions and Related Advantages}

\begin{itemize}
\item Recent work of Le Gall and Tamaki~\cite{LeGallTamaki26} shows that the threshold-set structure used in the Hastings--Dalzell analyses for MAX-E$k$-LIN2 and MAX-$k$-CSP~\cite{DalzellPCB23} also yields classical conditioning-and-local-search algorithms with better proven exponents than the published short-path bounds. 
The new search-and-condition algorithms of Le Gall and Tamaki do not directly dequantize the short-path guiding state, and do not apply directly to the generalized short-path framework of Chakrabarti et al.~\cite{ChakrabartiEtAl24}. It remains to be investigated whether these limitations are essential, or whether the short-path algorithm can indeed be fully dequantized. Conversely, it is possible that combining short-path techniques with the classical methods of Le Gall and Tamaki results in even larger speedups. 
\item The regimes of speedup must be validated more thoroughly, and supported by numerical evidence. Based on numerical investigations~\cite{DalzellPCB23,ChakrabartiEtAl24} the actual speedup offered by the algorithms is larger than that predicted by theory, and there may be polynomial speedups in settings where an end-to-end proof is not yet possible. A more significant benchmarking against classical methods is also required to comment on advantage for typical, rather than worst-case or general, instances.
\end{itemize}

\pagebreak
\section{Topological Properties}

\emph{Section contributed by:} Matthias Christandl, Pietro M. Posta, Robbie King

\subsection{Jones Polynomial}

\emph{Maximum Quantum Speedup: Exponential}\\
\emph{Evidence Basis: Rigorous quantum algorithm, complexity-theoretic evidence for classical hardness}

\paragraph{Problem Description}

The Jones polynomial is a knot invariant, which is known to be \#P-hard to exactly compute for arbitrary links~\cite{JaegerVertiganWelsh1990}. Still, it is a sensible question to ask whether an algorithm could efficiently approximate it.

Given a link $L$, typically presented either as the \emph{trace closure} or the \emph{plat closure} of a braid $\beta \in B_n$ on $n$ strands with $m$ crossings, and an integer $k$, the task is to compute (an approximation of) the value of the Jones polynomial $V(L, t)$ at the specific root of unity $t = e^{2\pi i/k}$.
The braid $\beta$ is provided as a product of the braid group generators, corresponding to swaps of adjacent strands.

Quantum computing can be of help here, based on the observation that, when evaluated on a root of unity, the Jones polynomial of the link corresponding to the closure of a braid $\beta\in B_n$ can be seen as (up to some known proportionality constant) either the Markov trace (trace closure version) or a specific matrix entry (plat closure version) of the matrix corresponding to $\beta$ in a \emph{unitary} representation of the braid group $B_n$. 
Unitarity is crucial here, as it allows us to map the generators of $B_n$ to quantum circuits, and (linearly in $m$) construct a quantum circuit implementing the representation of $\beta$.

\paragraph{Potential Application Spaces}

Knot theory is a foundational mathematical subject which resides in topology and geometry, drawing in particular on algebra when it comes to knot invariants. It has applications in a range of areas, notably in physics through statistical physics, quantum field theory and condensed matter physics~\cite{Kauffman2001,witten1989quantum}, and in chemistry and biology through tangled chains of polymers or DNA strands~\cite{Du95,Deweese2008, Micheletti2011polymers}.

The mathematical tools developed in order to understand the complexity of knots and their invariants, and in particular of the Jones polynomial, have turned out to be useful for the field of computing itself, most prominently through the development of the topological model of quantum computation~\cite{FreedmanP/NP, Lahtinen_2017}, which is closely tied to an experimental effort, as well as a new type of algorithm (approximating the Jones polynomial) and complexity theory~\cite{AharonovJonesLandau2009, freedman2000modularfunctoruniversalquantum}.
Since any quantum computation can be reduced to the approximation of the Jones polynomial of the plat closure of a braid, any further development on this problem would lead to consequences for the field of quantum computing as a whole.

\paragraph{Classical Complexity}

It is not entirely clear what the best classical algorithm for approximating the Jones polynomial is; however it is known that the problem of calculating an exact evaluation (and in fact even any \emph{value-distinguishing} approximation~\cite{kuperberg-ToC}) of the Jones polynomial is \#P-hard~\cite{JaegerVertiganWelsh1990}, even when restricting to alternating links.

Various methods for evaluating the Jones polynomial have been developed, either via direct expansion (such as via \emph{skein relations}) or by numerically computing the action on the \emph{skein space}, essentially resulting in the classical simulation of the best quantum algorithms. While both these families of algorithms exhibit exponential complexity in the input parameters, a heuristic argument given in ref.~\cite{kuperberg-ToC} suggests that this second class (in ref.~\cite{kuperberg-ToC} referred to as \emph{Morse} algorithms), which exhibits exponential complexity in $n$, the number of strands, but crucially not in $m$, the number of crossings, may be essentially optimal for classical computation. The reason behind this conjecture is that the evaluation of the Jones polynomial of a braid can encode any planar quantum circuit. Assuming certain search problems associated with linear-depth reversible planar (classical) circuits require exponential time on a classical computer, this problem
can be encoded in a braid such that the relevant input size would translate (up to some polynomial factor) to the number of strands $n$ of the braid. The problem of evaluating the Jones polynomial of this braid would then be equivalent to solving the search problem for the circuit, suggesting that it would require exponential time in $n$ via a classical algorithm.

\paragraph{Quantum Complexity}

The first efficient quantum algorithm for this specific problem was implicitly described in refs.~\cite{FreedmanKitaevWang2002, freedman2000modularfunctoruniversalquantum}, and then explicitly constructed in ref.~\cite{AharonovJonesLandau2009}. The algorithm works as follows:
a quantum computer can efficiently implement the action of each generator of the braid group (a swap of two adjacent strands) as a single quantum circuit acting on some high (exponential) dimensional Hilbert space. The size of the total quantum circuit associated with the representation of the input braid $\beta$ is therefore essentially linear in the number of crossings, $m$, and polynomial in the other parameters (number of strands, $n$ and specific root of evaluation, $k$).
The algorithm then implements the \emph{Hadamard test} to statistically approximate either the Markov trace (in the trace closure version) or a specific entry (in the plat closure version) of the matrix representation of the braid element, providing an \emph{additive} approximation (and therefore not subject to the hardness results in ref.~\cite{kuperberg-ToC}) of the Jones polynomial.
In particular, for the trace closure version of the problem, the algorithm, with an exponentially small (in $n,\ k$ and $m$) error probability, outputs a complex number $r$ such that \[
\left| r-V(L,t) \right|< \varepsilon d^{n-1},
\]
where $t:=e^{2\pi i/k},\ d:=-t^{1/2}-t^{-1/2}$ and $\varepsilon$ is inversely polynomial in $n,\ k$ and $m$. A similar additive approximation holds for the plat closure version.
The problem of getting additive approximations of the Jones polynomial is itself known to be either DQC1-hard (trace closure version) or BQP-hard (plat closure version)~\cite{shor2008estimatingjonespolynomialscomplete, AharonovJonesLandau2009}; therefore, the quantum algorithm provides a completeness result for these problems.

\paragraph{Regimes of Advantage}

\begin{itemize}
    \item The quantum algorithm works when the evaluation point of the Jones polynomial is a principal $k$-th root of unity. Since there exist efficient classical algorithms when $k\in\{1,2,3,4,6\}$, the advantage is obtained on non-lattice (i.e.,\ $k=5$ or $k\ge 7$) roots of unity.
\end{itemize}

While the hardness result for additive approximations shows worst-case time advantages (unless BPP=BQP), we would be interested in showing that for a general, random braid, its corresponding Jones polynomial is hard to compute (or approximate) classically. In this regard, as discussed earlier, a heuristic argument given in ref.~\cite{kuperberg-ToC} suggests that the best possible classical algorithm should be essentially equivalent to a classical simulation of the quantum algorithm, numerically computing the action of the braid element. Combining this with the fact that representations of sufficiently long random braids form an $\varepsilon$-approximate unitary $2$-design~\cite{mann2017complexityrandomquantumcomputations}, we would get that the best classical algorithm should essentially compute either the trace or a specific entry of an exponentially large, (close to) random, unitary matrix, strongly suggesting that for random inputs the problem is classically hard.
Moreover, based on a numerical analysis, ref.~\cite{QuantinuumJones} suggests a modified version of the quantum algorithm may result in a speedup over the best current classical algorithms on random braids with $\ge 2800$ crossings for 
the trace closure version of the problem, with
evaluation on fifth roots of unity.

\paragraph{Notes on Future Directions and Related Advantages}
\begin{itemize}
    \item Efficient quantum algorithms have been proposed for other topological invariants and related quantities, such as the Tutte polynomial~\cite{aharonovTutte2007}, other families of generalizations of the Jones polynomial~\cite{wocjanyard_2008}, Khovanov homology~\cite{schmidhuber2025} and Betti numbers (see Sec.~\ref{sec:TDA}).
    \item The quantum algorithm for approximating the Jones polynomial exemplifies the power of quantum computers to efficiently compute group-theoretic quantities. This advantage originates from the ability to implement unitary representations of said groups directly on a quantum computer, thereby saving time when compared to a classical simulation, as well as providing a new model of computation. Considering the symmetric group, for instance, leads to the permutational quantum computer~\cite{jordan2010permutational}.
    \item Recently, the quantum computational complexity of other classically hard representation-theoretic quantities, such as the Kronecker and plethysm coefficients, has been investigated \cite{christandl2015algcomp, bravyi2024, christandl2026plethysmbqp}, pointing to possible quantum advantage. 
\end{itemize}

\subsection{Topological Data Analysis}
\label{sec:TDA}
\emph{Maximum Quantum Speedup: Superpolynomial}\\
\emph{Evidence Basis: Rigorous quantum algorithm against the classical state of the art}

\paragraph{Problem Description}

Many real-world datasets are noisy samples from some high-dimensional manifold. Given such a dataset, there are a variety of ways to ask about the topology of the underlying manifold; this is the realm of \emph{topological data analysis} (TDA)~\cite{wasserman2018topological}. The Betti numbers $\{\beta_k\}$ are a topological invariant of particular interest, which intuitively encode the number of $k$-dimensional holes in the dataset. The problem then becomes: what is the complexity of computing $\beta_k$ from the initial classical dataset? Ref.~\cite{lloyd2016quantum} first proposed a quantum algorithm for this task, which was developed in refs.~\cite{gunn2019review, hayakawa2022quantum, mcardle2022streamlined, berry2024analyzing}.

\paragraph{Potential Application Spaces}

One of the most exciting applications of topological data analysis is to analyze neural networks in the brain~\cite{petri2014homological,giusti2016two,reimann:hal-01706964}, where higher-order connections could be important for certain functions in the brain. Due to the high degree of connectivity in the brain, high-dimensional Betti numbers could become relevant. As discussed below, the high-dimensional regime increases the margin of quantum advantage for TDA.

Another application is to probe large-scale cosmic structure~\cite{van2011alpha}. The galaxies are distributed according to a complicated web, and understanding this distribution in detail is a major problem in cosmology~\cite{chen2015cosmic}. However, the distribution of galaxies is fundamentally three-dimensional, which may not enter the regime of quantum advantage.

TDA has been applied to image recognition to extract global features that may not be visible to convolutional techniques~\cite{li2014persistence,carriere2015stable,bonis2016persistence}. Of course, the image recognition problem today has so much training data available that it is difficult to improve on existing machine learning methods.

Last but not least, TDA has been applied to analyze the structure of proteins~\cite{kovacev2016using}. Protein folding and protein binding are complex processes which underlie much of biology, and a better understanding would have great impact in drug design and medicine. Perhaps non-local topological features of proteins are important in determining their folding properties.

But in all cases, one would need to provide good evidence that the high order Betti number is extremely large in order to argue for a speedup, and also that the combinatorial Laplacian gap is not too small. It was shown in ref.~\cite{berry2024analyzing} that when the edge probability is chosen to maximize the Betti number of Erd\H{o}s-R\'enyi random graphs, then the Betti number is sufficiently large that the quantum algorithm gives an overall quartic speedup relative to classical algorithms.

\paragraph{Classical Complexity}

The classical complexity of computing $\beta_k$ scales roughly as $\mathcal{O}(N^k)$ where $N$ is the size of the data, reflecting the combinatorial growth in the number of simplices that must be considered.

\paragraph{Quantum Complexity}

The quantum speedup relies on the observation that the homology can be encoded as the groundspace of an operator---the combinatorial Laplacian---on an $N^k$-dimensional Hilbert space. In time
\begin{align*}
    T = O\left(\frac{N^3}{\epsilon \delta} \sqrt{\frac{N^k}{\beta_k}}\right)
\end{align*}
we can compute $\beta_k$ to multiplicative error $\epsilon$. Here $\delta$ is the spectral gap of the combinatorial Laplacian. It was shown in ref.~\cite{berry2024analyzing} that this gap could be constant as $N$ scales.

Notice the dependence on $N^k / \beta_k$. If $\beta_k$ is small, say constant, then $T \sim N^{k/2}$ and the speedup is merely quadratic. However, it was shown in ref.~\cite{berry2024analyzing} that $\beta_k$ could be as large as a constant times $N^k$, in which case $T \sim N^3$ and there is a large polynomial speedup. Moreover, in some applications $k$ could also scale with $N$, in which case we could reach a superpolynomial speedup. Note that if one seeks to compute $\beta_k$ to a fixed additive error then the speedup is at most quadratic~\cite{berry2024analyzing}.

\paragraph{Regimes of Advantage}

The quantum speedup is largest when the following conditions are met. 
\begin{itemize}
    \item The speedup is largest when we are computing a very high dimensional Betti number $\beta_k$. If the dimension $k$ is increasing with the dataset size $N$, we can hope for superpolynomial speedups. 
    \item The spectral gap of the combinatorial Laplacian must not be too small. For example, it should be inverse polynomial in the size of the dataset, ideally with no dependence on the dimension $k$. 
    \item The $\beta_k$ should be very large, ideally comparable to $N^k$. This involves the dataset in question having \emph{many} holes of dimension $k$.
    \item When translating estimates of normalized Betti numbers to additive error into a fixed additive-error estimate of the unnormalized Betti number, the required precision can become exponentially small in the problem size. Consequently, the presently identified regimes of super-quadratic quantum advantage for estimating unnormalized Betti numbers rely on relative (multiplicative) error and on Betti numbers that grow asymptotically.
\end{itemize}
The first condition is dependent on the particular application domain. Ref.~\cite{berry2024analyzing} gives examples where the second and third conditions are met. We also note that refs.~\cite{berry2024analyzing,schmidhuber2022complexity} show that whenever targeting a fixed additive error (as opposed to a fixed multiplicative error) in the Betti number, the speedup is only quadratic.

\paragraph{Notes on Future Directions and Related Advantages}

\begin{itemize}
\item Many applications of topological data analysis are concerned with \emph{persistent homology}~\cite{edelsbrunner2013persistent}, which investigates how the topology changes as we vary the length scale at which we view the dataset. Are there quantum algorithms which achieve large speedups on this problem?
\end{itemize}

\clearpage
\pagebreak 
\part*{Acknowledgements}
The concept for this project developed initially through conversations between Catherine Lefebvre and Ryan Babbush, as well as Elica Kyoseva, Barry C.\ Sanders and Matthias Troyer, as part of their responsibilities as Advisory Committee members for the Open Quantum Institute (OQI). 
The OQI---an initiative born at GESDA, hosted at CERN, with the support of UBS---gratefully acknowledges all members of its Advisory Committee for their valuable support throughout the pilot phase, 
with particular appreciation for the scientific expertise and contributions of R. Babbush, V. Bu\v{z}ek, P. Chomaz, M. Christandl, G. Crossman, R. Fazio, E. Kyoseva, S. Odeh, P. Osei, F. Petruccione, B. Sanders, V. Savona, U. Sinha, J. Thiele, and M. Troyer.
TOB thanks D. Wild for feedback on the Quantum simulation section.
PK thanks D. Sutter for insightful discussions and feedback on an early outline of the project. SJ was partially supported by the Amazon AI Fellowship.

\clearpage

\bibliographystyle{unsrt}
\bibliography{references_master}

\part*{Disclaimer}
During the preparation and revision of this manuscript, the authors used generative-AI tools, including OpenAI ChatGPT, Anthropic Claude, and Google Gemini, to obtain critical feedback on structure, exposition, and selected technical claims. All suggestions were independently reviewed and verified against the cited literature, and the authors take full responsibility for the final content.\\

This paper was prepared for informational purposes with contributions from the Global Technology Applied Research center of JPMorgan Chase \& Co. This paper is not a product of the Research Department of JPMorgan Chase \& Co. or its affiliates. Neither JPMorgan Chase \& Co. nor any of its affiliates makes any explicit or implied representation or warranty and none of them accept any liability in connection with this paper, including, without limitation, with respect to the completeness, accuracy, or reliability of the information contained herein and the potential legal, compliance, tax, or accounting effects thereof. This document is not intended as investment research or investment advice, or as a recommendation, offer, or solicitation for the purchase or sale of any security, financial instrument, financial product or service, or to be used in any way for evaluating the merits of participating in any transaction.

\end{document}